# Bipolar Ephemeral Active Regions, Magnetic Flux Cancellation, and Solar Magnetic Explosions

Ronald L. Moore[1,2], Navdeep K. Panesar[3,4], Alphonse C. Sterling[2], Sanjiv K. Tiwari[3,4]

[1]Center for Space Plasma and Aeronomic Research (CSPAR), UAH, Huntsville AL, 35805 USA; ronald.l.moore@nasa.gov

[2]NASA Marshall Space Flight Center, Huntsville, AL 35812, USA

[3]Lockheed Martin Solar Astrophysics Laboratory, 3251 Hanover Street Building 252, Palo Alto, CA 94304, USA

[4]Bay Area Environmental Research Institute, NASA Research Park, Moffett Field, CA 94035, USA; panesar@baeri.org

## Abstract

We examine the cradle-to-grave magnetic evolution of 10 bipolar ephemeral active regions (BEARs) in solar coronal holes, especially aspects of the magnetic evolution leading to each of 43 obvious microflare events. The data are from *Solar Dynamics Observatory*: 211 Å coronal EUV images and line-of-sight photospheric magnetograms. We find evidence that (1) each microflare event is a magnetic explosion that results in a miniature flare arcade astride the polarity inversion line (PIL) of the explosive lobe of the BEAR's anemone magnetic field; (2) relative to the BEAR's emerged flux-rope Ω loop, the anemone's explosive lobe can be an inside lobe, an outside lobe, or an inside & outside lobe; (3) 5 events are confined explosions, 20 events are mostly-confined explosions, and 18 events are blowout explosions, which are miniatures of the magnetic explosions that make coronal mass ejections (CMEs); (4) contrary to the expectation of Moore et al (2010), none of the 18 blowout events explode from inside the BEAR's Ω loop during the Ω loop's emergence; (5) before and during each of the 43 microflare events there is magnetic flux cancellation at the PIL of the anemone's explosive lobe. From finding evident flux cancellation at the underlying PIL before and during all 43 microflare events – together with BEARs evidently being miniatures of all larger solar bipolar active regions – we expect that in essentially the same way, flux cancellation in sunspot active regions prepares and triggers the magnetic explosions for many major flares and CMEs.

## 1. Introduction and Background

### 1.1. BMRs and BEARs

All of the magnetic field on the Sun evidently comes from Ω-loop stitches in horizontal flux ropes that are somehow generated in and/or at the bottom of the convection zone by the convection and the sunspot-cycle global dynamo process (Zwaan 1987; Fan 2009; Moore et al 2016, 2020; Charbonneau 2020). Each new bipolar magnetic region (BMR, alternatively called a bipolar active region) is evidently made by the emergence of a flux-rope Ω loop of the size of the BMR (Zwann 1987; van Driel-Gesztelyi & Green 2015; Moore et al 2020). The Ω-loop field emerges from below the photosphere and balloons up through the chromosphere and into the corona (van Direl-Gesztelyi & Green 2015). The continual evolution of the Sun's magnetic field in and above the photosphere is the consequence of emergence of new Ω-loop field, movement of the field's feet by convection in and below the photosphere, and flux-cancellation submergence of the field at polarity inversion lines (PILs) between converging opposite-polarity flux (e.g., Moore & Rabin 1985; Zwann 1987; van Ballegooijen & Martens 1989). By these processes, during and after emergence, the magnetic flux of a BMR Ω loop gradually spreads and gradually mixes and cancels with itself and with ambient flux until the BMR/active region is no longer distinguishable from its surroundings in photospheric magnetograms and in chromospheric and coronal images (e.g., Rabin et al 1984; van Driel-Gesztelyi & Green 2015). Thus, the empirical lifetime of a BMR is the time from when its emerging magnetic field first becomes discernible in magnetograms until the BMR/active region is no longer discernible (e.g., van Driel-Gesztelyi & Green 2015).

From the review of the evolution of solar bipolar active regions by van Driel-Gesztelyi & Green (2015), we expect that most BMRs that have magnetic explosions that make major flares and coronal mass ejections (CMEs) are large enough (have enough magnetic flux) to have one or more full-fledged umbra-with-penumbra sunspots when the flux-rope Ω loop finishes emerging. For BMRs that have full-fledged sunspots, the Ω loop's magnetic flux (the flux of either polarity in the BMR at the completion of emergence) is at least $5 \times 10^{21}$ Mx, and can be as much as $3 \times 10^{22}$ Mx (Wang & Sheeley 1989; van Driel-Gesztelyi & Green 2015). The evolution of a BMR's magnetic flux can be well tracked in detail in magnetograms only when the BMR is far enough from the limb that magnetogram degradation by near-limb projection effects is negligible, no more that about 45 heliocentric degrees from disk center (Falconer et al 2016). The lifetimes of BMRs large enough to have full-fledged sunspots range from weeks for the smallest of these BMRs to months for the largest (van Driel-Gesztelyi & Green 2015). So, for a BMR big enough to be expected to possibly have major CME/flare explosions, the evolution of its magnetic flux in present-day magnetograms (taken from on or near Earth) cannot be well tracked over the BMR's entire life, but only for the 1-week intervals in which the Sun's rotation carries the BMR across the central solar disk within about 45° of disk center.

The smallest sunspots are called pores. A pore is evidently a sunspot umbra that is too small (less than about 4,000 km in diameter) to have a penumbra, an outskirt that is brighter than the umbra but darker than the ambient photosphere (Bruzek 1977). BMRs for which the emerged Ω loop's magnetic flux is in the range $1 \times 10^{20}$ – $5 \times 10^{21}$ Mx are big enough (have enough flux) to have pores but too small to have full-fledged sunspots (van Driel-Gesztelyi & Green 2015). BMRs near the small-flux end of the flux range of BMRs with pores have flux of $1 \times 10^{20}$ – $3 \times 10^{20}$ Mx in the emerged Ω loop. From Table 1 of van Driel-Gesztelyi & Green (2015) and Figure 2 of Wang & Sheeley (1989), we estimate that these smaller of the pore-bearing BMRs typically have a lifetime of a few days, and a horizontal span of 20,000 – 30,000 km. So, the smaller of the pore-bearing BMRs typically span a large fraction of the 35,000 km span of a typical supergranule convection cell (e.g., Beckers 1977).

The flux in the emerged flux-rope **Ω** loop for a BMR that is too small to have a pore is less than $10^{20}$ Mx. Spotless BMRs having emerged-**Ω**-loop flux in the range 3 x $10^{18}$ – 1 x $10^{20}$ Mx are usually called ephemeral regions (Martres & Bruzek 1977; van Driel-Gesztelyi & Green 2015). The ten BMRs studied in this paper are ephemeral regions. They have no pores or larger sunspots. At the completion of emergence, the flux of either polarity in each of them is in the range $10^{19}$ - $10^{20}$ Mx, and the BMR spans about 10,000 km, less than half the diameter of a supergranule. As our BMRs illustrate, BMRs of this size have lifetimes of about a day (van Driel-Gesztelyi & Green 2015). Due to its short lifetime, the entire cradle-to-grave magnetic evolution of a BMR of this size that emerges on the central disk can be well tracked in magnetograms.

In this paper, we call ephemeral-region BMRs "bipolar ephemeral active regions" (BEARs) instead of the usual name "ephemeral regions," to emphasize that we take these little spotless BMRs to be instructive miniatures of the larger bipolar active regions that have full-fledged sunspots and sometimes have major CME/flare explosions. In our view, if the magnetic evolution of BEARs and the magnetic evolution of all larger BMRs are governed by physical processes that act the same way in BMRs of all sizes, then pre-magnetic-explosion magnetic evolution that we observe in BEARs could reveal the essence of the large-active-region magnetic evolution that leads to major flares and CMEs.

### 1.2. The Motivation and Strategy for Our Investigation

A CME and its source-region arcade of hot flare loops are made by a blowout explosion of closed magnetic field that is initially in the chromosphere and corona and is rooted in the photosphere (e.g., Rust et al 1980; Moore & LaBonte 1980; Kahler 1992; Moore et al 2001, 2007; Patsourakos et al 2020). The pre-explosion field is basically an arcade that arches over the PIL between two abutting opposite-polarity domains of photospheric magnetic flux. The arcade's core is its inner field rooted close to the PIL. Because the pre-explosion core field is greatly deformed from its completely relaxed zero-free-energy potential field configuration, it has a large store of free magnetic energy, ample for the whole arcade to undergo a CME/flare blowout eruption (e.g., Hagyard et al 1984; Moore 1988; Moore & Roumeliotis 1992; Canfield et al 1999). The core field is strongly sheared: its horizontal direction is nearly along the PIL instead of nearly orthogonal as it would be if the field were nearly potential. While it is well established that the usual pre-explosion field configuration for CME/flare explosions is an arcade with strongly sheared core field, how the Sun builds that configuration remains a controversial open question (Patsourakos et al 2020). This question is the fundamental motivation for the investigation reported here.

BEARs are near – but not at – the small-size, short-life, high-population end of the continuous size spectrum of solar BMRs. The smallest, briefest, and most-numerous BMRs emerge in about one out of ten granule convection cells (Ishikawa et al 2008). Each has magnetic flux ~ 5 x $10^{17}$ Mx in its emerged flux-rope **Ω** loop, and has about the span (~ 1000 km) and lifetime (~ 5 min) of the granule in which it emerges (Lites et al 2008; Ishikawa et al 2008; Ishikawa & Tsuneta 2010). The BMR grows in step with the emerging granule to span the granule as the emergence of both the BMR and the granule is completed. The observed progression gives the impression that the emerging BMR Ω loop is a flux-rope stitch that is formed in and by the MHD flow in the granule convection cell (Moore et al 2020).

A BMR at the large-size, long-life, low-population opposite end of the BMR size spectrum has, in each of its two opposite-polarity flux domains at the end of emergence, one or more full-fledged sunspots. The BMR has ~ $10^{22}$ Mx of magnetic flux in its newly emerged flux-rope Ω loop, spans ~ 200,000 km, and lasts for months (Wang & Sheeley 1989; van Driel-Gesztelyi & Green 2015). In the Miesch et al (2008) numerical hydrodynamic simulation of the global convection zone, the vertical extent of the largest convection cells, so called giant cells, is the entire 200,000 km vertical extent of the convection zone. The giant cells have horizontal diameters that range from ~ 100,000 km to ~ 200,000 km. That is, the simulated convection zone's largest giant cells and the largest observed BMRs have the same horizontal span, ~ 200,000 km.

Recently, Moore et al (2020) pointed out that the flux-rope Ω loop for granule-size BMRs appears to be made by the granule in which it emerges, and that the largest BMRs observed have the span of the largest giant cells in the Miesch et al (2008) simulated convection zone. In addition, Moore et al (2020) pointed out the following three things. First, for BMRs of all sizes, across the entire 1000 km – 200,000 km range of the BMR size spectrum, the emerged Ω-loop magnetic arch typically has more or less the same 3D form. The magnetic arch is an elongated dome covering a roughly elliptical area that is about twice as long as it is wide (Fan 2009; Ishikawa et al 2010; van Driel-Gesztelyi & Green 2015). Second, there is evidently a continuous size spectrum of continually evolving convection cells, ranging from granules (~ 1000 km across) to the largest giant cells (~ 200,000 km across), the tops of all of which are in the photosphere. This is mutually indicated by (1) the power spectrum of the Sun's photospheric convection flows measured from full-disk dopplergrams having 1-arcsecond spatial resolution and 45-second cadence (Hathway et al 2015), (2) simulations of the free convection (a) in the top 2500 km of the convection zone (Stein & Nordlund 1989), (b) in the top 20,000 km (Stein et al 2009), and (c) in the global convection zone at depths below 15,000 km (Miesch et al 2008). Third, Stein & Nordlund (2012) simulated the production of a BMR having sunspot pores and spanning ~ 20,000 km by a convection cell of that diameter. The BMR was made by the convection cell ingesting horizontal magnetic field placed at cell bottom. The convection cell thereby made a flux-rope Ω loop that made the BMR by emerging in the cell's central up flow and having its legs in the down flows at opposite edges of the cell.

Based on the above observations and simulations of BMRs and solar convection, Moore et al (2020) (1) proposed that the flux-rope Ω-loop that emerges to become any BMR is made by a convection cell of the Ω loop's size from initially horizontal magnetic field ingested into the cell at cell bottom, and (2) laid out how this way of making BMR Ω loops plausibly explains why the span and flux of the largest observed BMRs are ~ 200,000 km and $10^{22}$ Mx. More broadly, Moore et al (2020) infer that the flux-rope Ω loops of BMRs of all sizes are made by basically the same magnetoconvection processes in the convection cells from which they emerge. This inference strengthens our expectation that throughout the lives of BMRs of all sizes the evolution of each BMR's magnetic field is governed by basically the same magnetoconvection processes. Hence, we think that tracking the field evolution leading to microflare-making magnetic explosions in BEARs could shed light on how the field evolution in sunspot BMRs leads to magnetic explosions that make major flares and CMEs.

A sunspot-bearing BMR can emerge abutting or intermixed with one or more pre-existing BMRs of comparable or larger size and having field strength comparable to that of the new BMR. Instead of being simply a larger single-bipole active region, the multi-BMR combination is usually more complex, having a more complex arrangement of opposite-polarity flux domains and more than one major PIL (e.g., Machado et al 1988; Moore et al 1999). On the other hand, a sunspot-bearing BMR can emerge in isolation, well away from any pre-existing field strong enough to significantly influence whether the new BMR has a magnetic explosion that produces a CME and/or flare during or after the BMR's emergence (e.g. Moore et al 2001). In any BMR – isolated or not – for the field to undergo a CME-/flare-making explosion, it must have a large-enough store of free energy. The energy given to a CME and/or flare is evidently released via eruption of a flux rope that is built and released by reconnection of sheared field in the core of a BMR's explosive magnetic arcade (Moore 1988; Machado et al 1988; Moore et al 2001, Patsourakos et al 2020). The underlying open question is: When and how do explosive sheared core fields arise in the magnetic evolution of BMRs? Finding clues to this question is the primary objective of the investigation reported here.

The strategy of our investigation is to track the cradle-to-grave magnetic evolution of each of ten BEARs that emerge in coronal holes on the central disk. We seek to identify any aspects of the evolution that lead to obvious microflare-making explosions of initially closed magnetic field having either both feet or one foot rooted in a BEAR's magnetic flux. We take any obvious characteristic magnetic-evolution process leading to microflare-making magnetic explosions in BEAR magnetic fields to be a likely essential magnetic-evolution process that leads to major flare-/CME-making magnetic explosions in sunspot active regions.

### 1.3. Explosions in Magnetic Anemones

For an isolated sunspot-bearing BMR, a flare-producing flux-rope-eruption explosion of the sheared core field is either *ejective* or *confined* (Pallavicini et al 1977; Machado et al 1988; Moore & Roumeliotis 1992; Moore et al 1999, 2001). In an ejective explosion, the exploding field simultaneously produces both a flare arcade by internal reconnection and a CME by exploding out into the solar wind. In a confined explosion, a flare is again produced by internal reconnection of the exploding field, but no CME is produced because the explosion is arrested within the BMR's enveloping magnetic arch. Instead of being blown open, the BMR's magnetic arch is strong enough to confine the explosion and stay closed.

Now consider a sunspot-bearing BMR that, instead of emerging in isolation, has emerged in surrounding magnetic field that is of strength comparable to that of the BMR's field and that constitutes one foot and lower leg of a larger far-reaching magnetic loop. The foot and leg of the large loop surround and encompass the entire BMR. Within the foot of the large loop, including the BMR's flux, the majority of the flux is of the polarity of one foot of the BMR's emerged $\Omega$ loop. The only flux of the opposite polarity, the minority-polarity flux, is the BMR's domain of flux of the minority polarity. As is depicted in Figure 1 for this situation, the BMR's PIL encircles the BMR's minority-polarity flux, instead of being – if the BMR were isolated – only inside the BMR (that is, only between the BMR's two opposite-polarity flux domains). Relative to the BMR, the PIL has two intervals: (1) the PIL's interval inside the BMR, which interval we call the PIL's "inside" interval, and (2) the rest of the PIL, which we call the PIL's "outside" interval. As the BMR's $\Omega$ loop emerged, continual external reconnection of the minority-polarity leg of the $\Omega$ loop with adjacent opposite-direction far-reaching field transformed the $\Omega$ loop to a two-lobed magnetic anemone (Shibata el al 1994). Relative to the BMR, one lobe of the anemone is the "inside" lobe, the part of the $\Omega$ loop that has not been reconnected. The inside lobe envelops the PIL's inside interval. The anemone's other lobe is the "outside" lobe, consisting of loops that have been made by the external reconnection. The outside lobe envelops the PIL's outside interval. Above the anemone of closed field, there is a magnetic null point or small current sheet between the anemone field and the ambient far-reaching field, as in Figure 1.

For a sunspot-bearing BMR situated in comparable-strength far-reaching field as in Figure 1, when an explosion of sheared core field produces a flare, there are three possibilities for the location and extent of the core-field explosion along the PIL within the anemone. The explosion and flare brightening can occur (1) "inside," along only the PIL of the anemone's inside lobe, or (2) "outside," along only the PIL of the anemone's outside lobe, or (3) "inside & outside," partly along the PIL's inside interval and partly along the PIL's outside interval.

For a flare-producing explosion along any interval of the anemone's PIL, there are three possibilities for how the explosion plays out, depending on how much the explosion drives external reconnection of the enveloping anemone field with the ambient far-reaching field. (Figure 4 of Panesar et al 2016a depicts such reconnection for surges and jets from sunspot-bearing BMRs embedded in one foot of a far-reaching magnetic loop. That drawing follows the drawing by Sterling et al 2015 for the driving of the spire-making external reconnection in X-ray jets in coronal holes by a minifilament-carrying flux-rope eruption from sheared core field in a lobe of the magnetic-anemone base of the jet.) One possibility is that the flare is produced by a confined explosion that corresponds to a flare-producing confined explosion in an isolated BMR: the explosion is so strongly confined within the anemone that in drives no reconnection of the outside of the anemone with ambient far-reaching field. Another possibility is that the flare is produced by a mostly-confined explosion: the explosion drives some external reconnection of the anemone with impacted far-reaching field, but the explosion is arrested before the reconnection eats into much of the erupting flux rope inside the enveloping section of the anemone. In this case, the driven external reconnection produces a relatively weak surge/jet of multi-temperature plasma that shoots out along the reconnected far-reaching field, such as the weaker jets of Panesar et al (2016a). A mostly-confined flare-producing explosion in the anemone of a BMR encased in far-reaching field corresponds more closely to an isolated-BMR confined explosion than to an isolated-BMR ejective explosion. The third possibility is that the flare is produced by a blowout explosion that corresponds to an isolated-BMR flare-producing ejective explosion that becomes a CME: instead of being entirely or mostly arrested within the anemone, the erupting core-field flux rope blows out the enveloping section of the anemone and continues driving external reconnection of that section until the external reconnection eats into and opens most or all of the erupting flux rope. In this case, the driven external reconnection produces a relatively strong surge/jet of multi-temperature plasma that shoots out along the reconnected far-reaching field, such as the stronger jets of Panesar et al (2016a).

### 1.4. The Coronal-bright-point and Coronal-jet Background of Our Investigation

Beginning with *Skylab*, soft X-ray images and EUV images of the corona registered with magnetograms have shown that many so-called coronal bright points seen in coronal images are BEARs, and that BEARs are like all larger BMRs in having magnetic loops heated enough to be bright in coronal emission (Withbroe & Noyes 1977; Golub et al 1977; Habbal & Withbroe 1981). Coronal bright points often have sudden flare-like bursts of coronal X-ray and EUV emission, bursts known as X-ray bright-point flares or coronal bright-point flares (Golub et al 1974; Moore et al 1977; Habbal & Withbroe 1981). In this paper, we use the term "microflare" for such miniature flares observed in BEARs in coronal images.

Coronal images having spatial resolution of a few arcseconds or better show that in coronal holes a BEAR's magnetic field is a cluster of magnetic loops that displays anemone-like 3D form (e.g., Shibata et al 1994). Coronal images having such good resolution and having a cadence of ~ 1 min or faster show that, as a microflare occurs in a BEAR-anemone coronal bright point, often a jet of coronal-temperature plasma stems from the coronal bright point and shoots up into the corona along far-reaching magnetic field (Shibata et al 1992). In this paper, we often refer to the long body of the jet as the jet's spire to distinguish it from the coronal-bright-point anemone base of the jet. Coronal jets from BEARs occur all over the Sun wherever obvious coronal-bright-point BEARs occur, which includes all latitudes in coronal holes and quiet regions (Martres & Bruzek 1977; Cirtain et al 2007; Panesar et al 2016b, 2018; McGlasson et al 2019). An outstandingly brightest part of the microflare brightening usually is concentrated at an edge of the base of the jet, is much smaller than the jet's whole coronal-bright-point base, and is laterally offset from the jet's spire (Shibata et al 1992; Yokoyama & Shibata 1995; Moore et al 2010; Sterling et al 2015). The base-edge intense compact brightening is commonly called the jet-base bright point or jet bright point (JBP, e.g., Sterling et al 2015).

Coronal jets were discovered by Shibata et al (1992) in *Yohkoh* coronal X-ray images which had resolution of a few arcseconds and a cadence no faster than about 1 min (Acton et al 1992). These images showed that the base of a jet is a pre-existing coronal bright point, and that the spire is co-produced with an offset JBP at the edge of the base (Shibata et al 1992). Because it was known that many coronal bright points are small emerging or emerged bipolar magnetic regions, i.e., BEARs, it was plausible to suppose – and was plausibly demonstrated by MHD simulations – that (1) the spire and the JBP are, respectively, the upward and downward products of a burst of reconnection of the outside of the BEAR's emerging Ω loop with ambient oppositely-directed far-reaching field, and (2) the burst of reconnection is driven by ongoing emergence of the Ω loop (Shibata et al 1992; Yokoyama & Shibata 1995).

*Hinode* coronal X-ray images have 1-arcsecond pixels and routinely have 0.5 – 1 min cadence (Golub et al 2007). These images have revealed that most coronal jets seen in coronal X-ray images are one or the other of two different types: so-called standard X-ray jets and so-called blowout X-ray jets (Moore et al 2010). In a standard X-ray jet, the spire is a single strand of width much less than that of the coronal-bright-point base. Relative to the JBP at an edge of the base, brightening in the rest of the base is usually noticeably less intense. A blowout X-ray jet briefly begins like a standard jet, with a narrow single-stranded spire growing in synchrony with the base-edge JBP. Then, brightness of intensity comparable to that of the JBP rapidly turns on in the rest of the base as the spire becomes multi-stranded and about as wide as the base (Moore et al 2010).

In this paper, we assume that narrow-spire coronal jets and wide-spire coronal jets seen in coronal EUV images correspond to the narrow-spire jets (standard jets) and wide-spire jets (blowout jets) seen in coronal X-ray images.

From evidence suggesting that blowout jets are driven by small filament eruptions, but standard jets are not, Moore et al (2010) proposed that only standard X-ray jets are produced by emergence-driven external reconnection of an emerging Ω loop with ambient far-reaching field in the manner modeled by Yokoyama & Shibata (1995). In blowout X-ray jets, Moore et al (2010) envisioned that the JBP and initial single-strand spire are made by external reconnection of an emerging Ω loop, but this reconnection and the Ω loop's ensuing more widespread external reconnection that makes the rest of the multi-stranded wide spire are driven not by the Ω loop's emergence but by blowout eruption of the Ω loop. Moore et al (2010) proposed that the Ω loop's blowout is driven by the blowout eruption of a small filament-carrying flux rope from sheared field along the PIL between the two feet of the emerging Ω loop, in the manner of the arcade blowouts by larger filament-flux-rope eruptions that make CMEs. In the Moore et al (2010) scenario for blowout jets, internal reconnection of the Ω loop's legs that collapse together under the erupting flux rope makes a miniature flare arcade in the manner in which the flare arcade in a CME eruption is made. The (non-JBP) brightening seen inside the coronal-bright-point base of a blowout X-ray jet is taken by Moore et al (2010) to be the miniature flare arcade.

The *Solar Dynamics Observatory (SDO)* Atmospheric Imaging Assembly (AIA) takes full-disk EUV images of the corona and chromosphere-coronal transition region (Lemen et al 2012). These have 0.6-arcsecond pixels and 12 s cadence. The *SDO* Helioseismic and Magnetic Imager (HMI) takes full-disk line-of-sight magnetograms (Scherrer et al 2012). These have 0.5-arcsecond pixels and 45 s cadence, and a noise level of about 10 G (Schou et al 2012; Couvidat et al 2016). From AIA EUV images in combination with *Hinode* coronal X-ray images (Sterling et al 2015), and from AIA EUV images in combination with HMI magnetograms (Panesar et al 2016b, 2018a; McGlasson et al 2019), it has been found that at least many – if not most – standard jets and blowout jets in coronal holes and quiet regions do not fit either the Yokoyama & Shibata (1995) model or the Moore et al (2010) scenario for jet production. Sterling et al (2015) discovered (1) that in many standard jets and in many blowout jets the burst of jet-base-anemone external reconnection that makes the spire is driven by the eruption of a minifilament from the PIL of one of the two lobes of the base anemone, (2) that the JBP in the examined standard jets and blowout jets is not made by external reconnection of an emerging Ω loop with far-reaching field a la Yokoyama & Shibata (1995) and Moore et al (2010), but is the miniature flare arcade made by internal reconnection of the erupting lobe's legs under the erupting minifilament flux rope, and (3) that blowout jets are made by minifilament blowout eruptions, and standard jets are made by mostly-confined minifilament eruptions. Panesar et al (2016b, 2018a) and McGlasson et al (2019) found strong evidence that most coronal jets in quiet regions and coronal holes are driven by minifilament eruptions, and that in a large majority of jet-driving minifilament eruptions the eruption is prepared and triggered by flux cancellation at the PIL of the sheared core field from which the minifilament erupts, not by flux emergence at the PIL. Consistent with these findings for jets in quiet regions and coronal holes, Sterling et al (2017) found evidence that many jets at the edges of active regions are preceded and triggered by flux cancellation at the underlying PIL of the driving magnetic explosion.

From the above and other recent *SDO*-based observational studies of coronal jets (e.g., Panesar et al 2020; Harden et al 2021), it is evident that coronal jets in quiet regions and coronal holes are seldom, if ever, made by an emerging BEAR Ω loop's external reconnection that is directly driven by the Ω loop's emergence, as proposed by Shibata et al (1992) and numerically modeled by Yokoyama & Shibata (1995). On the other hand, these *SDO*-based studies have left open the question of whether some blowout jets in quiet regions and coronal holes are driven by a blowout minifilament eruption from inside a BEAR's emerging Ω loop, as proposed by Moore et al (2010) and numerically modeled by Archontis & Hood (2013) and Pariat et al (2015). A secondary objective of the investigation reported here is to address this question.

### 1.5. The Crux of this Paper

In this paper, we lay out our findings from examination of the magnetic evolution leading to each of 43 obvious microflares in the anemone magnetic fields of 10 BEARs in central-disk coronal holes. We tracked the magnetic evolution and microflaring of each BEAR in HMI magnetograms and AIA 211 Å coronal EUV images, from the beginning of the BEAR's emergence to its extinction. The observations indicate that blowout eruptions from inside *emerging* BEARs (in the manner proposed by Moore et al 2010 for blowout jets) rarely, if ever, happen. For each of the 43 microflares – regardless of whether the microflare was made by a confined magnetic explosion, a mostly-confined explosion, or a blowout explosion – we see evidence of pre-microflare flux cancellation at the PIL on which the microflare is seated. This result points to flux cancellation at the underlying PIL being the essential process that builds – and finally renders unstable – the sheared core fields that explode in sunspot active regions (and in decaying large BMRs that no longer have sunspots) to make major flares and CMEs.

## 2. Data and Methods

We chose 10 BEARs for scrutiny of (1) when and where in each BEAR's evolving magnetic field obvious microflares happened, and (2) for each microflare, whether the magnetic explosion that produced it was a confined explosion, a mostly-confined explosion, or a blowout explosion. For the following reasons, we chose BEARS that were in coronal holes on the central disk. One reason is that the open magnetic field that fills a coronal hole is unipolar; it is all rooted in the coronal hole's majority-polarity magnetic flux in the photosphere. Hence, a BEAR in a coronal hole is encased in unipolar open field, resulting in the BEAR's magnetic field having the anemone topology depicted in Figure 1. This field configuration facilitates our classification of the type of magnetic explosion that produces each microflare in a BEAR's anemone field. We classify a microflare-producing explosion (viewed in AIA 211 Å images) to be (1) a confined explosion if it produces no jet spire, (2) a partly-confined explosion if it produces a standard-jet single-strand narrow spire, or (3) a blowout explosion if it produces a blowout-jet multi-strand wide spire. Another reason is that, due to the absence of coronal-emission fog along the line of sight to a BEAR in a central-disk coronal hole compared to that for a BEAR in a quiet region, the type of the magnetic explosion producing a microflare in a BEAR is usually more easily discerned when the BEAR is in a central-disk coronal hole than when the BEAR is in a quiet region. Yet another reason is that, for a BEAR in a coronal hole, we expect the BEAR's minority-polarity flux is eventually all removed by merging and cancelling with the coronal hole's ambient majority-polarity flux and/or with the BEAR's majority-polarity flux. That is, in a central-disk coronal hole, we expect to track a BEAR's minority-polarity flux in HMI magnetograms until all or nearly all of it has been removed by cancellation, instead of soon losing track of it because of its merging with pre-existing ambient flux of the same polarity, as would more often occur in quiet regions in which there are more nearly equal amounts of ambient flux of each polarity than in a coronal hole. We chose BEARs that were no farther than about $45^{o}$ from disk center, so that, in HMI line-of-sight magnetograms, each BEAR's magnetic flux could be tracked practically as well as if the BEAR were at disk center.

We randomly selected each of our 10 BEARs as follows. Using Helioviewer to view a full-disk AIA 211 Å image taken when one or more coronal holes were on the central disk, we would spot a prospective BEAR as an anemone coronal bright point in a central-disk coronal hole. (On-disk coronal holes and BEARs in on-disk coronal holes stand out starkly in the full-disk images from AIA's 211 Å channel, at least as starkly as in the images from AIA's other coronal EUV channels.) With Helioviewer, we then zoomed in on the prospective-BEAR anemone bright point in the 211 Å image, and in a co-temporal co-aligned HMI line-of-sight magnetogram, to judge whether the anemone was apparently that of the complex mix of two or more BMRs or apparently that of a single-Ω-loop BMR. If the anemone was apparently rooted in a single BMR, then, with 15-minute time steps, we visually tracked the BMR backward in time to its birth. We take the start of the BMR's birth to be when the emerging BMR's minority-polarity flux becomes discernible in HMI magnetograms. The BEAR was selected for our study if it emerged as a single Ω loop with its minority-polarity flux far enough from other minority-polarity flux that the BEAR's minority-polarity flux could be tracked through the BEAR's emergence and far into the BEAR's decay to extinction. We take the end of a BEAR's emergence and the beginning of its decay to be when the BEAR's minority-polarity flux stops increasing, that is, when the BEAR's minority flux attains its maximum amount before starting to decline to extinction.

For each selected BEAR, we visually tracked the BEAR's minority-polarity flux forward in time with 15-minute time steps from birth to the end of its trackable life in Helioviewer zoomed-in HMI magnetograms. We take the end of a BEAR's life to be when, in Helioviewer zoomed-in HMI magnetograms, the BEAR's minority-polarity flux either becomes unrecognizable by mingling with separately emerged minority-polarity flux or has vanished, either by cancelling with encountered majority-polarity flux or, by spreading, has become too weak to be detected by HMI.

For each of the 10 selected BEARs, for a small region centered on the BEAR and spanning 1-2 arcminutes in heliocentric x (east-west) and y (north-south), we made a movie of AIA 211 Å images and a movie of HMI magnetograms. The JSOC cutout service (http://jsoc.stanford.edu/ajax/exportdata.html) was used to download the data for each movie. Because each AIA cutout and its co-temporal HMI cutout span the same x and y intervals of the full-disk AIA 211 Å image or full-disk HMI magnetogram to within the width of a pixel, each AIA cutout is nominally registered with its HMI cutout to within ±1 arcsecond in x and y. In our microflare events, the approximate centering of few-arcsecond-long JBP miniature flare loops in the 211 Å image cutouts on the underlying PIL in the co-temporal magnetogram cutouts indicates that any misalignment is usually about that small. In some cases, from offset of the middle of the JBP loops from being centered on the underlying PIL, the 211 Å cutout and the magnetogram cutout are evidently shifted from each other by as much as about 2 arcseconds. In all cases, the misalignment is adequately small for us to discern whether a microflare-making magnetic explosion occurs on an inside interval, on an outside interval, or on an inside & outside interval of the BEAR's PIL.

The intensity of each of the AIA 211 Å images in the movies and figures in this paper is normalized by the image's exposure time. Using SolarSoft routines (Freeland & Handy 1998), for each BEAR, we derotated the cutout images and magnetograms to a particular time, so that solar features in the movies show no westward drift from solar rotation. The two movies for each BEAR start an hour before the BEAR's minority-polarity-flux birth time and end about an hour after the BEAR's minority-polarity-flux end time, both of which times we visually estimated from HMI magnetograms viewed with Helioviewer. These two estimated times for each BEAR are given in Table 1. The 211 Å movies have 1-minute time steps, which is enough time resolution to show the type of magnetic explosion that produces each microflare. The magnetogram movies have 6-minute time steps, which is enough time resolution to adequately show the BEAR's minority-polarity flux's growth and decay and its merging with other flux.

From each BEAR's magnetogram movie, we made a time plot (such as in Figure 3, for BEAR BR2) of the BEAR's minority-polarity flux. To make this plot, we chose in the field of view of the magnetogram movie a stationary box that, when the BEAR's emergence ends, has about twice the span of the BEAR's magnetic bipole, but is centered more nearly on the BEAR's minority-polarity flux than on the middle of the bipole. The box size is only big enough that none of the BEAR's minority-polarity flux migrates out of the box during its cradle-to-grave evolution in the magnetogram movie. Except near the start and end of the magnetogram movie, the BEAR's minority-polarity flux is practically all of the minority-polarity flux in box. In the time-flux plot, each plotted flux value is the total minority-polarity flux in the box in a frame of the BEAR's magnetogram movie, plotted at the time of that frame. As is typical of sunspot-bearing larger BMRs, the time-flux plot for each of our 10 BEARs has basically the shape of a single saw tooth. That is, the flux grows rapidly to its maximum and then declines more slowly (that is, less steeply than the steepness of the rise) to the end of its trackable life. We take the time of maximum flux in each BEAR's minority-flux time-flux plot to mark the end of the BEAR's emergence and the start of the BEAR's decay to extinction.

From each BEAR's 211 Å movie, we made a time plot (such as in Figure 3, for BEAR BR2) of the BEAR's 211 Å luminosity. To make this plot, in each 211 Å image for the BEAR's 211 Å movie, we measured the 211 Å luminosity in the same sub-field-of-view from which the minority-polarity flux was measured for the BEAR's time-flux plot. The BEAR's 211 Å movie shows any obvious microflares that are produced in the BEAR's anemone magnetic field during its cradle-to-grave evolution. Each obvious microflare in the BEAR's 211 Å movie produces an obvious narrow spike in the BEAR's time-luminosity plot (such as in Figure 3, for BEAR BR2).

By stepping through a microflare in the movie, we found the frame in which we judge the microflare brightening and any accompanying jet spire, in combination, to be the most visible. The time of this event-maximum-visibility frame relative to the time of the BEAR's minority-flux maximum says whether the microflare explosion occurred during the BEAR's emergence or during the BEAR's decay. From whether the microflare brightening in the BEAR magnetic anemone is accompanied by a jet spire, and from the width of the spire, we classify the microflare-producing magnetic explosion to be a confined explosion, a mostly-confined explosion, or a blowout explosion. For each microflare, from the 211 Å movie together with the magnetogram movie, we observe whether the PIL interval of the exploding core field and microflare brightening is an interval of only the BEAR's inside PIL, an interval of only the BEAR's outside PIL, or an interval that is inside PIL on one end and outside PIL on the other end. This tells us, in particular, for a microflare produced by a blowout magnetic explosion during the BEAR's emergence, whether the explosion was from inside the BEAR's emerging magnetic arch as proposed by Moore et al (2010) for blowout jets.

Each BEAR's magnetogram movie shows the pre-microflare arrangement and migration of the opposite-polarity flux bracketing the PIL interval on which each microflare occurs. In the magnetogram movie, before and during a microflare, if those two fluxes are impacted against each other so that the PIL between them is sharp in the magnetograms, and/or if either flux is discernibly diminishing, and/or if the two flux patches are discernibly converging on the PIL, we take this to be evidence that before and during the onset of the microflare-making magnetic explosion, flux was probably canceling at the underlying PIL.

## 3. Results

### 3.1. *Table 1*

Table 1 lists for each of the 10 BEARs the basic information for the cradle-to-grave evolution of the anemone field, in terms of the BEAR's minority-polarity flux and the microflares produced by the evolving field. The BEARs are numbered and listed chronologically. Each BEAR's number name is in the first column of Table 1. For each BEAR's minority-polarity flux, columns 2 – 8 give, respectively, (2) the polarity, (3) the birth place on the solar disk, (4) the birth date and time, (5) the maximum flux of the BEAR's minority-flux flux-time plot, (6) the date and time of the time-flux plot's maximum, (7) the end time, and (8) whether or not the minority-flux's trackable life ends by cancellation, and, if by cancellation, the interval of the anemone's PIL at which the minority flux vanishes. For each of 43 obvious microflares that happen in the magnetic anemones of the 10 BEARs, the last four columns (9 – 12) of Table 1 give, respectively, (9) the date and time of the 211 Å movie frame in which we judge the microflare event to be most outstanding, (10) the interval of the anemone's PIL on which the microflare occurs, (11) the type (i.e., class) of the microflare's driving magnetic explosion, and (12) our yes-or-no judgement of whether, in the magnetogram movie, we see evidence for flux cancellation at the underlying PIL before and during the microflare.

In the remainder of this Section, we first present as examples three selected BEARs (BR2, BR7, and BR9) and nine selected microflare events that happened in them. We selected these nine microflare events to show one example of each of the nine possible combinations of the PIL location and type of the driving magnetic explosion: (1) inside, confined, (2) inside, mostly confined, (3) inside, blowout, (4) outside, confined, (5) outside, mostly confined, (6) outside, blowout, (7) inside & outside, confined, (8) inside & outside, mostly confined, or (9) inside & outside, blowout. After presenting in detail the example BEARs and microflare events, we consider and interpret the data in Table 1 as a whole.

### *3.2. BEAR BR2*

As Table 1 indicates, BEAR BR2 emerged about 20° north northeast of disk center in a coronal hole that was filled with open field rooted in negative-polarity flux. BR2MOVIE, the animation accompanying Figure 2, spans the cradle-to-grave evolution of BR2 in HMI magnetograms and AIA 211 Å images. Figure 2 is one frame of BR2MOVIE, at a time during the peak of BR2's positive-polarity flux. In BR2's magnetogram movie in BR2MOVIE, BR2 starts emerging at about 21:30 UT on 2012 June 30. It emerges for about eight hours. For about a day thereafter, BR2's positive flux can be tracked as it decays, until it is joined by newly-emerging non-BR2 positive flux.

The magnetogram in Figure 2 is also the upper left panel of Figure 3. The closed red curve in this magnetogram in Figures 2 and 3 encloses practically all of BR2's positive and negative flux detected at 05:01:25 UT on 2021 July 1. The magnetogram shows that the polarity of the majority of the ambient flux in which BR2 emerges is negative (black). Nearly all of the positive (white) flux in the white box in the magnetogram in Figure 3 (all except the positive flux near the south side of the box) is BR2's emerged positive flux. The magnetogram also shows that the BR2 bipole is tilted about 45° counterclockwise from east-west; BR2's emerged leading flux is negative and is in the northwest corner of the white box. As BR2 emerges, its positive and negative fluxes spread apart, resulting in its positive flux moving southeast to come into contact and gradually cancel with an ambient patch of negative flux, as is happening in the magnetogram in Figures 2 and 3.

The 211 Å image in Figure 2 is also the lower left panel of Figure 3. The white box in the 211 Å image in Figure 3 is the same as that in the magnetogram in Figure 3; it covers the same area of the Sun's surface in both.

The upper right panel of Figure 3 is the time-flux plot for BR2's minority-polarity (positive) flux. This plot shows, for the area in the white box in both left panels, the progression of the amount of positive flux found in the box from the sequence of frames in BR2's entire magnetogram movie in BR2MOVIE. The plot starts about an hour before BR2's positive flux becomes discernible in the movie and continues through the flux's peak and trackable decay. The black vertical dashed line marks 05:40 UT on 2012 July 1, the time of the movie frame having the most positive flux in the box. Each of the other vertical lines marks the maximum-visibility time (in the 211 Å images in BR2MOVIE) of one of BR2's five microflares. That is, the five microflare lines mark the times of BR2's five microflare events listed in Table 1. (Here and in the time-flux plots in Figures 7 and 14 for BR7 and BR9, each microflare line's style indicates the type class of the microflare explosion, and the line's color indicates the BEAR-anemone PIL interval on which the microflare explosion occurs. Dotted lines are for confined explosions; dot-dash lines are for mostly-confined explosions; solid lines are for blowout explosions. Red is for a microflare that is seated only on the PIL's inside interval; green is for a microflare that is seated only on the PIL's outside interval; amber is for a microflare that is seated partly on the PIL's inside interval and partly on the PIL's outside interval.) Because the five microflares occur on intervals of BR2's outside PIL, their lines are green. The dotted line marks the time of the confined-explosion microflare that occurs during BR2's emergence, and the four solid lines mark the four blowout-eruption microflares that occur during BR2's decay.

The lower right panel of Figure 3 is the 211 Å time-luminosity plot for BR2. This plot shows, for the area in the white box in both left panels, the progression of the area integral of the 211 Å intensity in the box from the sequence of frames in BR2's entire 211 Å movie in BR2MOVIE. The vertical lines mark the same times as the corresponding vertical lines in BR2's time-flux plot in the upper right panel. In the time-luminosity plot, each of the five microflare lines falls in the luminosity spike that is made by the microflare that at the time marked by the line has its maximum visibility in the 211 Å images.

BR2MOVIE shows that during the three hours before BR2's first microflare event, as the southeast outer edge of BR2's positive flux is canceling with encountered ambient negative network flux at the outside PIL on which that microflare occurs, BR2 increases its positive flux by emerging more positive flux than is lost by cancellation at the outside PIL. A consequence of that continual dominance of emergence over cancellation is the continual rise in BR2's positive-flux time profile during the three hours before the confined-explosion microflare's dotted vertical line in Figure 3. In BR2MOVIE, starting about half an hour after the confined-explosion microflare, BR2's emergence in completed by a burst of emergence that lasts for an hour and a half, ending by 05:00 UT. That burst of field emergence gives BR2's time-flux plot in Figure 3 the steep final rise that starts about 03:00 UT and ends about 04:30 UT. By 04:30 UT, the advancing southeast edge of BR2's positive flux from the burst of emergence is near the ambient negative-flux network lane at which flux cancellation preceded BR2's confined-explosion microflare. By 04:30 UT, BR2's positive flux is plausibly canceling with that negative flux fast enough to begin slowing the rise in BR2's positive-flux time profile in Figure 3. The ending of the burst of flux emergence and the continuing cancellation at the PIL between the negative network lane and BR2's encroaching positive flux presumably result in the continual decline of the BR2's positive-flux time profile after 06:00 UT.

BR2MOVIE shows that BR2's second microflare event, a blowout-explosion microflare, is an explosion of the core of the BR2 anemone's outside lobe that envelops the PIL of the cancellation of BR2's positive flux with the encountered negative network lane. By the time of that microflare event (08:35 UT), half or more of the network lane's flux has been canceled by BR2's positive flux. The cancellation continues until, by 12:00 UT, all of that negative-flux network lane has been canceled, and only about half remains of the positive flux BR2 had when BR2 finished emerging. For the rest of its trackable life, BR2's positive flux appears to gradually cancel with a negative network lane to its northeast and with negative network flux patches to its south. BR2's third microflare event, a blowout explosion, explodes from the core of the BR2 outside lobe section that envelops BR2's outside PIL to the south. BR2's fourth and fifth microflare events, each a blowout explosion, explode from both the core of that southern section of the anemone's outside lobe and the core of the anemone's outside lobe section that envelops BR2's outside PIL to the northeast.

### *3.2.1. Example Confined-explosion Microflare on a BEAR's Outside PIL*

BR2's confined-explosion microflare is our chosen example of a confined-explosion microflare that occurs on a BEAR's outside PIL. That microflare is shown in Figure 4. Panels (a), (b), and (c) of Figure 4 are the images for frames of BR2's 211 Å movie, respectively at an hour before the micro flare, at the microflare's maximum visibility, and half an hour later. Panels (g), (h), and (i) are the magnetograms for the frames of BR2's magnetogram movie nearest in time to the three 211 Å images. The closed red curve in the panel (g) magnetogram encloses what we judge from the magnetogram movie to be practically all of BR2's emerged positive and negative flux at that time, 01:43:25 UT on 2012 July 1. The black or white saturation level of the magnetograms in this paper's movies and figures is 50 G. In panels (d), (e), and (f) of Figure 4, the 40 G contours (red for positive, blue for negative flux) from each magnetogram are superposed on the corresponding 211 Å image. The black worm-like curved line in panel (e) traces what we judge from the magnetogram in panel (h) and the bright microflare arcade in the 211 Å image in panel (b) to be the path of the PIL underlying the microflare.

Because the microflare's 211 Å image in Figure 4 and its images in BR2's 211 Å movie show no discernible accompanying jet spire stemming from BR2, this microflare is evidently made by a magnetic explosion that is confined inside the enveloping growing external lobe of BR2's anemone magnetic field. In Figure 4, from close inspection of either the magnetograms in panels (g) and (h) or the magnetogram contours in panels (d) and (e), it is discernible that flux cancellation occurs at the microflare's underlying PIL during the hour before the microflare. The flux cancellation is confirmed by BR2's magnetogram movie in BR2MOVIE. The magnetogram movie shows that, for three hours leading up to this microflare, while BR2 is still emerging, BR2's emerged positive flux has migrated southeast far enough that it is in contact and cancelling with encountered ambient negative network flux, along the PIL on which the microflare finally occurs. Thus, Figure 4 and BR2MOVIE show clear evidence that flux cancellation leads to this confined-explosion microflare.

### *3.2.2. Example Blowout-explosion Microflare on a BEAR's Outside PIL*

Our chosen example of a blowout-explosion microflare that occurs on a BEAR's outside PIL is BR2's first blowout microflare event. That event is shown in Figure 5. As in Figure 4 for BR2's confined microflare event, the three pairs of co-temporal 211 Å images and magnetograms in Figure 5 show BR2 an hour before the blowout event, at the event's maximum visibility, and half an hour later. The closed red curve in the panel (g) magnetogram encloses what we judge from the magnetogram movie to be BR2's evolved positive and negative flux at that time, 07:37:25 UT on 2012 July 1. As in panel (e) of Figure 4, the worm-like black curve in panel (e) of Figure 5 traces what we judge from the magnetogram in panel (h) and the JBP microflare arcade in the 211 Å image in panel (b) to be the path of the PIL underlying the JBP.

The 211 Å image and magnetogram in panels (a), (d), and (g) show a north-south dark minifilament that runs along BR2's eastern outside PIL between BR2's eastwardly advancing emerged positive flux and the encountered negative network flux. The minifilament is evidence of strong shear in the PIL-tracing core field in which the minifilament's cool dark plasma is suspended. As can be seen in BR2MOVIE, by the time of the event-maximum-visibility panels (b), (e), and (h) in Figure 5, the minifilament-carrying sheared core field has undergone a blowout explosion. In panel (b), the explosion has produced both the wide blowout-jet spire that extends northwest from BR2 and the JBP miniature flare arcade that sits on the PIL from which the minifilament erupted in the explosion. From comparison of the two magnetograms in panels (g) and (h), it can be seen that over the one-hour interval positive and negative flux has merged and canceled along that PIL. The flux cancellation is confirmed by BR2's magnetogram movie. In BR2MOVIE, merging and cancellation of positive flux with negative flux along that PIL evidently starts at least a few hours before the microflare event, and the minifilament signature of sheared core field at that PIL appears a couple of hours before the microflare eruption. In addition, Panesar et al (2017) show a time-distance plot (from HMI magnetograms of this interval of BR2's outside PIL) of the convergence and cancellation of positive and negative flux that leads to the minifilament's formation and eruption. Thus, there is compelling evidence that flux cancellation at the underlying PIL led to this blowout-explosion microflare event.

### *3.3. BEAR BR7*

BEAR BR7 emerged about 10° northwest of disk center in a negative-polarity coronal hole**.** BR7MOVIE, the animation accompanying Figure 6, spans the cradle-to-grave evolution of BR7 in HMI magnetograms and AIA 211 Å images. Figure 6 is one frame of BR7MOVIE, at a time during the first hour after BR7's maximum minority-polarity positive flux. In BR7's magnetogram movie in BR7MOVIE, BR7 emerges with its leading (western) flux negative and impacted against a nearly stationary clump of negative network flux. As BR7 emerges, BR7's negative flux melds with that network flux clump while most of BR7's rapidly emerging positive flux moves eastward and a smaller amount moves southwest and south.

BR7 is noticeably shorter-lived and smaller than BR2. BR7's minority-polarity positive flux emerges to maximum in an hour and a half, and lasts for only three quarters of a day; BR2's minority-polarity positive flux takes eight hours to reach its maximum and lasts for one and a third days. At the end of emergence, BR7's bipolar magnetic flux region spans about 10,000 km and has positive flux of about $10^{19}$ Mx, compared to about 20,000 km and about 4 x $10^{19}$ Mx for BR2. The migration of BR7's minority-polarity flux starkly differs from that of BR2's minority-polarity flux. As BR2's minority-polarity (positive) flux decreases, it continues to migrate away from BR2's majority-polarity (negative) flux. In contrast, as BR7's minority-polarity (positive) flux decreases, it migrates back toward BR7's majority-polarity (negative) flux and finally vanishes by merging and canceling with it.

The magnetogram and the 211 Å image in Figure 6 are the two left panels of Figure 7. The closed red curve in this magnetogram in Figures 6 and 7 encloses BR7's positive and negative flux seen in BR7MOVIE at this time, 07:30:27 UT on 2013 February 27. The two right panels of Figure 7 are the time profiles of the minority-polarity (positive) flux and the 211 Å luminosity measured, from the magnetograms and 211 Å images for BR7MOVIE, in the white-box area outlined in the two left panels.

Listed in Table 1 and marked by vertical lines in the minority-flux time profile and in the 211 Å luminosity time profile in Figure 7 are 12 obvious microflare explosions that happen in BR7's evolving anemone magnetic field in BR7MOVIE. As the microflare-marking vertical lines in BR7's time-flux plot in Figure 7 indicate, the first two microflare events explode entirely from BR7's outside PIL. Each of these is a mostly-confined explosion from the northeast front of BR7's eastward-moving emerging positive flux, where the advancing positive flux is canceling encountered barely detected faint negative flux. Two other microflare events explode entirely from BR7's inside PIL. Both of these happen during the decline of BR7's positive flux. The first of these two is a confined explosion, the second a blowout. BR7's other eight microflare events all explode from PIL intervals that are partly an interval of the inside PIL of the positive flux and partly an interval of the outside PIL. One of these eight is a confined explosion, one is a blowout, and six are mostly-confined explosions. Five of our nine chosen example microflare events happen in BR7.

### *3.3.1. Example Mostly-confined-explosion Microflare on a BEAR's Outside PIL*

Our example of a mostly-confined-explosion microflare event that occurs on a BEAR's outside PIL is the first microflare event that happens in BR7, early in BR7's rapid emergence. That event is shown in Figure 8. The three pairs of co-temporal 211 Å images and magnetograms in Figure 8 show BR7 about 10 minutes before, at the microflare event's maximum visibility, and about 20 minutes later. In the event-maximum-visibility 211 Å image in panel (b), the downward arrow points to the JBP and the slanted upward arrow points to the event's narrow jet spire. The narrow spire is evidence that this microflare event is a mostly-confined magnetic explosion.

In Figure 8, the magnetograms in panels (g), (h), and (i) span 30 minutes, starting about 15 minutes after BR7 starts emerging. They show that BR7 emerges with its negative flux melded with the stationary negative network flux clump pointed to by the lower of the two horizontal arrows in panel (g). The negative flux pointed to by the other horizontal arrow is the next network flux clump north of BR7. The almost-closed red curve in panel (g) encloses what we judge from the magnetogram movie to be BR7's emerged positive and negative flux at that time, 05:30:27 UT on 2013 February 27. The magnetogram panels and BR7MOVIE show that BR7's positive flux rapidly emerges in a growing crescent pattern, the outside of which moves eastward and somewhat northward. In panel (e), the 40 G magnetic flux contours on the panel-(b) 211 Å image show that the JBP microflare brightening evidently sits on an interval of BR7's outside PIL, between BR7's emerging positive flux and the negative network flux clump pointed to by the upper horizontal arrow in panel (g). The black curved line in panel (e) traces what we judge from the magnetogram in panel (h) and the JBP brightening in the 211 Å image in panel (b) to be the path of that interval of BR7's outside PIL.

In Figure 8, the slanted arrow in panels (g), (h), and (i) points to a faintly detected stationary negative flux clump. In panel (g), the vertical downward arrow points to yet fainter possible negative flux nearer to the encroaching crescent of emerging positive flux. In panels (h) and (i), the front of the positive flux crescent approaches and then touches the weak stationary negative flux clump. In a negative-polarity coronal hole, the polarity of any diffused flux that is hidden by HMI's 10 G noise level is more likely to be negative than positive. Hence, we take the observed eastward and northward advance of BR7's emerging crescent of positive flux into a region of hardly detected negative flux, and the crescent's observed cancellation of such negative flux, to be evidence for probable explosion-preparing flux cancellation at the PIL underlying this mostly-confined-explosion microflare event on the outside of BR7's emerging minority-polarity flux. On this basis, we deem that this event's entry in the last column of Table 1 is yes (there is evidence for flux cancellation at the underlying PIL) rather than no.

### *3.3.2. Example Mostly-confined-explosion Microflare on an Inside & Outside Interval of a BEAR's PIL*

Our example of a mostly-confined-explosion microflare event that occurs on an inside & outside interval of a BEAR's PIL is BR7's third microflare event. That event is shown in Figure 9. The three pairs of co-temporal 211 Å images and magnetograms in Figure 9 show BR7 12 minutes before, at the event's maximum visibility, and 12 minutes after. The closed red curve in panel (g) encloses what we judge from the magnetogram movie to be BR7's emerged positive and negative flux at that time, 06:06:27 UT on 2013 February 27. In the event's maximum-visibility 211 Å image in panel (b), the vertical arrow points to the JBP miniature flare arcade and the slanted arrow points to the short narrow spire produced by the microflare-making magnetic explosion. From the jet spire being narrow, we infer that the magnetic explosion is mostly confined by the enveloping lobe of BR7's anemone magnetic field in which it occurs.

Northeast of the JBP in panel (b), there is brightening in much of the rest of BR7. In BR7MOVIE, this extensive brightening turns on in step with the JBP and jet spire. We assume that the extensive brightening is from reconnection-heated plasma in new outer loops added to the passive lobe of BR7's anemone magnetic field by external reconnection of the anemone's lobe that mostly confines the sheared-core-field flux-rope-eruption magnetic explosion that drives the lobe's external reconnection a la Sterling et al (2015). That is, the new outer hot loops of the passive lobe are the downward product of the explosion-enveloping lobe's explosion-driven burst of external reconnection, and the upward product of that reconnection is the jet-spire outflow along reconnected open field. We assume that the JBP is made by internal reconnection under the erupting flux rope, that is, by reconnection of opposite legs of the anemone's confining-lobe field that implode against each other under the erupting flux rope.

In Figure 9, the horizontal arrow in the magnetogram panels points to a small patch of positive flux. From panels (e) and (h) together, it appears that the JBP arch in panels (b) and (e) has its north foot in that positive flux patch and its south foot in negative network flux. The 211 Å image in panels (b) and (e) is the frame of BR7's 211 Å movie in BR7MOVIE that best shows both the JBP and the jet spire. In each of the 211 Å movie's two frames right before the one shown in panels (b) and (e), the JBP arcade extends farther north along its core-field PIL on the west side of the positive flux patch, and it appears that the arcade's northern loops arch from the positive flux patch westward to negative flux that appears to be some of BR7's emerged flux. Thus, it appears that the exploding core field in this microflare event is along a PIL interval that has its north end inside BR7 and its south end outside BR7. The black curved line in panel (e) traces what we judge to be the path of this inside & outside PIL interval underlying the JBP of this mostly-confined-explosion microflare event.

BR7MOVIE shows that the positive flux patch pointed to by the horizontal arrow in panels (g), (h), and (i) of Figure 9 emerges next to BR7's negative flux starting at 05:48 UT and becomes stronger as it moves south to where it is at 06:06 UT in panel (g). It appears to be depleted some by colliding and partly canceling with the negative network flux on its south side, before being replenished by further emergence to become as it is at 06:18 UT in panel (h) of Figure 9. In BR7MOVIE, it appears that at the time of panel (h) the positive flux patch is canceling at its south-side outside PIL with the negative network flux and at its west-side inside PIL with BR7's negative emerged flux. Thus, there is apparent flux cancellation at the underlying PIL before and during this microflare event.

As BR7's time-flux plot in Figure 7 shows, this inside & outside mostly-confined-explosion microflare event occurs early in BR7's hour-long peak in its positive flux, after BR7 has completed its rapid emergence. Occurring half an hour before BR7's time of maximum measured minority-polarity flux, this event occurs near the end of BR7's emergence phase, after nearly all of BR7's flux has emerged. As Table 1 shows, of our 43 microflare events, this is the only one that occurs either partly or entirely on its BEAR's inside PIL before the time of its BEAR's maximum minority-polarity flux.

### *3.3.3. Example Confined-explosion Microflare on a BEAR's Inside PIL*

Our example of a confined-explosion microflare event that occurs entirely on a BEAR's inside PIL is BR7's fourth microflare event. As is seen in BR7's time-flux plot in Figure 7, this microflare even is the first to happen in the decay of BR7's minority-polarity positive flux. This event is shown in Figure 10. In Figure 10, the three pairs of co-temporal 211 Å images and magnetograms are a half hour before, at the event's maximum visibility, and a half hour after. The closed red curve in panel (g) encloses BR7's positive and negative flux that is detected in BR7MOVIE's magnetogram at this time, 07:00:27 UT on 2013 February 27.

From its lack of any jet spire in its 211 Å images in BR7MOVIE, we conclude that this microflare-making magnetic explosion is a confined explosion. In panel (e) of Figure 10, the 40 G contours of positive and negative flux show that the microflare 211 Å emission is from a magnetic arch over BR7's inside PIL. The black curved line in panel (e) traces what we judge from the magnetogram in panel (h) to be the path of the interval of BR7's inside PIL underlying the microflare arch. The east foot of the arch is in BR7's positive flux and the west foot is in BR7's negative flux.

From the magnetogram movie in BR7MOVIE and from comparison of the magnetograms in panels (g), (h), and (i) of Figure 10, we see that continually through the hour centered on the event, BR7's positive and negative flux diffuse toward each, presumably canceling at the PIL between them. This cancellation plausibly results in much of the decline during that hour in BR7's time-flux plot in Figure 7. Thus, there is clear evidence that flux cancellation at the underlying PIL occurs before and during this confined magnetic explosion.

### *3.3.4. Example Confined-explosion Microflare on an Inside & Outside Interval of a BEAR's PIL*

Our example of a confined-explosion microflare that occurs on an inside & outside interval of a BEAR's PIL is BR7's fifth microflare event. This event is shown in Figure 11. As BR7's time-flux plot shows, this event is BR7's second microflare event in the decay phase of its minority-polarity positive flux and occurs about four hours after the first one, the one shown in Figure 10.

In BR7MOVIE and from comparison of panels (b) and (e) in Figure 11 with panels (b) and (e) in Figure 10, we see that the microflare in Figure 11 is a discernibly bigger version of the one in Figure 10. They are both made by confined magnetic explosions, i.e., have no jet spire, but the microflare arcade in Figure 11 is "fatter." Compared to the microflare in in Figure 10, which appears to be a single loop, the microflare in Figure 11 appears to be an arcade of loops that has greater extent along the underlying PIL.

In Figure 11, the three pairs of co-temporal 211 Å images and magnetograms are 10 minutes before, at the event's maximum visibility, and half an hour after. The closed red curve in panel (g) encloses what we judge from the magnetogram movie in BR7MOVIE to be BR7's positive and negative flux in this magnetogram, taken by HMI at 11:54:27 UT on 2013 February 27. The black curved line in panel (e) traces what we judge to be the path of BR7's inside & outside PIL interval underlying the microflare arcade. In BR7MOVIE, during the four hours between the microflare in Figure 10 and the microflare in Figure 11, neighboring network flux – north of the combination of BR7's negative flux and the negative network flux clump that BR7's negative flux melded with as BR7 emerged – drifts south and accumulates against the northside of BR7's negative flux. From that, we judge that the north end of the microflare arcade in panels (b) and (e) of Figure 11 arches over the interval of BR7's outside PIL that is between BR7's positive flux and the negative network flux that is against the north side of BR7's negative flux, and that the south end of the microflare arcade arches over BR7's inside PIL between BR7's positive and negative flux. On this basis, we take this confined-explosion microflare to be an example of one made by a core-field explosion from a PIL interval that is an inside & outside interval of a BEAR's PIL.

From BR7MOVIE and from comparison of the three magnetograms in Figure 11, we see that for half an hour before and half an hour after the microflare, along the inside & outside interval of BR7's PIL that is under the microflare arcade in panel (e) of Figure 11, positive flux and negative flux merge together and apparently cancel. That is, there is evident flux cancellation at the underlying PIL before and during the confined-explosion microflare event of Figure 11.

### *3.3.5. Example Blowout-explosion Microflare on an Inside & Outside Interval of a BEAR's PIL*

The last of our five example microflare events from BR7 is BR7's seventh microflare event, which is made by a blowout magnetic explosion from another inside & outside interval of BR7's PIL. This event is shown in Figure 12. It happens two hours after BR7's microflare event shown in Figure 11. In panel (g) of Figure 12, the closed red curve encloses what we judge from the magnetogram movie in BR7MOVIE to be BR7's positive and negative flux in this magnetogram taken by HMI at 13:42:27 UT on 2013 February 27. In BR7MOVIE, during the two hours from the event in Figure 11 to the event in Figure 12, most of BR7's remaining positive flux drifts south and comes into contact with non-BR7 negative network flux on its south and southwest sides.

In Figure 12, the three pairs of co-temporal 211 Å and magnetograms are half an hour before, at the event's maximum visibility, and roughly 10 minutes after. The 211 Å image in panel (b) of Figure 12 is the frame of BR7's 211 Å movie in BR7MOVIE that we judge best shows both the JBP "flare arcade" and the jet spire. Because the spire has multiple strands and is wide enough, we judge it to be a blowout-jet spire, and hence deem that the magnetic explosion that makes this microflare event is a blowout explosion. The black curved line in panel (e) traces in the contoured magnetogram what we judge from the magnetogram in panel (h) to be the path of BR7's inside & outside PIL interval that has its outside interval between BR7's positive flux and the negative network flux patch on the southwest side of the positive flux, and has its inside interval between BR7's positive flux and BR7's negative flux on the northwest side of the positive flux. In panel (e), presumably due to the AIA 211 Å image being shifted about two arcseconds to the northeast relative to the magnetogram, the bright JBP miniature flare loop only touches the PIL between BR7's remaining positive flux and the impacted small negative network flux patch on the southwest side of the positive flux, instead of being centered on that PIL. In the next two frames of BR7MOVIE, the JBP is more extensive and reaches over that interval of BR7's outside PIL and over BR7's inside PIL between the positive flux and BR7's negative flux on the northwest side of the positive flux. From this evidence, we conclude that this microflare event is made by a blowout explosion of core field from along this inside & outside interval of BR7's PIL.

BR7's magnetogram movie in BR7MOVIE and comparison of the three magnetograms in Figure 12 show that, in the time interval spanned by the three magnetograms, flux from BR7's main patch of positive flux merges with BR7's negative flux to the northwest and with negative network flux to the south and southwest. We consider this to be good evidence of flux cancellation at the underlying inside & outside PIL before and during this blowout-explosion microflare event.

### *3.4. BEAR BR9*

BEAR BR9 emerged about 15° northeast of disk center in a positive-polarity coronal hole. BR9MOVIE, the animation accompanying Figure 13, spans the cradle-to-grave evolution of BR9 in HMI magnetograms and AIA 211 Å images. Figure 13 is one frame of BR9MOVIE, at an hour after BR9's maximum minority-polarity negative flux. The magnetogram in Figure 13 shows that the polarity of the majority of the ambient flux in which BR9 emerges is positive (white), that the flux in the leading end of BR9's emerged bipole is negative (black), and that the emerged bipole is tilted clockwise from east-west by about 60°.

In BR9MOVIE, BR9 emerges in about the same manner as BR2 and BR7, that is, roughly in the manner expected for the emergence of a single flux-rope Ω loop. In contrast to BR2 and BR7, which each emerge well isolated (far) from any other BEAR, BR9 emerges north of a nearby east-west BEAR, starting after that previously-emerged BEAR had finished emerging. In terms of the span of the bipole at the end of emergence, BR9 is about the size of BR2 and twice bigger than BR7. The time-flux plots in Figures 3, 7, and 14 show that BR9's maximum flux is roughly that of BR2 and several times that of BR7. In duration of emergence and cradle-to-grave duration, BR9 is between BR7 and BR2. BR9's minority-polarity flux emerges to its maximum in about 4 hours and lasts for about 24 hours, compared to about 2 hours and 18 hours for BR7 and about 8 hours and 32 hours for BR2.

The magnetogram and 211 Å image in Figure 13 are the two left panels of Figure 14. The closed red curve in the magnetogram in Figures 13 and 14 encloses BR9's positive and negative flux seen in BR9MOVIE at this time, 05:59:51 UT on 2013 May 2. The two right panels of Figure 14 are the time profiles of the minority-polarity (negative) flux and the 211 Å luminosity measured from all the magnetograms and 211 Å images shown in BR9MOVIE, in the area outlined by the white box in the two left panels.

BR9MOVIE starts at 01:00 UT on 2013 May 2. It shows that BR9 actually started emerging at 01:00 UT instead of at 03:00 UT, which is the start time that we estimated from viewing HMI magnetograms on Helioviewer and that is BR9's start time given in Table 1. In BR9MOVIE, BR9 emerges about 15 arc seconds north of the previously-emerged BEAR. In the beginning of BR9MOVIE, the previously-emerged BEAR is seen as an anemone coronal bright point in the 211 Å images. BR9's time-flux plot in Figure 14 shows that at the start of BR9's emergence at 01:00

UT, the previously-emerged BEAR's negative flux was about 2 x $10^{19}$ Mx. The emergence of BR9's negative flux from 01:00 UT through about 01:30 UT presumably gives the initial rise in the time-flux plot. Then, from about 01:30 UT through about 02:00 UT, as BR9's negative flux continues to emerge, there is a plateau in the time-flux plot. The plateau is presumably due to the previously-emerged BEAR's negative flux decreasing – by spreading and by canceling with encountered positive flux including some of the previously-emerged BEAR's positive flux – as much as BR9's negative flux increases. During the half-hour plateau, an obvious blowout-explosion microflare event occurs in the previously-emerged BEAR's anemone magnetic field. That eruption is so extensive that it involves much of BR9's emerging bipole. In BR9MOVIE, the blowout eruption appears to perhaps be triggered by BR9's emerging negative flux encountering some of the previously-emerged BEAR's positive flux, but the blowout explosion definitely erupts from the previously-emerged BEAR's magnetic anemone. Therefore, we do not count this microflare event as one of BR9's microflare events.

In BR9MOVIE, after about 02:00 UT, BR9's negative flux emerges in two main clumps, an eastern clump and a western clump. The eastern clump grows by continued emergence of the negative flux that perhaps triggered the previously-emerged BEAR's blowout-explosion microflare event. The western clump emerges about 10 arc seconds west of the eastern clump. Both clumps grow and move south southwest in step with each other. By the time of the magnetogram in Figures 13 and 14 (06:00 UT), an hour after negative-flux maximum in the white box in Figure 14, BR9's western negative flux clump has reached the previously-emerged BEAR's remaining negative flux, and about half of BR9's eastern negative flux clump has canceled with the previously-emerged BEAR's positive flux and with positive network flux adjacent to – or intermixed with – the previously-emerged BEAR's positive flux. By 07:00 UT, an hour after the magnetograms in Figures 13 and 14, BR9's western negative flux has melded with the previously-emerged BEAR's negative flux, and by 07:45 UT, BR9's eastern negative flux has disappeared by continuing its cancellation with the positive flux clump of network flux intermixed with the previously-emerged BEAR's positive flux. After possibly triggering the previously-emerged BEAR's blowout-explosion microflare event very early in BR9's emergence, the cancellation of BR9's eastern negative flux clump does not lead to any microflare events.

In BR9MOVIE, after the total cancellation of BR9's eastern negative flux clump, BR9's western negative flux clump merges with the previously-emerged BEAR's negative flux, and the combined clump of negative flux drifts eastward and gradually cancels with the same positive network flux clump with which BR9's eastern flux clump has canceled. By 22:00 UT, we judge that probably all of the previously-emerged BEAR's negative flux has canceled. That is, we judge that the small remaining clump of negative flux is probably mostly from BR9. As that cancellation continues, a microflare event occurs that produces a complex broad blowout-jet spire along with a JBP microflare brightening on that outside flux-cancellation PIL and on an inside PIL between what we take to be scattered remnants of BR9's negative flux and remnants of BR9's positive flux that are moving south and encountering that scattered weak negative flux. On this basis, BR9's first microflare event is a second example of an inside & outside blowout-explosion microflare event. The maximum-visibility time of this event is marked by the amber solid line in BR9's time-flux and time-luminosity plots in Figure 14. Further cancelation of BR9's main remnant negative flux clump on its north side with what we assume is BR9's southward-moving remnant positive flux occurs before and during each of BR9's final two microflare events. The maximum-visibility times of these two microflare events are marked by the two red vertical lines in BR9's time-flux and time-luminosity plots in Figure 14. Each of these two microflare events explode from the north-side flux-cancellation PIL of BR9's main clump of remnant negative flux. These two events are the last of our nine selected example microflare events.

### *3.4.1. Example Blowout-explosion Microflare on a BEAR's Inside PIL*

The second of BR9's three microflare events is our example of a blowout-explosion microflare event that occurs on a BEAR's inside PIL. That event is shown in Figure 15. The three pairs of co-temporal magnetograms and 211 Å images in Figure 15 are half an hour before, at the event's maximum visibility, and half an hour after. The closed red curve in the magnetogram in panel (g) encloses what we judge from BR9's magnetogram movie in BR9MOVIE to be BR9's remaining positive and negative flux at this time, 00:35:51 UT on 2013 May 3. From the spire's broad width and complex structure and from the spire's eruption in BR9MOVIE, this microflare event is obviously a blowout jet made by a blowout explosion seated in BR9's magnetic anemone.

In panel (e) of Figure 15, the panel (h) magnetogram's 40 G contours on the panel (b) 211 Å image show that the miniature-flare-arcade JBP sits on the PIL between BR9's main remnant clump of minority-polarity negative flux and what we take to be a small clump of BR9's remnant positive flux that is encountering the north side of the negative flux clump. On this basis, we take this PIL to be an inside interval of the decayed BR9's PIL. The black curved line in panel (e) traces what we judge from the magnetogram in panel (h) to be the path of that PIL interval.

In the three magnetograms in Figure 15, the upward slanted arrow points to the negative flux clump on the south side of the microflare JBP's underlying PIL, and the downward slanted arrow points to the positive flux clump on the north side of that PIL. From BR9MOVIE and from comparison of the three magnetograms in Figure 15, it appears that during the one-hour interval centered on the event's time of maximum visibility, these two opposite-polarity flux clumps gradually merge and cancel while additional BR9 remnant positive flux from the north approaches that PIL. We take this to be credible evidence that flux cancellation occurred at the underlying PIL before and during this inside-PIL blowout-explosion microflare event.

### *3.4.2. Example Mostly-confined-explosion Microflare on a BEAR's Inside PIL*

Our ninth example microflare event is BR9's third and final microflare event. This is an example of a mostly-confined-explosion microflare event that occurs on a BEAR's inside PIL. This event is shown in Figure 16. The three pairs of co-temporal magnetograms and 211 Å images are half an hour before, at the event's maximum visibility, and half an hour after. The closed red curve in the magnetogram in panel (g) encloses what we judge from BR9's magnetogram movie in BR9MOVIE to be BR9's remaining positive and negative flux at that time, 01:29:51 UT on 2013 May 3. The 211 Å image in panel (b) shows that this event produces a jet spire and that the spire is a single narrow strand that along most of its extent has width much less that the width of the jet's base. Hence, we take this event's driving magnetic explosion to be mostly confined.

BR9MOVIE and comparison of panel (e) in Figures 15 and 16 show that the JBP of BR9's mostly-confined-explosion event of Figure 16 occurs on the same PIL as the JBP of BR9's blowout-explosion microflare event of Figure 15, after another hour of flux cancellation at that inside PIL. The black curved line in panel (e) of Figure 16 traces what we judge from the magnetogram in panel (h) to be the path of that PIL interval underlying the JBP of this mostly-confined-explosion microflare event.

In the first magnetogram in Figure 16, the upward slanted arrow points to the remnant BR9 negative flux clump and the downward slanted arrow points to a BR9 remnant positive flux clump that is merging and canceling with the negative flux clump at the forthcoming event's underlying PIL. In the second magnetogram in Figure 16, at the event's time of maximum visibility, the two arrows point to the same two flux clumps, which have by now nearly canceled each other. As BR9MOVIE shows, because these two tiny flux clumps cancel out during the next half hour after the second magnetogram in Figure 16, they are no longer present in the third magnetogram. Thus, there is good evidence that flux cancellation occurred at the underlying PIL before and during this inside-PIL mostly-confined-explosion microflare event.

### *3.5. Results from Table 1 as a Whole*

#### *3.5.1. The 10 BEARs*

A measure of a BEAR's size at the end of emergence, in terms of the BEAR's magnetic flux content, is the BEAR's maximum minority flux value in Table 1. Each listed value is the maximum of the time-flux plot of the BEAR's minority-polarity flux. (Figures 3, 7, and 14 show the flux-time plots for our three example BEARs.) Figure 17 is a histogram showing the distribution of the maximum flux values of the 10 BEARs. It shows at a glance that the BEAR magnetic-flux-content size in our small random sample ranges from about 1 x $10^{19}$ Mx to about 9 x $10^{19}$ Mx, and is roughly evenly distributed over that range. The average BEAR maximum minority-polarity flux is 5.3 x $10^{19}$ Mx.

Our measure of a BEAR's lifetime is the length of the interval from the birth time to the end time of the BEAR's minority-polarity flux. For each BEAR, these two times are given in Table 1. Figure 18 is a histogram of the lifetimes of our 10 BEARs. It shows that the lifetimes range above and below 1 day by about three quarters of a day and are roughly evenly distributed over that range. The average lifetime of our BEARs is 1.0 day. This is an indication that our BEARs are normal ephemeral regions of the flux-content size of our BEARs (~ 5 x $10^{19}$ Mx). That is, ephemeral regions of this size typically last for about a day (van Driel-Gesztelyi & Green 2015).

We define a BEAR's emergence time to be the length of the interval from the BEAR's minority-flux-birth time to the BEAR's minority-flux-maximum time (in Table 1). Figure 19 is a histogram of the emergence times of our 10 BEARs. It shows that the emergence times range below 0.5 day. Only two BEARs (BR7 and BR9) emerge in less than 0.1 day. The average emergence time is 0.21 day. More meaningful than the emergence time alone is the ratio of the emergence time to the BEAR's lifetime. Figure 20 is a histogram of the (emergence time/lifetime) ratios for the 10 BEARs. It shows that each BEAR's flux emerges in less than half the BEAR's lifetime. This reflects that each BEAR's flux time-plot is like that of most BMRs of larger size in having the shape of a saw tooth. That is, the rise to maximum is faster than the fall after maximum, as in the three example flux-time plots in Figures 3, 7, and 14. The average (emergence time/lifetime) ratio for the 10 BEARs is 0.24. This value is twice or more than the (emergence time/lifetime) ratio for many much-larger BMRs that have emergence times of a week or two and have large sunspots at the end of emergence (van Driel-Gesztelyi & Green 2015). A plausible reason for this is that our BEARs emerge in coronal holes in which there is ample nearby network flux to cancel the BEAR's minority-polarity flux in about a day. In contrast, lone BMRs with large sunspots have much more flux than the nearby surrounding quiet-region network in which they emerge. Presumably for this reason, a lone large-sunspot BMR, compared to our BEARs, takes longer relative to its emergence time for its flux to decay by cancellation and spreading enough that the BMR is no longer discernible.

In Table 1 we also list for each BEAR whether the BEAR's minority-polarity flux is extinguished by (1) cancellation at an inside interval of the minority-flux-encircling PIL of the BEAR's magnetic anemone, or (2) by cancellation at an inside & outside interval of that PIL, or (3) by cancellation at an outside interval of that PIL, or (4) not by evident cancellation ("End PIL Place" category "None" in column eight of Table 1), but either by becoming so dispersed that it is no longer discernible in HMI magnetograms or by becoming confused/intermixed with new separately emerged minority-polarity flux. For our 10 BEARs, Figure 21 is the histogram for the four categories of how the BEAR's minority-polarity flux meets its end. It graphically displays that the end of the minority flux is by cancellation for only six of the 10 BEARs. Of these six, the final cancellation is at an inside PIL for two (BR1 and BR10), at an inside & outside PIL for one (BR7), and at an outside PIL for three (BR3, BR4, and BR9). For four BEARs, even though much of the minority-polarity flux is evidently lost by cancellation, the end of the minority flux is not by observed cancellation. In three of these (BR2, BR5, and BR6), the trackable life of the BEAR's minority flux ends by the BEAR's remnant minority flux becoming mixed with newly emerged minority flux. In one (BR8), the minority flux disappears by dispersing. Thus, as Figure 21 displays, among our 10 BEARs there is no outstanding preference for the manner of the final demise of the BEAR's minority-polarity flux. This is plausibly due to the randomness of the convection flows that govern the evolution of both a BEAR's flux and the nearby network flux.

Our 10 BEARs produce 43 obvious microflare events during their cradle-to-grave evolution. Figure 22 is the histogram for the number of microflare events a BEAR produces in its lifetime. It shows that, among our 10 BEARs, the total number of events a BEAR produces ranges from 0 to 12, and is unevenly distributed over that range. The average whole number of events each BEAR produces is 4. Six BEARs each produce less than four events, and four BEARs each produce more than four events. Of particular significance, no event happens during the life of one BEAR (BR1) even though its minority-polarity flux is clearly consumed by cancellation, mostly at the BEAR's Inside PIL. This indicates that flux cancellation at a BEAR's minority-flux-encircling PIL does not always eventually cause a microflare-producing magnetic explosion. The absence of any microflare events in BR1 and the wide range of the number of microflare events produced by the other nine BEARs are indications that the number of microflare events that happen in a BEAR depends on other conditions or processes in addition to enough flux cancellation at the underlying PILs. Nevertheless, our observations indicate that flux cancellation at the underlying PIL is usually a necessary prerequisite for the occurrence of a microflare in a BEAR's magnetic anemone.

### *3.5.2. The 43 Microflare Events*

Table 2 gives the distribution of our 43 microflare events among the nine possible pairings of the type of event and the place of the event's underlying PIL with respect to the BEAR's anemone magnetic field. The number in each of the nine slots in the three-slot by three-slot interior of Table 2 is the number of microflare events having that slot's pairing of event type and category of PIL place. From the numbers in the nine slots we have the following seven results. (1) Some events of each type occur on underlying PILs of each category of PIL place. (2) Only a small minority (5/43) of the events are confined events, and these are roughly evenly distributed among the three place categories of their underlying PILs. (3) The other 38 events are roughly evenly divided between mostly-confined events (20) and blowout events (18). (4) Only a small minority (2/20) of the mostly-confined events occur on inside PILs, a larger minority (6/20) occur on inside & outside PILs, and a small majority (12/20) are on outside PILs. (5) Only a small minority (2/18) of the blowout events are on inside PILs. The other 16 blowout events occur roughly equally often on either inside & outside PILs (7/16) or outside PILs (9/16). (6) The uneven distribution of the mostly-confined events and blowout events among the three categories of PIL place results in a similarly uneven distribution of the total number of events among the three categories of PIL place: a small minority (6/43) sit on inside PILs, a larger minority (14/43) sit on inside & outside PILs, and a small majority (23/43) sit on outside PILs. (7) On the other hand, the 43 events are nearly evenly split between 23 that erupt entirely from outside PILs and 20 that erupt at least partly from an inside PIL, i.e., from either an inside PIL or an inside & outside PIL.

In the same way as for the lack of obvious preference for a BEAR's minority-flux end-of-life PIL place (shown in Figure 21), we think the lack of preference of microflare-event occurrence place between outside PILs and either inside or inside & outside PILs is not surprising because it could reasonably result from the randomly changing convection flows that govern the evolution of both a BEAR's flux and neighboring network flux. The following simple possible evolution is consistent with the roughly even split between events on outside PILs and events on either inside or inside & outside PILs. As it emerges, the BEAR's minority-polarity flux moves away from the BEAR's majority-polarity flux and could collide and partly cancel with encountered majority-polarity network flux. After emergence, if the BEAR's majority-polarity flux is melded with a patch of majority-polarity network flux (as it usually is in our 10 BEARs), and if that patch of network flux is at the edge of a supergranule convection cell, then the supergranule's horizontal flow into the downflow sink on which the network flux patch presumably sits could sweep any remainder of the BEAR's minority flux – the BEAR's minority flux that has not already canceled with majority-polarity network flux – back to cancel with the BEAR's majority-polarity flux and/or with the network flux patch that is melded with the BEAR's majority-polarity flux.

From Table 1, of our 43 microflare events, 8 happen while the BEAR's flux is emerging and 35 happen after the end of the BEAR's flux emergence. This is roughly consistent with the average (emergence time/lifetime) ratio of the 10 BEARs being 0.24. From that value, the expected number of events during emergence (0.24 x 43) is 10, assuming that the chance of an event happening is constant throughout the life of a BEAR.

We define the lag time of a microflare event to be the time from the BEAR's minority-flux birth time to the event's maximum-visibility time. For each microflare event, these two times are listed in Table 1. Figure 23(a) is a histogram showing the distribution of the (event lag time/BEAR lifetime) ratio values of all 43 of our microflare events. It shows that our microflare events do not occur with roughly the same frequency throughout their BEAR's life. Instead, they occur more often in the last half of their BEAR's life than in the first half. Figure 23(a) shows that of the 43 microflare events, 31 occur in the last half of their BEAR's life and 12 occur in the first half.

Figures 23(b), 23(c), and 23(d) are the (event lag time/BEAR lifetime) histograms for (b) confined events, (c) mostly-confined events, and (d) blowout events. The five confined events are roughly evenly split between happening in the last half of their BEAR's life (3) or in the first half (2). Similarly to the confined events, the 20 mostly-confined events are also roughly evenly split between happening in the last half of their BEAR's life (11) or in the first half (9). As the histogram in Figure 23(d) shows, of the 18 blowout events, 17 happen in the last half of their BEAR's life and only 1 happens in the first half. Thus, most of the first-half/last-half imbalance in the histogram of all the events in Figure 23(a) comes from the blowout events.

A plausible reason for the strong first-half/last-half imbalance for when blowout events occur is the following. The farther into a BEAR's life, the greater the time for flux cancellation to possibly build up an explosive sheared-core-field/flux rope along and low above a cancellation PIL and to simultaneously reduce the amount of flux left in the outer part of the anemone lobe field that overarches and acts to confine an explosion of its core field. This suggests that the later in a BEAR's life a microflare-event magnetic explosion occurs, the more likely the explosion will be a blowout explosion instead of a confined explosion or a mostly-confined explosion.

As Table 1 shows, none of the 18 blowout-jet-making magnetic explosions happen on a BEAR's inside PIL during the BEAR's emergence. That is, none of the 18 blowout jets is made in the way envisioned by Moore et al (2010) for blowout jets.

The most striking result in Table 1 is the last column. It has a Yes for each of the 43 microflare events. The Yes means that before and during the microflare event the HMI magnetogram movie shows evidence of magnetic flux cancellation at the PIL underlying the miniature flare arcade. Specifically, before and during each microflare event – as we have pointed out in the nine example events – the magnetogram movie shows one or more of the following three indications of flux cancellation at the PIL interval on which the microflare arcade sits: (1) majority-polarity flux and some of the BEAR's minority-polarity flux are impacted against each other at the microflare's underlying PIL, (2) one or both of these two flux patches is discernibly shrinking, (3) the two flux patches are discernibly converging on that PIL. Another striking result shown in Table 1 is that BEAR BR1 has no microflare event, even though nearly all of BR1's minority-polarity flux is consumed by cancellation with majority-polarity flux. (Albeit BR1's minority-polarity flux mostly cancels with BR1's majority-polarity flux at BR1's Inside PIL, three other BEARs – BR7, BR9, and BR10 – produce a total of six microflare events on inside cancellation PILs.) From these two striking results we infer (1) that flux cancellation at the underlying PIL prepares and finally triggers each microflare-event explosion of the core field, but (2) if the field enveloping the cancellation PIL lacks some required initial condition that enables the ensuing cancellation to gradually build up the core-field's free energy and eventually trigger the explosion, then no microflare explosion occurs no matter how much of the BEAR's minority-polarity flux cancels at that PIL. In Section 4, we speculate that the required initial condition is that the BEAR's anemone-lobe magnetic arcade enveloping the cancellation PIL have, when the cancellation ensues, initial shear above some threshold amount that enables the cancellation to amplify the field shear in the core of the arcade.

## 4. Summary and Discussion

Using HMI magnetograms and AIA 211 Å images, we examined in detail the cradle-to-grave magnetic field evolution and microflare production of 10 bipolar ephemeral active regions (BEARs) in coronal holes on the central disk. Each BEAR is evidently made by the emergence of a magnetic flux-rope Ω loop from below the photosphere. Each BEAR's magnetic field emerges in less than half a day and decays to become indiscernible in less than two days after the start of emergence. At the end of emergence, each BEAR's bipolar magnetic flux spans ~10,000 km and the amount of flux of either polarity is in the range 1 – 9 x $10^{19}$ Mx, too little to make even the smallest sunspot (i.e., a dark pore the size of a granule, ~1000 km across). Even so, because solar bipolar magnetic regions of all sizes – from the smallest ones that have the size and lifetime of a granule to the largest ones that have large sunspots, span ~200,000 km, and last several solar

rotations – are evidently made basically same way by the emergence of a flux-rope Ω loop of their span, we take our BEARs to be representative of all larger bipolar magnetic regions in terms of their magnetic evolution. That is, we expect BEARs to have cradle-to-grave magnetic field evolution that is essentially similar to that of all larger bipolar active regions. Specifically, we expect that a BEAR's magnetic field evolution leading to microflare-producing magnetic explosions has essential similarities to the less continuously trackable magnetic field evolution leading to flare/CME-producing magnetic explosions in large-sunspot bipolar active regions. The primary goal of our investigation was to identify any evolutionary magnetic process in BEARs that usually directly precedes microflare-producing magnetic explosions. Finding an obvious evolutionary precursor process for microflares in BEARs would imply that the same process should prepare many flare/CME magnetic explosions in large-sunspot active regions.

As a BEAR's flux-rope Ω loop emerges in a coronal hole, the polarity of one foot of the emerging magnetic arch is the majority polarity (the polarity of the majority of the magnetic flux network in the coronal hole) and the flux in the other foot has the opposite polarity, the minority polarity. As the magnetic arch emerges, reconnection of its outside with the coronal hole's ambient majority-polarity open field makes the BEAR's magnetic field into an anemone-shape circular magnetic arcade that connects the BEAR's minority-polarity flux to ambient coronal-hole majority flux as well as to the BEAR's majority-polarity flux. The BEAR anemone magnetic arcade encloses in its core the polarity inversion line (PIL) that encircles the BEAR's minority-polarity flux, as in Figure 1. Each microflare-producing magnetic explosion is presumably an explosion of the anemone's sheared core field along an interval of the BEAR's minority-flux-encircling PIL. The PIL interval of each microflare explosion is one of three categories: (1) an inside PIL interval between the BEAR's two domains of opposite-polarity flux, (2) an outside PIL interval between the BEAR's minority-polarity flux and ambient coronal-hole majority-polarity flux, or (3) an inside & outside PIL interval, part of which is an inside interval and the rest of which is an outside interval. Because the BEAR anemone is encased in surrounding coronal-hole open magnetic field, as in Figure 1, each microflare-producing magnetic explosion can be assigned to one of three categories for the degree to which the explosion is confined inside the anemone's lobe in which it occurs: (1) confined, (2) mostly confined, or (3) blowout. A confined microflare-producing explosion drives no external reconnection of the enveloping lobe that produces a discernible jet spire in the BEAR's AIA 211 Å movie. Only a miniature flare arcade centered on the explosion's PIL interval is produced, presumably by internal reconnection of the erupting core field. A mostly-confined explosion drives a moderate amount of external reconnection, enough to produce only a single-strand narrow jet spire. The magnetic explosion also produces a miniature flare arcade arching over the explosion's PIL interval. The miniature flare arcade is the jet event's jet bright point (JBP), presumably made by internal reconnection of the legs of the erupting core field a la Sterling et al (2015). A blowout explosion makes a multi-strand broad jet spire of width comparable to that of the anemone, and also produces a JBP microflare arcade seated on the explosion's PIL interval, presumably by internal reconnection a la Sterling et al (2015).

A secondary purpose of our investigation was to observe how often the core-field magnetic explosion that makes a blowout jet is seated entirely on the BEAR's inside PIL during the BEAR's emergence, in the manner proposed by Moore et al (2010). The idea of Moore et al (2010) for the production of blowout jets is that the core of a BEAR's emerging magnetic arch is often so strongly sheared and twisted before and during its emergence that it undergoes a blowout explosion as soon as it has emerged enough. Of the 43 microflare-making core-field explosions in our 10 BEARs, 18 make blowout jets. None of these 18 explosions explode from either an inside PIL or an inside

& outside PIL during their BEAR's emergence. This is evidence that the core of a BEAR's emerging magnetic arch seldom, if ever, undergoes a blowout explosion. A much larger sample of emerging BEAR's, say 5 or 10 times larger, might have some blowout explosions that do explode from their emerging BEAR's inside PIL. The 18 blowout explosions that happen during the lives of our 10 BEAR's indicate no more than ~ 5% of blowout-jet-making magnetic explosions in BEARs are seated on the inside PIL while the BEAR's magnetic arch is emerging. That is far less often than Moore et al (2010) anticipated.

The primary and most significant finding of our study is that before and during each of the 43 microflare-producing magnetic explosions (5 confined, 20 mostly confined, 18 blowout) that occur in the cradle-to-grave evolution of the 10 BEARs in central-disk coronal holes, the HMI magnetogram movie shows discernible evidence of flux cancellation at the PIL interval on which the microflare arcade sits. In the magnetogram movie, before and during each microflare, opposite-polarity flux patches bracketing the microflare's PIL interval are impacted against each other at the PIL and/or either or both flux patches are discernibly diminishing and/or the two flux patches are discernibly converging on the PIL. This is evidence for flux cancellation at the underlying PIL being the basic process that builds the explosiveness of the core field enveloping the PIL and eventually triggers the explosion, regardless of whether the explosion is confined, mostly confined, or blowout. We therefore expect that for many confined-flare and blowout-flare/CME magnetic explosions in sunspot active regions, flux cancellation at the PIL of the pre-explosion core field prepares and triggers the explosion in essentially the same way as for the microflare-producing core-field explosions in coronal-hole BEARs.

Our evidence that flux cancellation prepares and triggers microflare magnetic explosions in BEARs is in agreement with the observational results of Tiwari et al (2014). They studied subflares in a large active region's sheared core field. They showed that each subflare was evidently triggered by an underlying microflare magnetic explosion that was centered on a compact site of ongoing flux cancellation at the subflare's underlying PIL.

Our results are consistent with observations reported by Tiwari et al (2019) of fine-scale explosive energy release in the core of a recently-emerged bipolar active region. See also Tiwari et al. (2022) for fine-scale bright dots in an emerging flux region (X-ray/coronal bright point). These explosive events are several times smaller in lateral extent than any of our 43 microflare magnetic explosions in BEARs. Because they sit on sharp PILs, they are plausibly magnetic explosions that are prepared and triggered by flux cancellation in the same way as for our microflare events in BEARs. Sudden UV and EUV brightenings at the base of jetlets at edges of magnetic network lanes are of the width of the explosive energy releases of Tiwari et al (2019), and many are observed to be at similarly small sites of flux cancellation (Panesar et al 2018b, 2019, 2020, 2021). The observed flux cancellation suggests that network jetlets, and all of the solar campfire events examined by Panesar et al (2021) (whether or not the event produces a jet spire), are made by magnetic explosions that are like the confined explosions, mostly-confined explosions and blowout explosions in BEARs but of still smaller scale.

Our observations are consistent with the way in which the flux cancellation builds the sheared/flux-rope core field being the way proposed by van Ballegooijen & Martens (1989) for flux cancellation in core field that has some initial shear. In that scenario, reconnection of the sheared core field low above the cancellation PIL is essential for amplifying the shear and building a flux rope above the PIL. Nine of our 10 BEARs have at least one microflare-producing magnetic explosion during the lifetime of the BEAR's minority-polarity flux, but one BEAR (BR1) has none. This suggests that the canceling core field needs to have some minimum degree of initial shear/twist for flux cancellation to make the field able to explode. Evidently, most BEARs either emerge with sufficient magnetic twist or their magnetic evolution gives their field sufficient twist and/or shear for flux cancellation to eventually cause some of the BEAR's core field to explode. Perhaps the flux cancellation in BR1 does not lead to any microflares because the initial magnetic shear is below some critical degree, so that the flux cancellation does not entail the reconnection envisioned by Rabin et al (1984), van Ballegooijen & Martens (1989), and Moore & Roumeliotis (1992), but instead is simply the submergence of shrinking field loops as their two feet approach each other at the PIL (e.g., Rabin et al 1984).

Only 6 of the 43 microflare events in our 10 BEARs exploded entirely from an inside interval of the BEAR's PIL. That is, 37 (86%) exploded from either an outside interval or an inside & outside interval of the BEAR's PIL. This result is similar to the observational result found by Mackay et al (2008): 92% of filaments in and around solar active regions reside entirely or partly on a PIL interval that is outside of the active region, i.e., only 8% reside entirely on a PIL interval that is entirely inside the active region. This observed similarity between the PIL placement of microflare events in BEARs and the PIL placement of filaments in and around active regions is reasonable if (1) the sheared field and flux rope that we suppose erupts to drive our microflare events is built by convection-driven flux cancellation at the underlying PIL (a la van Ballegooijen & Martens 1989) and (2) – for which there is observational evidence (Martin 1998; Panesar et al 2017) – the sheared field and flux rope that holds a filament is likewise built by convection-driven flux cancellation at the underlying PIL. Hence, we take the finding of Mackay et al (2008) to be further support for our premise that solar active regions of all sizes should have basically similar magnetoconvection-driven evolution of their magnetic field, and for our inference that convection-driven flux cancellation at the underlying PIL prepares and triggers many major flare and CME magnetic explosions in sunspot active regions.

As we have tried to make clear in Section 1.4, our view of the generation of either a so-called standard jet (narrow-spire jet) or a so-called blowout jet (wide-spire jet) in this paper is that of Sterling et al (2015), namely that in either case the driver of the jet event is an eruption of a flux rope (often carrying a minifilament) seated in the base of the jet. We suppose that in our narrow-spire jet events, the flux-rope eruption is mostly confined in the jet's closed-field anemone base. We suppose that in our wide-spire jet events, the flux-rope eruption blows out the enveloping lobe of the anemone base, thereby driving widespread external reconnection of first the envelope field and then the flux rope. We suppose that because the external reconnection in a blowout flux-rope eruption is more widespread than in a mostly-confined flux-rope eruption that makes a narrow-spire jet, it makes a wide jet spire instead of a narrow jet spire.

As is depicted by the schematic drawing in Figure 8(b) of Shibata (1999), Shibata (1999) proposed that a coronal jet spire could be the product of external reconnection of an erupting magnetic flux rope with encountered far-reaching magnetic field. At first glance, it might be thought that this proposal is essentially the same as the proposal of Sterling et al (2015) for how coronal jet spires are made: by external reconnection of erupting-flux-rope-driven closed field with encountered far-reaching field (external reconnection that starts at the outside of the anemone lobe enveloping the erupting flux rope and often progresses into the flux rope). Actually, the Sterling et al (2015) scenario for jet spire production fundamentally differs from that of Shibata (1999). In the Sterling et al (2015) scenario, before the flux rope's eruption drives the enveloping arcade's external reconnection that makes the spire, the minifilament-holding flux rope is already present low above the PIL in the core of that arcade. In the Shibata (1999) scenario, the erupting flux rope is made by reconnection in the current sheet between the outside of an emerging magnetic arcade and encountered far-reaching ambient field. We take this difference between the Sterling et al (2015) scenario and the Shibata (1999) scenario to be a clear fundamental difference: in the Sterling et al (2015) scenario the erupting flux rope forms in the sheared core field *inside* a lobe of the jet-base magnetic anemone and, before and during eruption onset, traces that lobe's PIL (at which PIL we now know flux is canceling), whereas in the Shibata (1999) scenario the erupting flux rope forms in a reconnection current sheet at the *outside* of a lobe of the jet-base anemone, which lobe is an emerging Ω loop. Furthermore, in the Sterling et al (2015) scenario the JBP is made by internal reconnection of the legs of the magnetic arcade that envelops the erupting flux rope, whereas in the Shibata (1999) scenario the JBP is made by an emerging arcade's external reconnection that also makes the erupting flux rope. [In the Sterling et al (2015) scenario and in the Shibata (1999) scenario, the base of the jet is a two-lobed magnetic anemone having one lobe smaller than the other. In both scenarios, the jet-base anemone is encased in far-reaching unipolar ambient magnetic field (as in our Figure 1). Thus, the topology of the pre-jet magnetic field is the same in both scenarios. Even so, the two scenarios are fundamentally different as follows. In the Sterling et al (2015) scenario, (i) the larger lobe may or may not be emerging during the production of the jet, (ii) the minifilament flux rope originates inside the smaller lobe, (iii) the JBP is made by reconnection inside the smaller lobe (by internal reconnection of the smaller lobe's field under the erupting flux rope), (iv) and the spire is made by the external reconnection of the erupting smaller lobe, which lobe envelopes and erupts along with the erupting minifilament flux rope. In contrast, in the Shibata (1999) scenario, (i) the larger lobe is assumed to be emerging to drive its jet-spire-making external reconnection, (ii) the erupting flux rope originates in the external-reconnection current sheet between the emerging lobe and ambient far-reaching field, (iii) the JBP is the arcade of new reconnection-heated loops added to the outside of the smaller lobe by the emerging larger lobe's external reconnection, and (iv) the same external reconnection makes the erupting flux rope that escapes into the far-reaching field to make the jet spire.] In addition, none of our 18 blowout events happen during the emergence of the BEAR's Ω loop. This result would be highly unlikely if blowout jets were typically made in the manner of the Shibata (1999) scenario, but is not at odds with the Sterling et al (2015) scenario. As we have tried to lay out in Section 1.4, many recent published observational studies of coronal jets present conclusive evidence for the Sterling et al (2015) scenario. We know of no published observational studies of coronal jets that present conclusive evidence for the Shibata (1999) scenario.

Shimojo et al (2007) present, in Hinode/XRT coronal X-ray images, two coronal jets. Each jet occurs near the limb and is made by the blowout eruption of what appears to be an erupting expanding flux rope, which eruption drives the onset and growth of both the JBP and jet spire as well as the brightening and growth of the non-JBP lobe of the jet's magnetic anemone base. Shimojo et al (2007) argue that the XRT observations are nicely consistent with the Shibata (1999) scenario for jet production by an erupting flux rope. However, from even very close scrutiny of the Shimojo et al (2007) XRT images, we cannot discern whether the erupting flux rope formed in and erupted from a current sheet at a site of external reconnection between ambient far-reaching field and the outside of the putative emerging lobe's leg that is closest to the JBP as in the Shibata (1999) scenario, or instead formed and started erupting from low above the PIL inside the JBP lobe of the anemone as in the Sterling et al (2015) scenario. In addition, in the XRT images, the non-JBP lobe rapidly brightens and grows larger *as the erupting flux rope erupts and the JBP and jet spire turn on and grow*. The Shibata (1999) scenario schematic shown in Shimojo et al (2007) indicates the opposite, namely that the external reconnection of the putatively-emerging non-JBP lobe (*during that reconnection's production of the JBP and jet spire*), by progressively opening outer loops of that lobe, plausibly should make that lobe progressively smaller, not progressively larger as is observed. In contrast, in the Sterling et al (2015) scenario, the spire-making external reconnection of the lobe enveloping the erupting minifilament flux rope simultaneously progressively adds reconnection-heated loops to the outside of the non-JBP lobe, making that lobe progressively larger, which is consistent with what the XRT images show. Therefore, because the XRT images appear to us to be at least as consistent with the Sterling et al (2015) scenario as with the Shibata (1999) scenario, we judge that the Shimojo (2007) evidence for the Shibata (1999) scenario is not conclusive.

Speculation about and discussion of the physics of the spire-making external reconnection of a BEAR's exploding anemone lobe having an erupting minifilament flux rope in its core is beyond the scope of this paper. We think that the empirical results reported in this paper are true regardless of the reconnection physics governing the spire-making external reconnection in our jet events.

In any of our 38 jet-spire-making microflare eruptions, we cannot prove that the jet spire was not made essentially according to the Shibata (1999) scenario, although, as we have discussed, we doubt that any were made that way. Whether the spire was made essentially according to the Shibata (1999) scenario, or essentially according to the Sterling et al (2015) scenario, or by any other conceivable way of driving the BEAR anemone's external reconnection that makes the spire, makes no difference to our empirical evidence that each of our 43 microflare eruptions was prepared and triggered by flux cancellation at the PIL underlying the microflare arcade, and makes no difference to the implication of this result that flux cancellation at the underlying PIL should be expected to prepare and trigger many major flare/CME eruptions in sunspot active regions.

Chintzoglou et al (2019) have shown that in some sunspot active regions the underlying PIL in magnetic explosions that make major flares and CMEs is between the magnetic flux in one foot (foot A) of an emerging or emerged flux-rope Ω loop and an encroaching opposite-polarity foot (foot B) of another emerging or emerged flux-rope Ω loop. As one or the other or both Ω loops emerge, reconnection in the corona connects foot-A flux to foot-B flux by a magnetic arcade over the PIL between foot A and foot B. As foot A and foot B approach each other and collide, their relative motion is often not nearly head-on but has a large component of shear. The relative motion thereby shears the PIL-enveloping magnetic arcade, giving it corresponding free magnetic energy that makes an explosion of the arcade possible. The findings of Chintzoglou et al (2019) are consistent with our view that flux cancellation at the PIL further prepares and finally triggers the sheared field to explode.

Our observations suggest that the essential process for the buildup and triggering of the magnetic explosion for either a major confined flare or a major CME and its flare is enough flux cancellation at the PIL of a magnetic arcade that has sufficient initial magnetic shear, not how the arcade acquires that initial shear. Our observations of the enveloping magnetic arcade and the flux cancellation leading to microflare-making magnetic explosions of all three types (confined, mostly confined, and blowout) from the inside PIL of BEARs suggest that shearing of the pre-explosion magnetic arcade by the collision of opposite-polarity feet from two separately emerged Ω loops of a multi-bipolar active region is not necessary for the preparation and triggering of magnetic explosions for major flares and CMEs. From our observations, we expect that enough flux cancellation at the inside PIL of a single emerged flux-rope Ω loop for a bipolar active region that is big enough to have large sunspots can – in the absence of shearing flows – prepare and trigger a magnetic explosion for either a major confined flare or a major CME and its flare, provided the emerged flux-rope Ω loop has sufficient initial magnetic twist.

This expectation is bolstered by the observations of Sterling et al (2018). They follow over several days the cradle-to-eruption magnetic evolution of each of two separate isolated small bipolar active regions that are bigger than our BEARs, big enough to have small sunspots when the active region has finished emerging. In each active region, a CME-making blowout explosion explodes from the inside PIL of the active region's emerged Ω-loop magnetic field after the two feet of the Ω loop converge on the PIL and much of the Ω loop's flux has cancelled at the PIL. There is no obvious relative shearing motion as the two opposite-polarity flux domains approach each other and merge. The magnetic evolution from emergence to blowout explosion suggests that each active region's Ω loop emerged with enough twist for the flux cancellation to build from that twist – a la van Ballegooijen & Martens (1989) – the explosive sheared core field/flux rope along and above the inside PIL.

The reported results are empirical results from our particular random set of 10 BEARs in central-disk coronal holes. We are confident that the main results from this pilot study will not be negated by a much larger sample of, say, 50 – 100 BEARs in coronal holes. That is, we expect that the main trends found from our small sample of BEARs might well be found to be of somewhat different strength, but the same tendencies will be found from any random sample of central-disk coronal-hole BEARs that is at least as large as ours.

Acknowledgements

The reported research was supported by the Heliophysics Division of NASA's Science Mission Directorate through the Heliophysics Guest Investigators program, the Heliophysics System Observatory Connect program, and the Solar Dynamics Observatory (SDO) mission. SDO is a mission of the Heliophysics Division's Living With a Star program. We thank the SDO/AIA and SDO/HMI teams for providing the AIA and HMI data. The work used Helioviewer, JHelioviewer, NASA ADSABS, and Solar Software. The anonymous referee helped us restructure the paper for greater clarity and led us to point out the independence of the paper's results on the specific scenario for making the jet spire in our microflare eruptions that have a jet spire.

**Table 1. Ten Coronal-Hole Cradle-to-Grave Bipolar Ephemeral Active Regions (BEARs) and their 43 Microflare Events**

| BEAR | | | | | | | | Microflare Event | | | |
|---|---|---|---|---|---|---|---|---|---|---|---|
| No. | Minority-Polarity Magnetic Flux | | | | | | | Max. Vis. Time (m/d UT) | PIL Place (Inside or Outside or Inside & Outside) | Type (Blowout or Mostly Confined or Confined) | Evident Pre-Event Flux Canc. at PIL (Yes/ No) |
| | Pol. (Pos./ Neg.) | Birth Place (x”, y”) | Birth Time (y m/d UT) | Max. ($10^{19}$ Mx) | Max. Time (m/d UT) | End Time (m/d UT) | End PIL Place (Inside or Outside or Inside & Outside or None) | | | | |
| BR1 | Pos. | -70, 104 | 2012 04/09 16:40 | 2.6 | 04/09 20:54 | 04/10 06:00 | Inside | No Event | No Event | No Event | No Event |
| BR2 | Pos. | -122, 290 | 2012 06/30 21:30 | 5.4 | 07/01 05:40 | 07/02 06:30 | None (New Emerg.) | 07/01 02:45 | Outside | Confined | Yes |
| | | | | | | | | 07/01 08:35 | Outside | Blowout | Yes |
| | | | | | | | | 07/01 15:00 | Outside | Blowout | Yes |
| | | | | | | | | 07/01 19:14 | Outside | Blowout | Yes |
| | | | | | | | | 07/02 02:20 | Outside | Blowout | Yes |
| BR3 | Neg. | -495, -276 | 2012 08/13 13:00 | 5.5 | 08/13 16:18 | 08/14 20:30 | Outside | 08/14 17:30 | Inside & Outside | Blowout | Yes |
| | | | | | | | | 08/14 19:10 | Inside & Outside | Blowout | Yes |
| | | | | | | | | 08/14 19:48 | Inside & Outside | Blowout | Yes |
| BR4 | Pos. | -581’ 396 | 2013 01/31 10:15 | 3.3 | 01/31 13:48 | 01/31 21:00 | Outside | 01/31 11:38 | Outside | Mostly Confined | Yes |
| | | | | | | | | 01/31 13:23 | Outside | Mostly Confined | Yes |
| | | | | | | | | 01/31 13:40 | Outside | Mostly Confined | Yes |
| | | | | | | | | 01/31 13:55 | Outside | Mostly Confined | Yes |
| | | | | | | | | 01/31 15:31 | Outside | Mostly Confined | Yes |
| | | | | | | | | 01/31 17:34 | Outside | Mostly Confined | Yes |
| | | | | | | | | 01/31 17:44 | Outside | Mostly Confined | Yes |
| | | | | | | | | 01/31 18:40 | Outside | Blowout | Yes |
| | | | | | | | | 01/31 20:26 | Outside | Blowout | Yes |
| BR5 | Pos. | 155, 596 | 2013 02/05 09:00 | 7.5 | 02/05 12:00 | 02/06 00:20 | None (New Emerg.) | 02/05 09:41 | Outside | Mostly Confined | Yes |
| | | | | | | | | 02/05 19:03 | Inside & Outside | Mostly Confined | Yes |
| BR6 | Neg. | 273, 592 | 2013 02/13 00:30 | 4.1 | 02/13 07:12 | 02/13 23:00 | None (New Emerg.) | 02/13 14:38 | Outside | Mostly Confined | Yes |
| | | | | | | | | 02/13 17:01 | Outside | Mostly Confined | Yes |
| | | | | | | | | 02/13 19:17 | Outside | Blowout | Yes |
| | | | | | | | | 02/13 19:31 | Outside | Blowout | Yes |
| | | | | | | | | 02/13 21:01 | Outside | Blowout | Yes |
| BR7 | Pos. | 107, 153 | 2013 02/27 05:15 | 1.8 | 02/27 06:48 | 02/27 18:00 | Inside & Outside | 02/27 05:38 | Outside | Mostly Confined | Yes |
| | | | | | | | | 02/27 05:52 | Outside | Mostly Confined | Yes |
| | | | | | | | | 02/27 06:19 | Inside & Outside | Mostly Confined | Yes |
| | | | | | | | | 02/27 07:36 | Inside | Confined | Yes |
| | | | | | | | | 02/27 12:00 | Inside & Outside | Confined | Yes |
| | | | | | | | | 02/27 12:46 | Inside & Outside | Mostly Confined | Yes |
| | | | | | | | | 02/27 14:10 | Inside & Outside | Blowout | Yes |
| | | | | | | | | 02/27 14:37 | Inside & Outside | Mostly Confined | Yes |
| | | | | | | | | 02/27 14:51 | Inside & Outside | Mostly Confined | Yes |
| | | | | | | | | 02/27 14:59 | Inside & Outside | Mostly Confined | Yes |
| | | | | | | | | 02/27 15:13 | Inside | Mostly Confined | Yes |
| | | | | | | | | 02/27 15:24 | Inside | Blowout | Yes |
| BR8 | Neg. | -350, -375 | 2013 03/01 20:00 | 4.2 | 03/02 07:48 | 03/03 06:30 | None (Dis-persed) | 03/02 15:29 | Inside & Outside | Blowout | Yes |
| | | | | | | | | 03/02 18:39 | Inside & Outside | Blowout | Yes |
| | | | | | | | | 03/02 20:45 | Outside | Confined | Yes |
| BR9 | Neg. | -272, 163 | 2013 05/02 03:00 | 7.0 | 05/02 05:00 | 05/03 03:00 | Outside | 05/02 22:46 | Inside & Outside | Blowout | Yes |
| | | | | | | | | 05/03 O1:03 | Inside | Blowout | Yes |
| | | | | | | | | 05/03 02:00 | Inside | Mostly Confined | Yes |
| BR10 | Pos. | -374, 101 | 2014 02/23 01:00 | 8.8 | 02/23 07:30 | 02/24 14:30 | Inside | 02/24 07:47 | Inside | Confined | Yes |

| **Table 2. Distribution of 43 Microflare Events by Type and PIL Place** | | | | |
|---|---|---|---|---|
| | | | | |
| **Type** | **PIL Place** | | | **Total** |
| | Inside | Inside & Outside | Outside | |
| Confined | 2 | 1 | 2 | 5 |
| Mostly Confined | 2 | 6 | 12 | 20 |
| Blowout | 2 | 7 | 9 | 18 |
| **Total** | 6 | 14 | 23 | 43 |

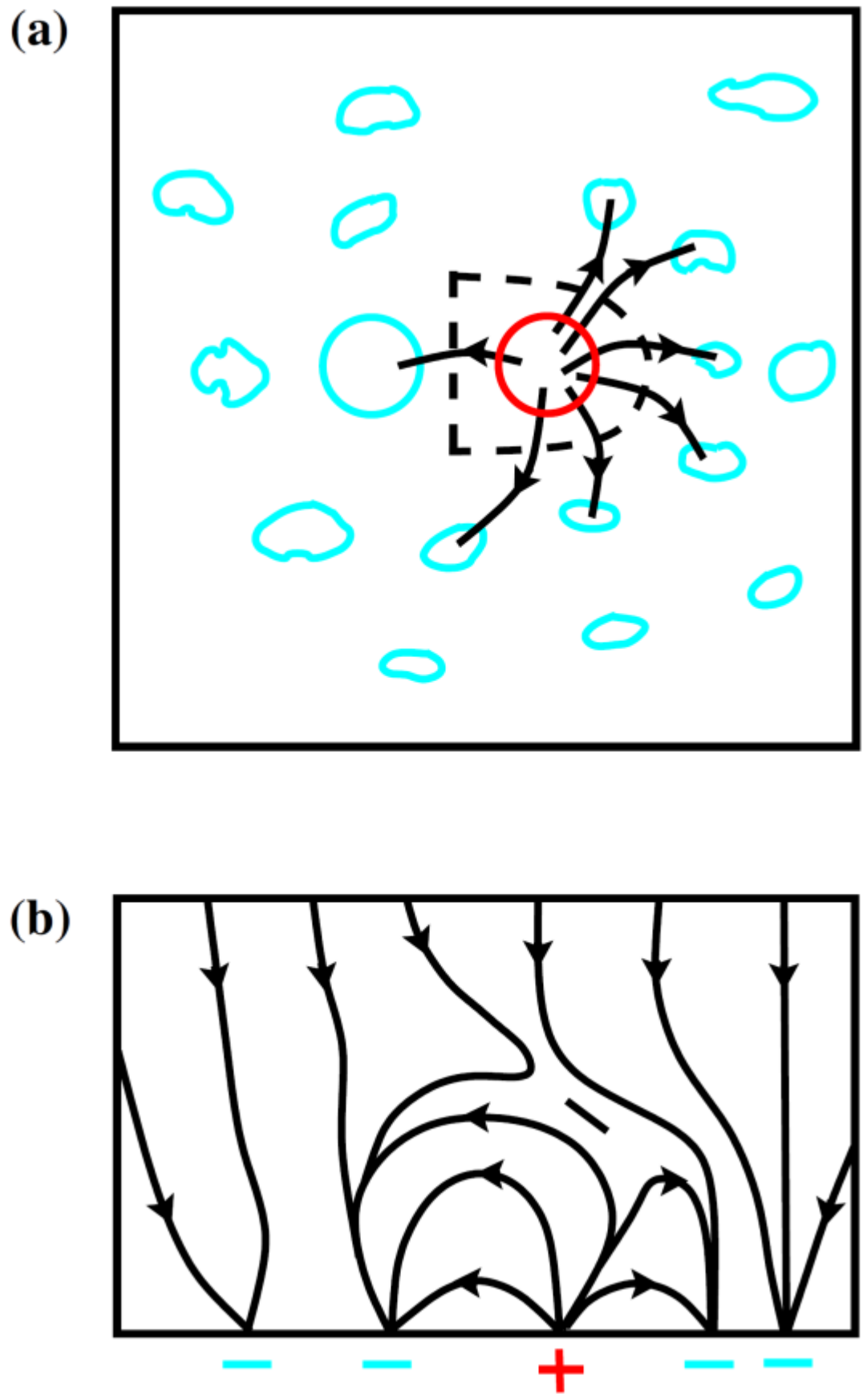

**Figure 1.** Topology of the 3D magnetic field in and around a BEAR in a coronal hole in which the open magnetic field stems from negative polarity flux in the photosphere. Blue closed curves and minus signs mark majority-polarity negative flux clumps. The red circle and the plus sign mark the BEAR's minority-polarity positive flux. Black curves are magnetic field lines. (a) Top view. In this drawing and in the magnetograms and coronal images of BEARs in this paper solar north is up and solar west is to the right. The blue circle is the BEAR's majority-polarity flux. The irregular closed blue curves are nearby majority-polarity flux clumps of the magnetic network. The D-shape dashed black curve is the PIL that encircles the BEAR's minority-polarity flux. The north-south straight segment of the PIL is the PIL's inside interval, between the BEAR's opposite-polarity flux domains. The rest of the PIL is the PIL's outside interval, between the BEAR's minority-polarity flux and network majority-polarity flux. The black field lines show the anemone form of the closed magnetic field that connects the BEAR's minority-polarity flux to the BEAR's majority-polarity flux and to majority-polarity network flux. (b) South side view of the 2D magnetic field in the east-west vertical plane through the centers of the BEAR's two opposite-polarity flux circles in panel (a). The plus sign marks the BEAR's minority-polarity flux domain, the first minus sign left of the plus sign marks the BEAR's majority-polarity flux, and the other minus signs mark majority-polarity network flux clumps. The closed black curves are lines of the closed field that constitutes the BEAR's magnetic anemone. The open black curves are lines of the open field encasing the anemone closed field. The black dash sandwiched between open and closed field is the magnetic-null current sheet at which the outside of the BEAR's emerging magnetic arch reconnects with ambient open field to both open the outside of the arch and connect the BEAR's outer minority-polarity flux to nearby majority-polarity network flux to make the anemone's outside lobe.

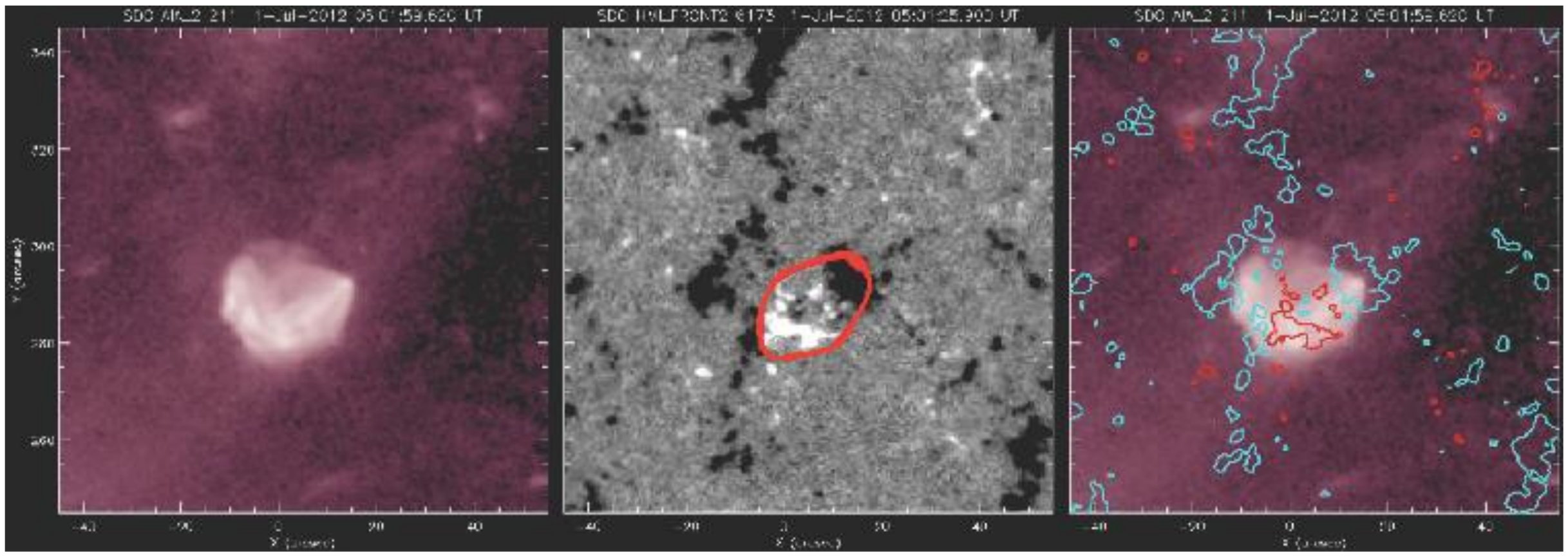

**Figure 2**. A single frame from BR2MOVIE, during the peak in BEAR BR2's minority-polarity (positive) flux, on 2012 July 1. Left panel: AIA 211 Å image of BR2 and its coronal-hole neighborhood, at 05:01:59 UT. Middle panel: HMI line-of-sight magnetogram of BR2 and its neighborhood, at 05:01:25 UT in the field of view of the left-panel 211 Å image. Here and in the other magnetograms in this paper the gray scale's field-strength saturation level is 50 G for either positive (white) flux or negative (black) flux. The closed red curve encloses BR2's positive and negative flux. Right panel: 40 G contours of the middle-panel magnetogram (red for positive flux, blue for negative) superposed on the left-panel 211 Å image.

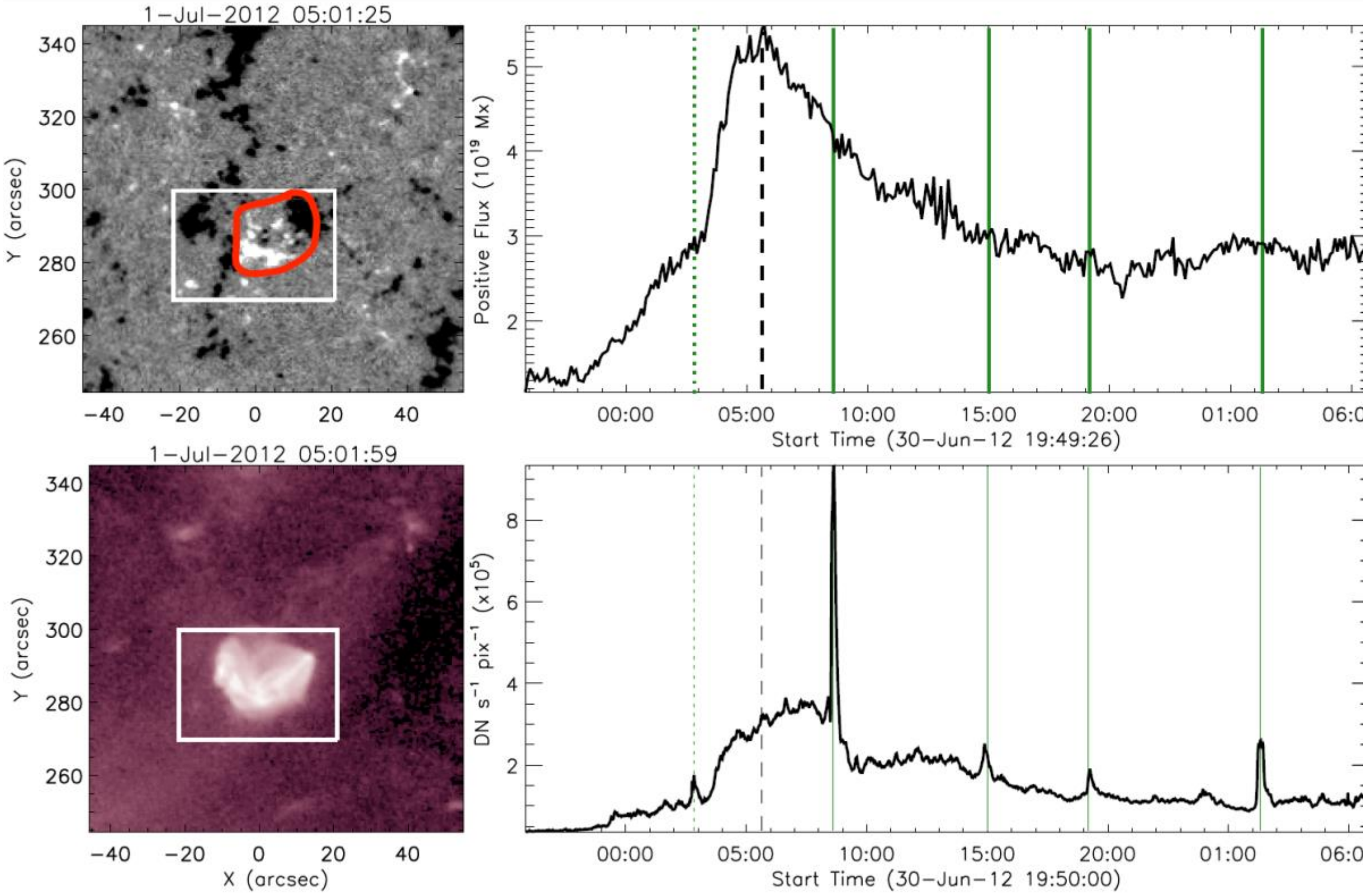

**Figure 3.** Time-flux plot and time-luminosity plot for BEAR BR2. Upper left panel: the magnetogram of Figure 2. The closed red curve encloses BR2's positive and negative flux. Lower left panel: the 211 Å image of Figure 2. The white box in the left two panels outlines the area from which the minority-polarity (positive) flux for BR2's time-flux plot was measured from the magnetograms of BR2's six-minute-cadence magnetogram movie, and from which the 211 Å luminosity for BR2's time-luminosity plot was measured from the 211 Å images for BR2's one-minute-cadence 211 Å movie. Upper right panel: BR2's minority-polarity-flux time-flux plot. The vertical black dashed line marks the time of BR2's maximum minority-polarity flux. The five vertical green lines mark the times of maximum visibility of BR2's five microflare events in BR2's one-minute-cadence 211 Å movie. The color green means that the event's miniature flare arcade is seated on an outside interval of BR2's PIL. The dotted green line marks a confined-explosion microflare event. The four solid green lines mark blowout-explosion microflare events. The last three blowout-explosion events happen when most of BR2's minority-polarity flux has canceled with majority-polarity network flux. Lower right panel: BR2's 211 Å time-luminosity plot. Each vertical line marks the same time marked by the corresponding line in the time-flux plot. Each green line sits in the luminosity spike from the microflare that has its maximum visibility at the line's time in BR2's 211 Å movie.

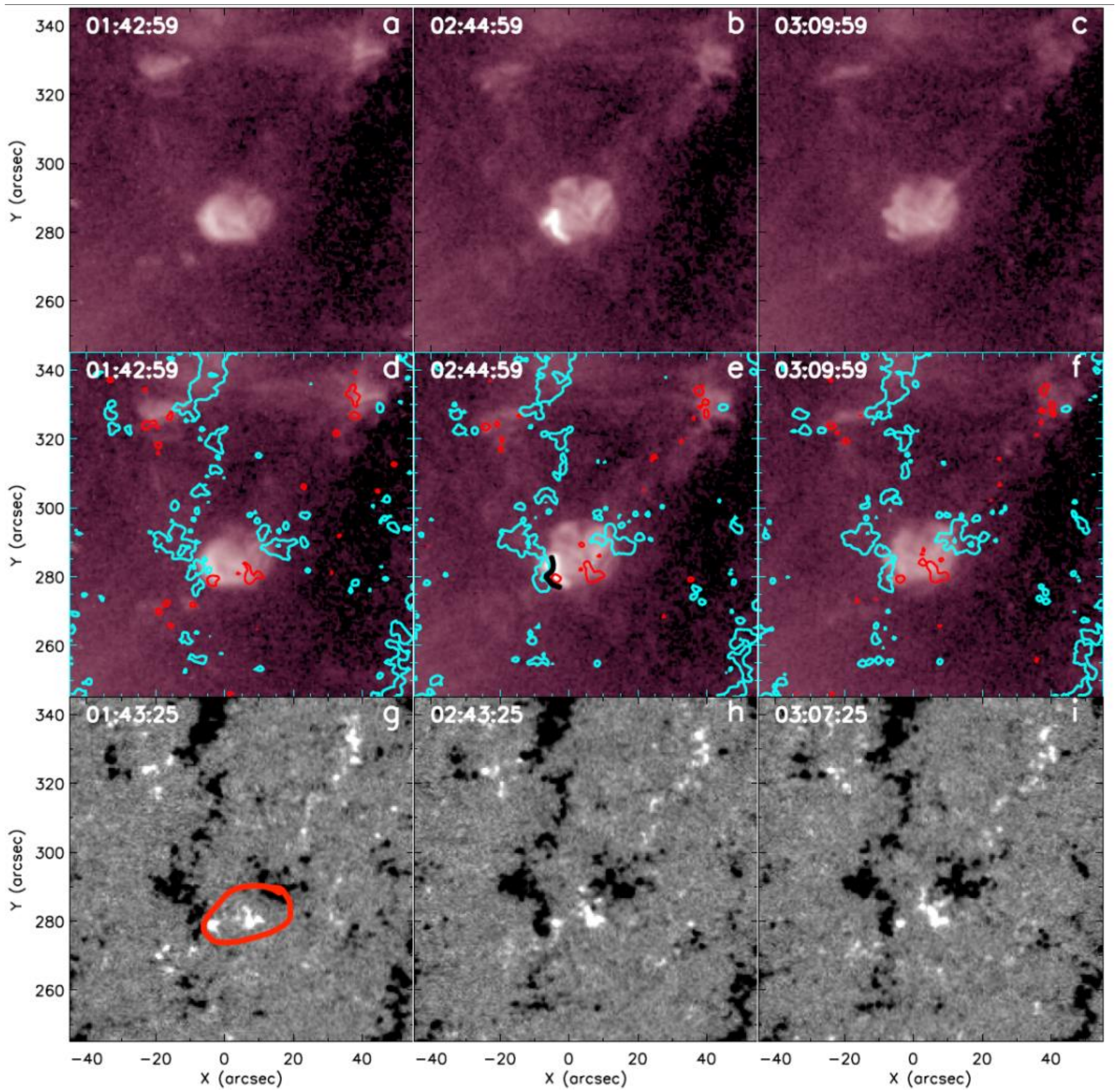

**Figure 4.** Confined-explosion microflare event that occurred on BR2's outside PIL during BR2's emergence. Here and in the nine-panel figures for our other eight example microflare events, the Universal Time of each panel is in the panel's upper left corner. Here in Figure 3, the top three panels are AIA 211 Å images of BR2 (a) an hour before, (b) at, and (c) half an hour after the time of the event's maximum visibility. Each of the bottom three panels (g, h, or i) is an HMI magnetogram that is co-aligned and co-temporal with the corresponding 211 Å image (a, b, or c) in the top row. In panel (g), the closed red curve encloses BR2's positive and negative flux. In the middle row (panels d, e, f), the 40 G contours of each magnetogram are overlaid on the corresponding 211 Å image. The 40 G contours are red or blue for positive or negative flux. We judge this event to be a confined magnetic explosion because it has no jet spire. The maximum-visibility bright microflare arcade sits on BR2's outside PIL between BR2's advancing outer minority-polarity positive flux and encountered majority-polarity negative network flux with which it is evidently cancelling in these magnetograms and in the magnetogram movie in BR2MOVIE (BR2.mp4) before and during the microflare. The worm-like black curved line in panel (e) is the path of the PIL through the microflare bright arcade.

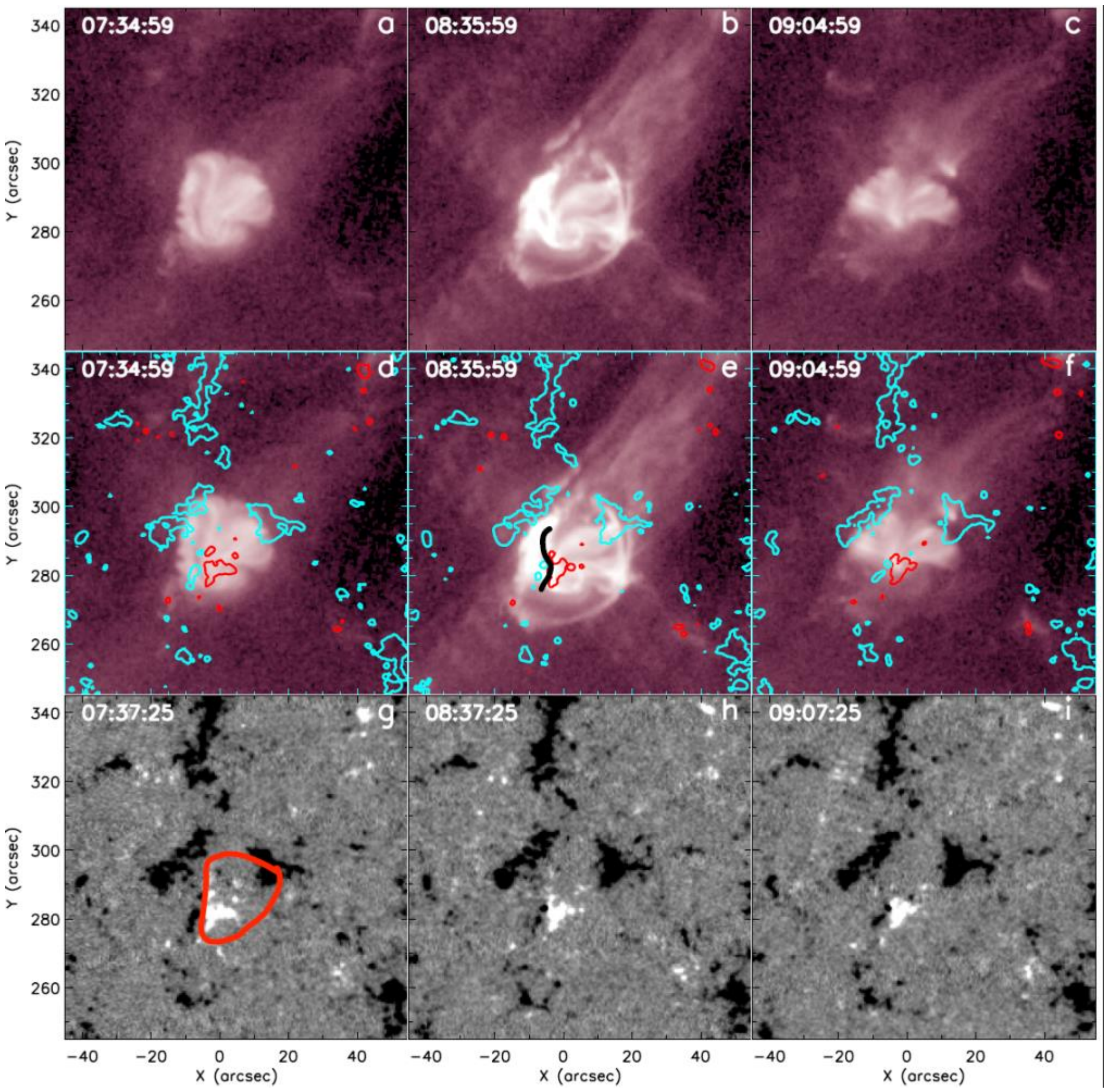

**Figure 5.** First of BR2's four blowout-explosion microflare events, each of which explode from BR2's outside PIL during BR2's decay phase. The format is the same as in Figure 4. The three AIA 211 Å images (a, b, c) and their co-temporal HMI magnetograms (g, h, i) are (a, g) an hour before, (b, h) at, and (c, i) half an hour after the time of the event's maximum visibility. In panel (g) the closed red curve encloses BR2's positive and negative flux. In panels (b) and (e), the broad, complex, blowout-jet spire extends northwest from its base, which is the anemone magnetic arcade that encircles BR2's minority-polarity positive flux. Panel (e), the three magnetograms, and BR2MOVIE (BR2.mp4) show that the event's bright JBP microflare arcade sits on BR2's outside PIL interval along which flux is cancelling before and during the blowout eruption of a minifilament from along the cancellation PIL. The worm-like black curved line in panel (e) is the path of that PIL interval. The pre-eruption dark minifilament tracing the cancellation PIL is discernible in panels (a) and (d).

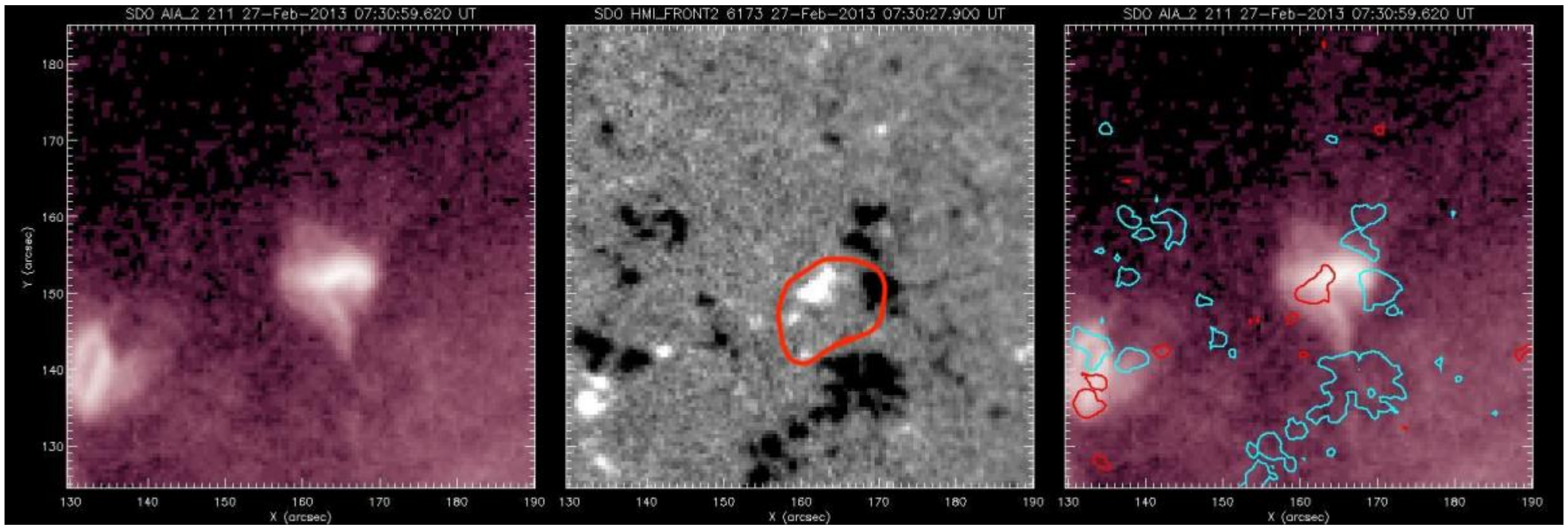

**Figure 6.** A single frame from BR7MOVIE, three-quarters of an hour after the maximum in BEAR BR7's minority-polarity (positive) flux, on 2013 February 27. Left panel: AIA 211 Å image of BR7 and its coronal-hole neighborhood, at 07:30:59 UT. Middle panel: HMI line-of-sight magnetogram of BR7 and its neighborhood, at 07:30:27 UT in the field of view of the left-panel 211 Å image. The closed red curve encloses BR7's positive and negative flux. Right panel: 40 G contours of the middle-panel magnetogram (red for positive flux, blue for negative) superposed on the left-panel 211 Å image.

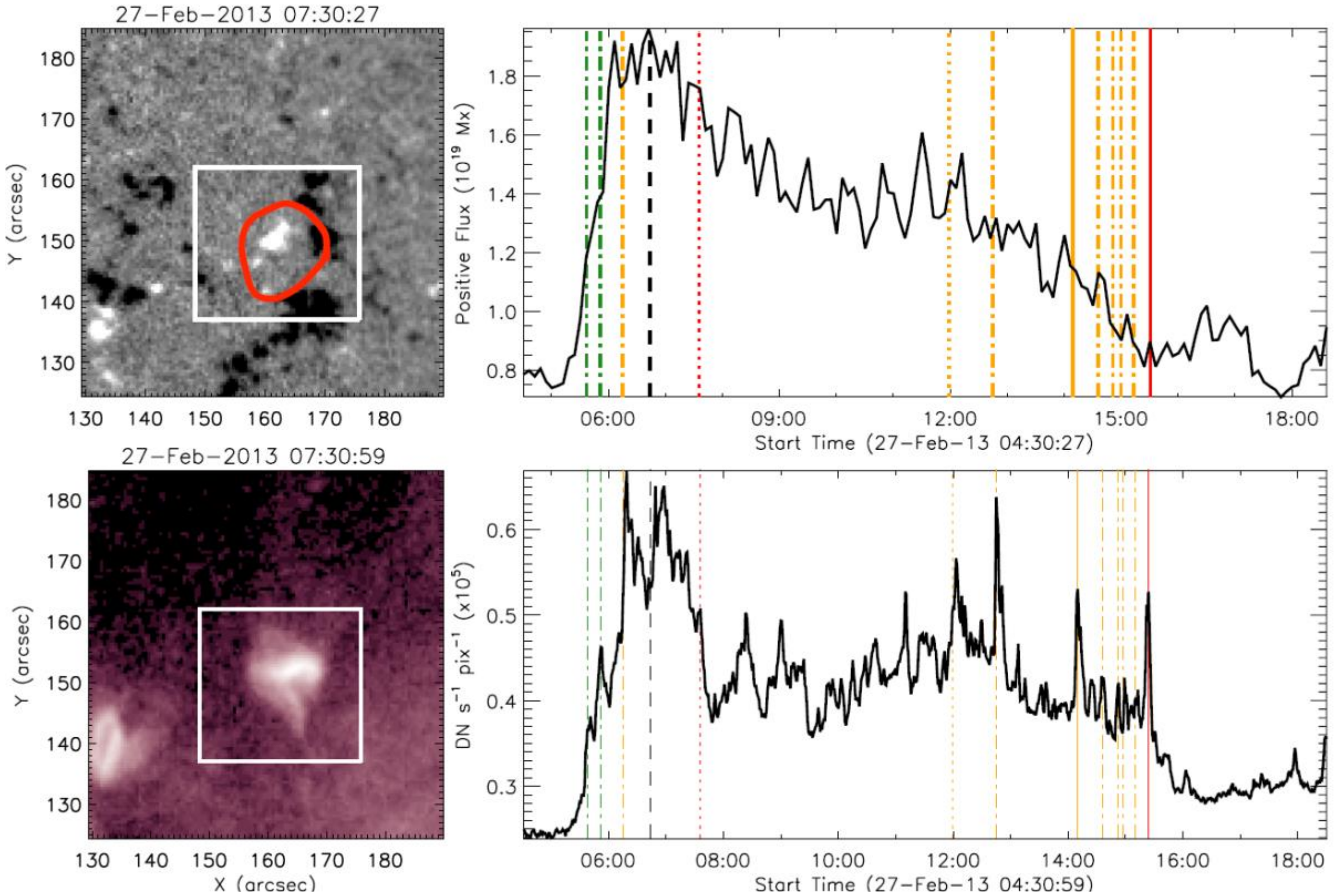

**Figure 7.** Time-flux plot and time luminosity plot for BEAR BR7. Upper left panel: the magnetogram of Figure 6. The closed red curve encloses BR7's positive and negative flux. Lower left panel: the 211 Å image of Figure 6. The white box in the left two panels bounds the area from which the minority-polarity (positive) flux for BR7's time-flux plot was measured from the magnetograms for BR7MOVIE, and from which the 211 Å luminosity for BR7's time-luminosity plot was measured from the 211 Å images for BR7MOVIE. Upper right panel: BR7's minority-polarity-flux time-flux plot. The black vertical dashed line marks the time of BR7's maximum minority-polarity flux. The twelve colored vertical lines mark the maximum-visibility times of BR7's twelve microflare events. The first three microflare events happen in the rise of BR7's time-flux plot, and the last six events, including BR7's two blowout-explosion events, happen when only a little of BR7's minority flux remains. Dotted lines mark confined-explosion events; dot-dash lines mark microflare events made by mostly-confined magnetic explosions; solid lines mark blowout-explosion events. Red lines mark events that explode from BR7's inside PIL; amber lines mark events that explode from an inside & outside interval of BR7's PIL; green lines mark events that explode from an outside interval of BR7's PIL. Lower right panel: BR7's 211 Å time-luminosity plot. Each vertical line marks the same time as its corresponding line in the time-flux plot. Each microflare-event line sits in the luminosity spike made by that event's microflare.

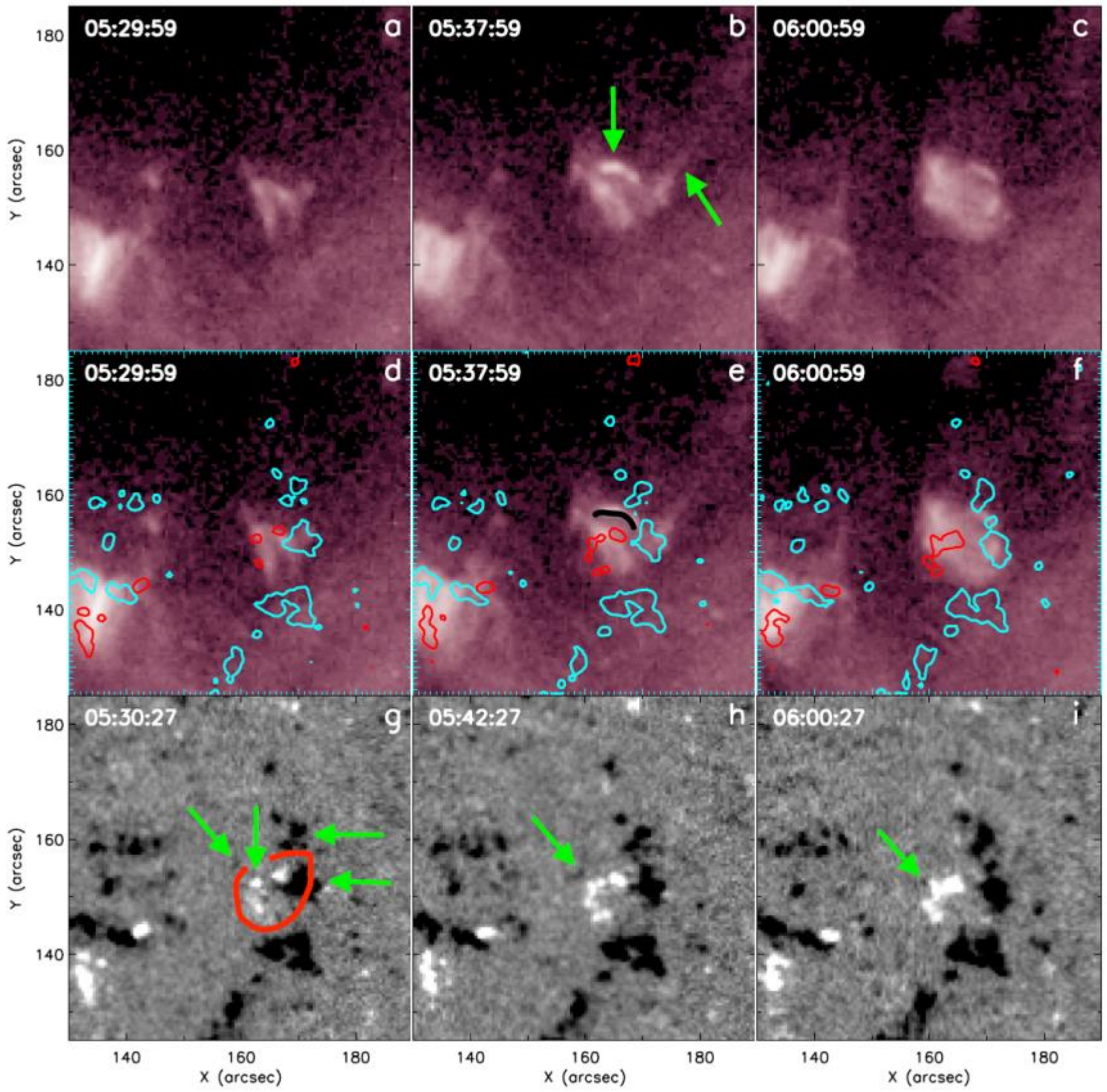

**Figure 8.** First of BR7's twelve microflare events. The format is the same as in Figure 4. This event is evidently a mostly-confined magnetic explosion from a flux-cancellation interval of BR7's outside PIL. It happens about 20 minutes after BR7 starts emerging. The three AIA 211 Å images (a, b, c) and their co-temporal HMI magnetograms (g, h, i) are (a, g) about 10 minutes before, (b, h) at, and (c, i) about 20 minutes after the time of the event's maximum visibility. The almost-closed red curve in panel (g) encloses BR7's positive and negative flux. In panel (b), the slanted arrow points to the event's single-strand narrow jet spire, and the downward vertical arrow points the bright JBP microflare arcade. These features are the evidence that this event is a mostly-confined magnetic explosion. Panels (e) and (h) show that the JBP sits on BR7's outside PIL interval on the north side of BR7's advancing arc of emerging minority-polarity positive flux. The black curved line in panel (e) is the path of that PIL interval. The arc of positive flux is plausibly encountering and canceling negative flux that is near or below the 10 G noise level of the HMI magnetograms, but is expected to be there in this negative-flux-dominated coronal hole. In panel (g), the lower east-pointing horizontal arrow points to the west side of a clump of negative network flux against the east side of which BR7's negative flux is emerging. The upper east-pointing horizontal arrow points to a negative clump of network flux near the north side of BR7. In panels (g), (h), and (i), the downward slanted arrow points to a small faint patch of negative flux encountered by BR7's advancing arc of emerging positive flux. The downward vertical arrow in panel (g) points to prospective yet-weaker negative flux that is apparently being canceled by BR7's advancing positive flux with which it is now in contact. We take the above aspects of these magnetograms and of the magnetogram movie in BR7MOVIE (BR7.mp4) to be discernible evidence and circumstantial evidence that flux cancellation on BR7's north-side outside PIL underlying the JBP occurs before and during this microflare event's magnetic explosion.

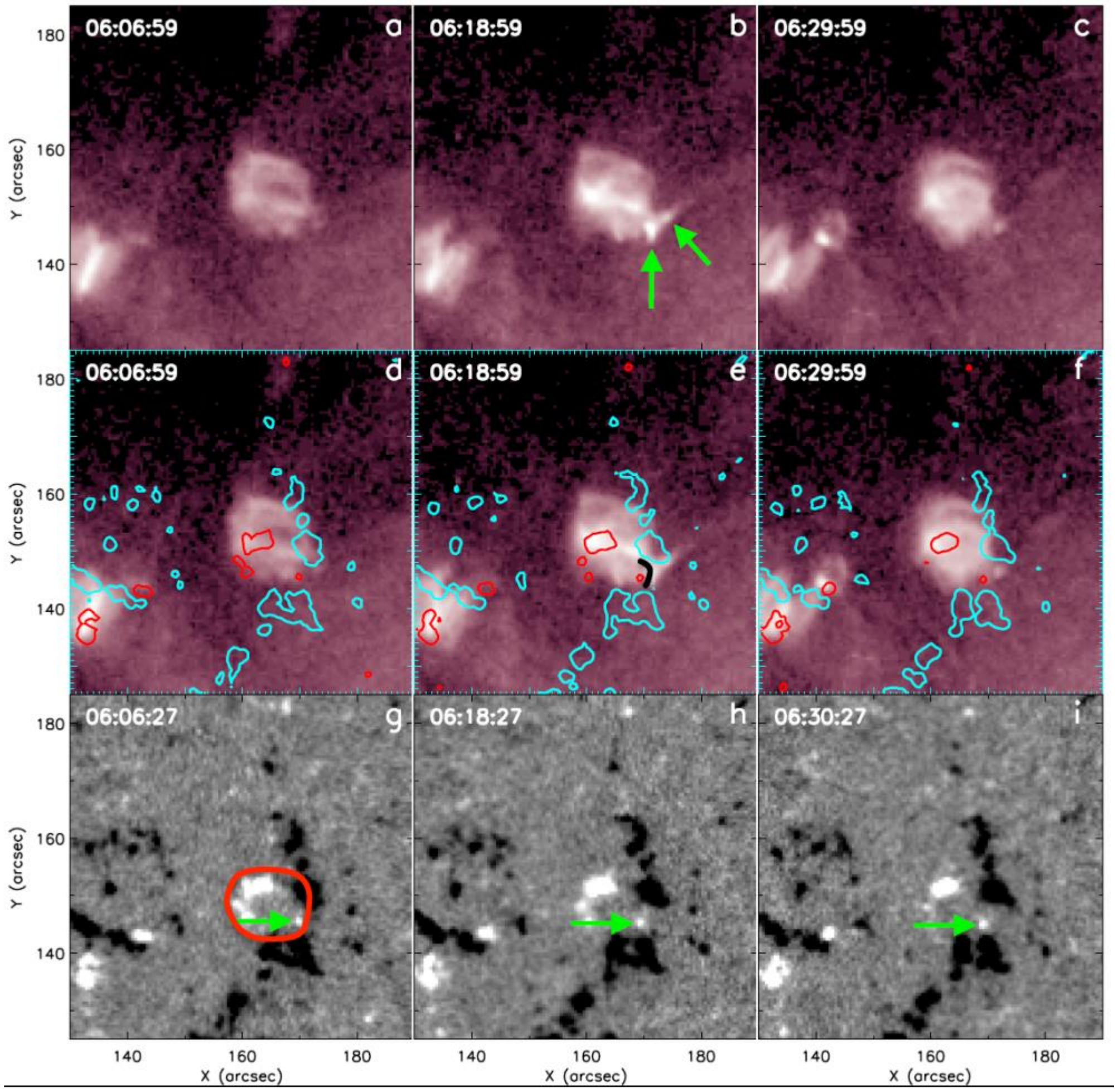

**Figure 9.** Last of BR7's three microflare events that happen before the time of BR7's maximum minority-polarity positive flux. The format is the same as in Figure 4. This event is a mostly-confined magnetic explosion from an inside & outside interval of BR7's PIL. The three AIA 211 Å images (a, b, c) and their co-temporal HMI magnetograms (g, h, i) are (a, g) 12 minutes before, (b, h) at, and (c, i) 12 minutes after the time of the event's maximum visibility. In panel (g), the closed red curve encloses BR7's positive and negative flux. In panel (b), the slanted arrow points to the event's short single-strand narrow jet spire, and the vertical arrow points to the bright JBP microflare arcade. These two features are the evidence that this event is a mostly-confined magnetic explosion. In panels (g), (h), and (i), the horizontal arrow points to a small patch of emerging positive flux that moves south southeast to encounter negative network flux. In panel (h), there is weak positive flux on the north northwest side of the stronger positive flux pointed to by the arrow. From the magnetogram movie in BR7MOVIE (BR7.mp4), we judge that this weak positive flux cancels with BR7's negative flux during and after the microflare event. Thus, before and during the microflare event, the magnetograms show evidence of flux cancellation at BR7's inside & outside PIL interval that runs around the west side of the positive flux. The PIL's inside end is between the positive flux and BR7's negative flux, and the PIL's outside end is between the positive flux and the negative network flux on the south side of the positive flux. Panels (e) and (h) show that the JBP sits on the inside & outside PIL. The black curved line in panel (e) is the path of that inside & outside PIL interval.

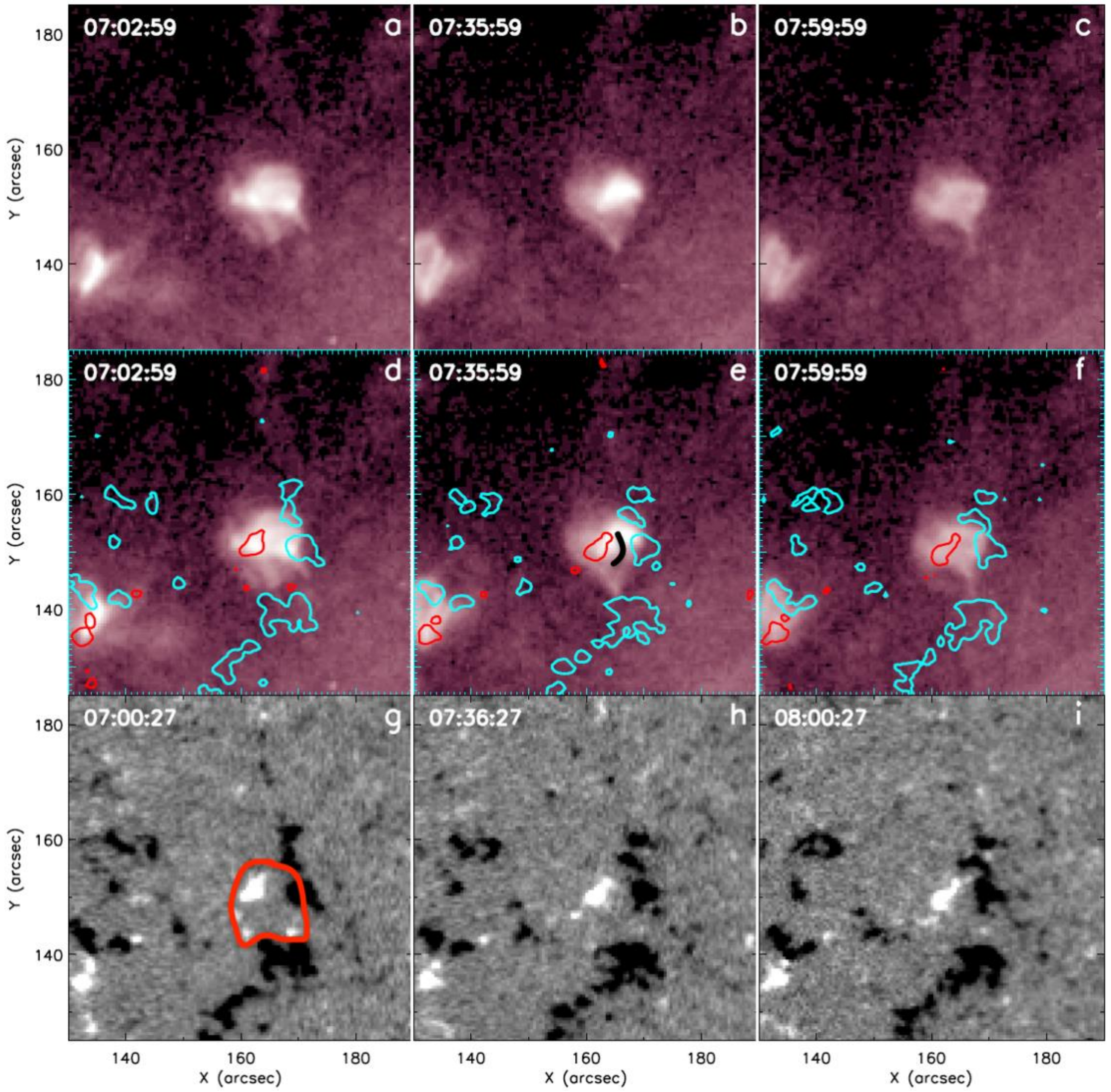

**Figure10.** BR7's fourth microflare event. The format is the same as in Figure 4. This event is a confined magnetic explosion from BR7's inside PIL. It is BR7's first microflare event during the decline of BR7's minority-polarity positive flux. The three AIA 211 Å images (a, b, c) and their co-temporal HMI magnetograms (g, h, i) are (a, b) half an hour before, (b, h) at, and (c, i) half an hour after the time of the event's maximum visibility. In panel (g), the closed red curve encloses BR7's positive and negative flux. We judge this event to be a confined magnetic explosion because it has no jet spire. Panels (e) and (h) show that one foot of the bright microflare arcade is in the middle and northern part of BR7's largest clump of positive flux and the other foot is in the northern part of BR7's negative flux. Panels (g), (h), and (i) and the magnetogram movie in BR7MOVIE (BR7.mp4) show that before, during, and after this microflare event, flux sheds westward from the northern half of the positive flux clump to apparently cancel with BR7's negative flux. This is the evidence that before and during this microflare magnetic explosion there is flux cancellation at the underlying PIL. The black curved line in panel (e) is the path of the inside PIL interval under the microflare arcade.

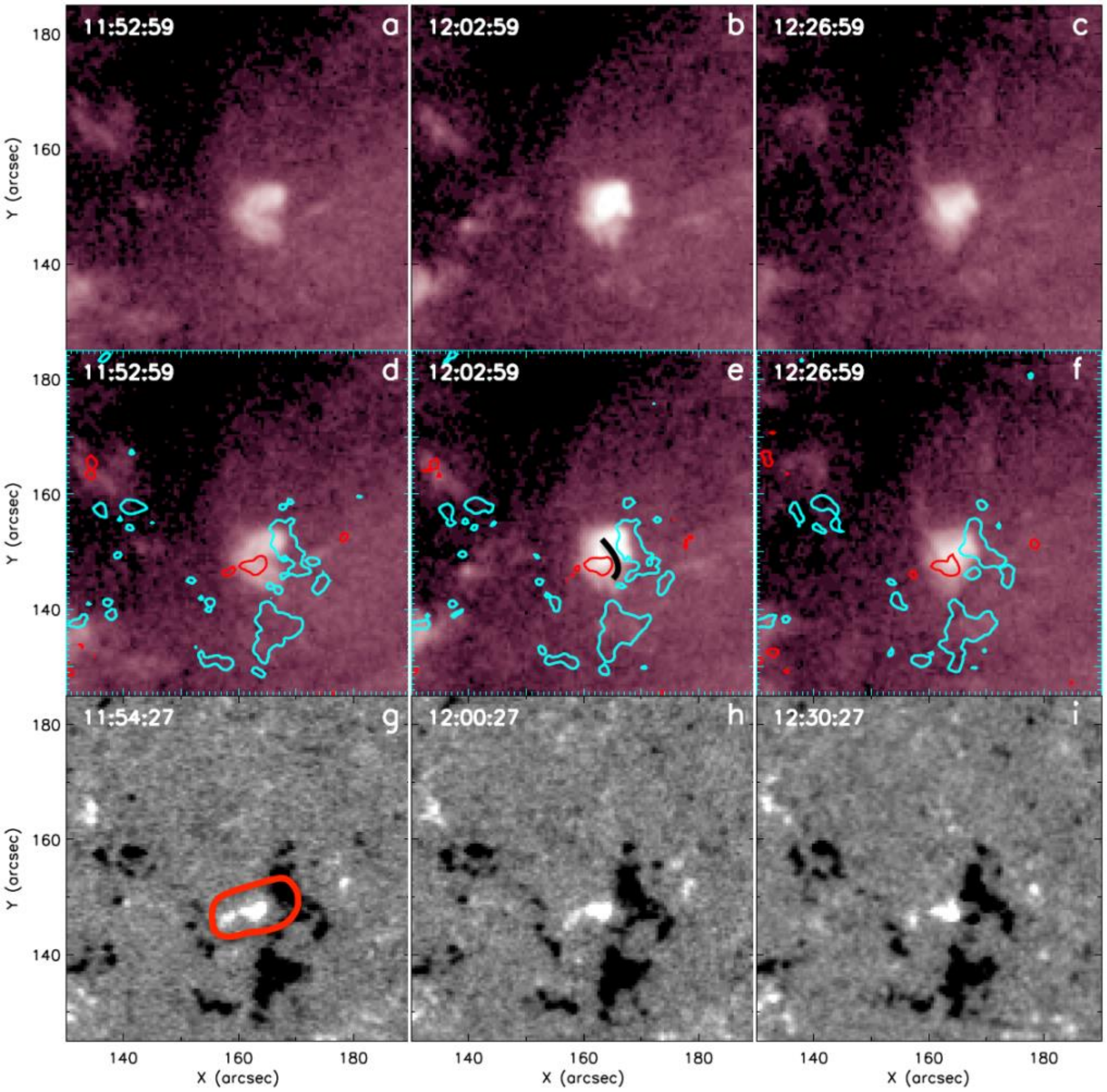

**Figure 11.** BR7's second microflare event during the decline of BR7's minority-polarity positive flux. The format is the same as in Figure 4. This event is a confined magnetic explosion seated on an inside & outside interval of BR7's PIL. The three AIA 211 Å images (a, b, c) and their co-temporal HMI magnetograms (g, h, i) are (a, g) 10 minutes before, (b, h) at, and (c, i) half an hour after the time of the event's maximum visibility. The closed red curve in panel (g) encloses BR7's positive and negative flux. We judge this event to be a confined magnetic explosion because it has no jet spire. In panel (e) the north side of the microflare bright arcade has one foot in BR7's biggest clump of positive flux and the other foot in negative network flux that has merged with the north side of BR7's negative flux. The south side of the microflare arcade also has one foot in BR7's positive flux clump, but the other foot is in what we judge is BR7's negative flux. From the magnetograms in Figures 7, 8, 9, and 10 as well as from the magnetogram movie in BR7MOVIE (BR7.mp4), we judge hat BR7's negative flux is immediately northwest of BR7's big clump of positive flux. On this basis, we judge that BR7's PIL interval under the microflare arcade is an inside & outside interval. The black curved line in panel (e) is the path of that inside & outside PIL interval. Panels (g), (h), and (i) and the magnetogram movie in BR7MOVIE (BR7.mp4) show evidence of flux cancellation at the underlying PIL – positive and negative flux merge at the PIL – before and during this microflare magnetic explosion.

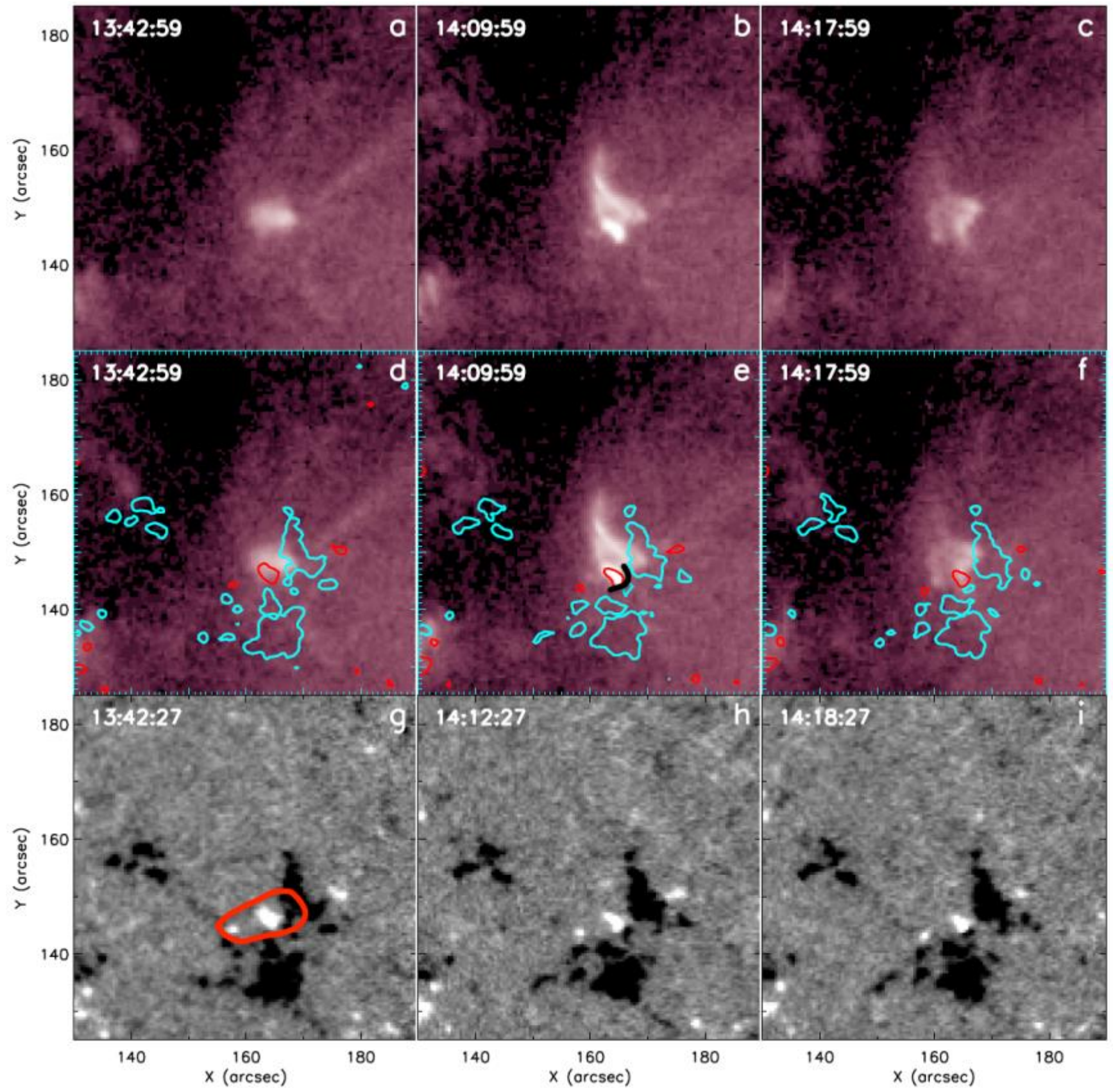

**Figure 12.** Seventh of BR7's twelve microflare events. The format is the same as in Figure 4. This event is a blowout magnetic explosion from another inside & outside interval of BR7's PIL. The three AIA 211 Å images (a, b, c) and their co-temporal HMI magnetograms (g, h, i) are (a, g) half an hour before, (b, h) at, and (c, i) about 10 minutes after the time of the event's maximum visibility. The closed red curve in panel (g) encloses BR7's positive and negative flux. Panel (b) shows that this event is a blowout jet. The jet's multi-strand wide spire extends northeast and the jet's JBP is at the south edge of the base. This is the evidence that this event is a blowout magnetic explosion. In panel (e) the JPB microflare arcade touches but does not bridge the PIL at its southwest end. We assume this indicates that the 211 Å image and the magnetogram have a small misalignment, the 211 Å image being offset northeast by perhaps ~ 2 arcseconds. In the next two frames of the 211 Å movie in BR7MOVIE (BR7.mp4), the microflare arcade has grown enough that it reaches over both that PIL interval and the inside PIL interval between the positive flux clump and BR7's negative flux northwest of the positive flux. From the magnetogram movie in BR7MOVIE (BR7.mp4) we judge that the negative flux on the south and southwest sides of the positive flux clump is network flux, and hence that the PIL interval between the positive flux clump and that negative flux is an outside interval of BR7's PIL. We therefore judge that the PIL interval on which this event's blowout magnetic explosion is seated is an inside & outside interval of BR7's PIL. The black curved line in panel (e) is that inside & outside PIL interval. The three magnetograms in panels (g), (h), and (i) and the magnetogram movie show that before and during this microflare event there is flux cancellation at both the outside portion and the inside portion of the magnetic explosion's underlying interval of BR7's PIL.

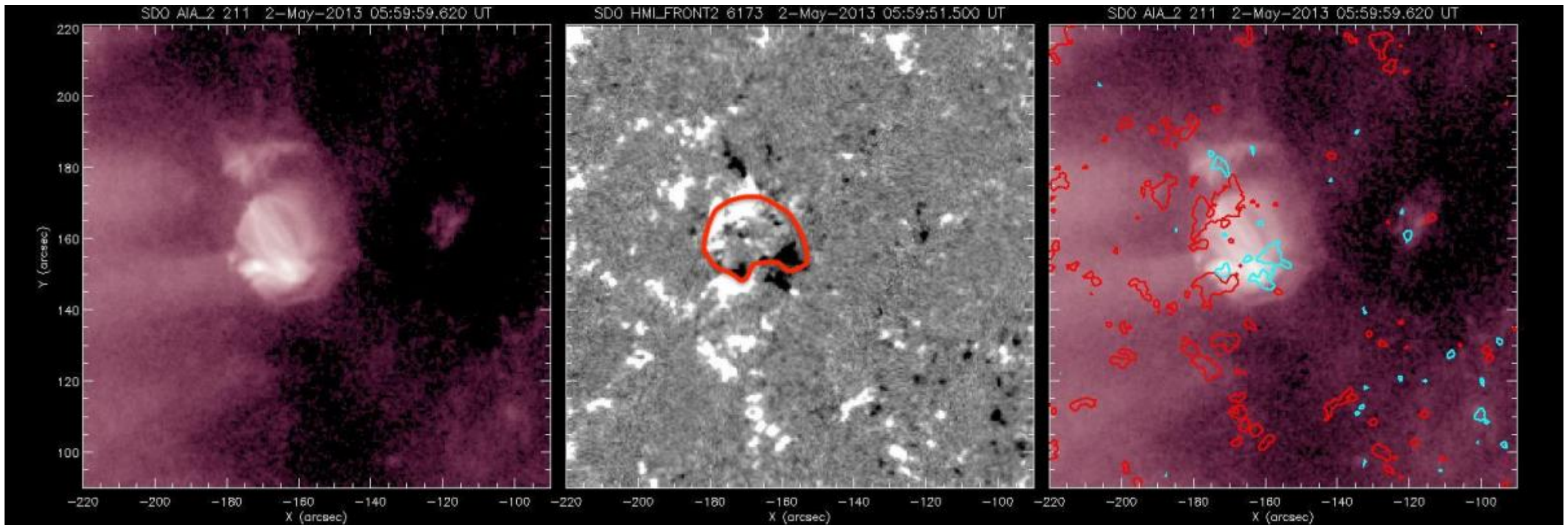

**Figure 13.** A single frame from BR9MOVIE, an hour after the maximum in BEAR BR9's minority-polarity (negative) flux, on 2013 May 2. Left panel: AIA 211 Å image of BR9 and its neighborhood, at 05:59:59 UT. Middle panel: HMI line-of-sight magnetogram of BR9 and its neighborhood, at 05:59:51 UT in the field of view of the left-panel 211 Å image. The closed red curve encloses BR9's positive and negative flux. Right panel: 40 G contours of the middle-panel magnetogram (red for positive flux, blue for negative) superposed on the left-panel 211 Å image.

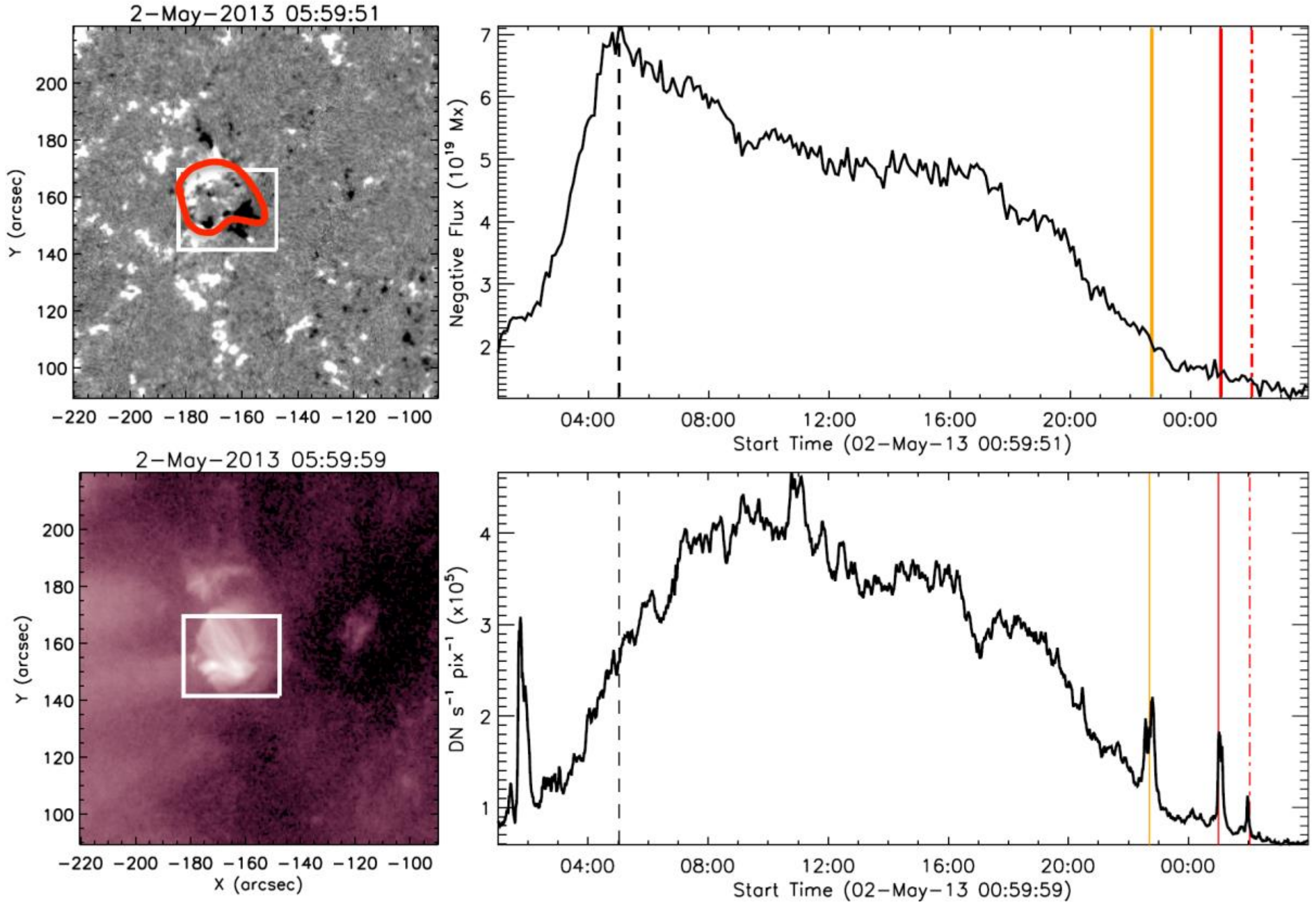

**Figure 14.** Time-flux plot and time-luminosity plot for BEAR BR9. Upper left panel: the magnetogram of Figure 13. The closed red curve encloses BR9's positive and negative flux. Lower left panel: the 211 Å image of Figure 13. The white box in the left two panels bounds the area from which the minority-polarity (negative) flux for BR9's time-flux plot was measured from the magnetograms for BR9MOVIE, and from which the 211 Å luminosity for BR9's time-luminosity plot was measured from the 211 Å images for BR9MOVIE. Upper right panel: BR9's minority-polarity-flux time-flux plot. The black vertical dashed line marks the time of BR9's maximum minority-polarity flux. The three colored vertical lines mark the maximum-visibility times of BR9's three microflare events, each of which happens near the end of BR9's life, when all but a little of BR9's minority-polarity negative flux has been canceled by encountered positive flux. The lines for the first two events are solid to signify that the magnetic explosion in each of these events is a blowout explosion, and the line for the last event is dot-dash to signify that the magnetic explosion in this event is a mostly-confined explosion. The amber color of the first line signifies that the event's magnetic explosion is from an inside & outside interval of BR9's PIL. The red color of each of the last two lines signifies that the event's magnetic explosion is from an inside interval of BR9's PIL. Lower right panel: BR9's 211 Å time-luminosity plot. Each vertical line makes the same time as its corresponding line in the time-flux plot. Each microflare-event line sits in the luminosity spike made by that event's microflare.

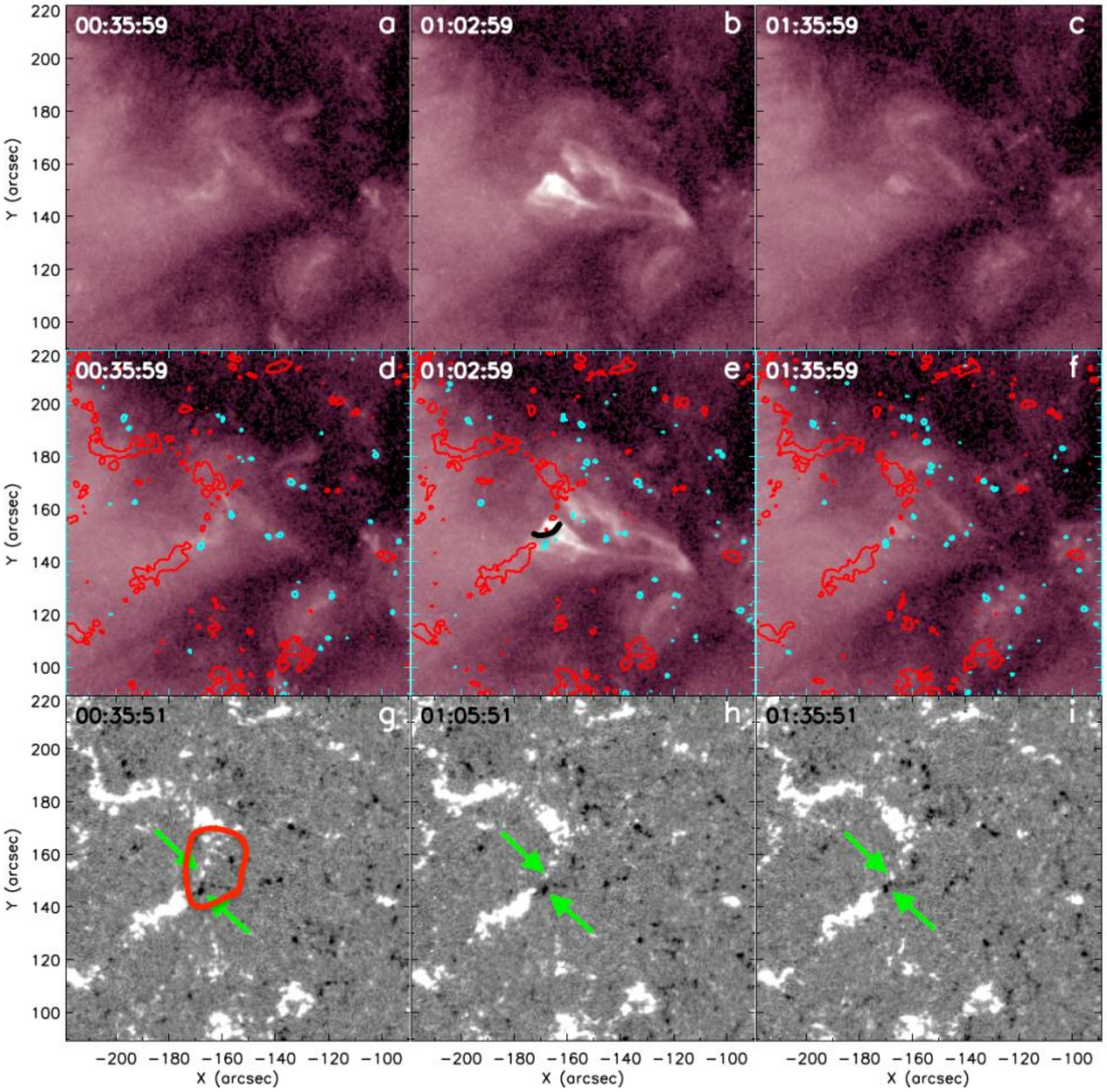

**Figure 15.** Second of BR9's three microflare events. The format is the same as in Figure 4. This event is a blowout magnetic explosion from an inside interval of BR9's PIL. The three AIA 211 Å images (a, b, c) and their co-temporal HMI magnetograms (g, h, i) are (a, g) half an hour before, (b, h) at, and (c, i) half an hour after the time of the event's maximum visibility. The closed red curve in panel (g) encloses BR9's positive and negative flux. Panel (b) shows that this event is a blowout jet having a complex multi-strand wide spire extending westward and having its bright JBP microflare arcade in the south half of the spire's base. This is the evidence that this event is a blowout magnetic explosion. Panel (e), along with the magnetogams in panels (g), (h), and (i), shows that the JBP microflare arcade sits on BR9's inside PIL interval having small patches of negative flux on its southwest side and small patches of positive flux on its northeast side. The magnetogram movie in BR9MOVIE (BR9.mp4) shows that those negative flux patches are remnants of BR9's negative flux and the positive flux patches are fragments of BR9's positive flux that are migrating southward. This is the evidence that the PIL interval underlying this event's magnetic explosion is an inside interval of BR9's PIL. The black curve in panel (e) is that interval of inside PIL. The two arrows in each of the three magnetograms point to a negative flux clump and to a positive flux clump that converge and partly cancel before and during this microflare event. In this way, the magnetograms here and in the magnetogram movie show good evidence that there is flux cancellation at the underlying PIL before and during this microflare-making magnetic explosion.

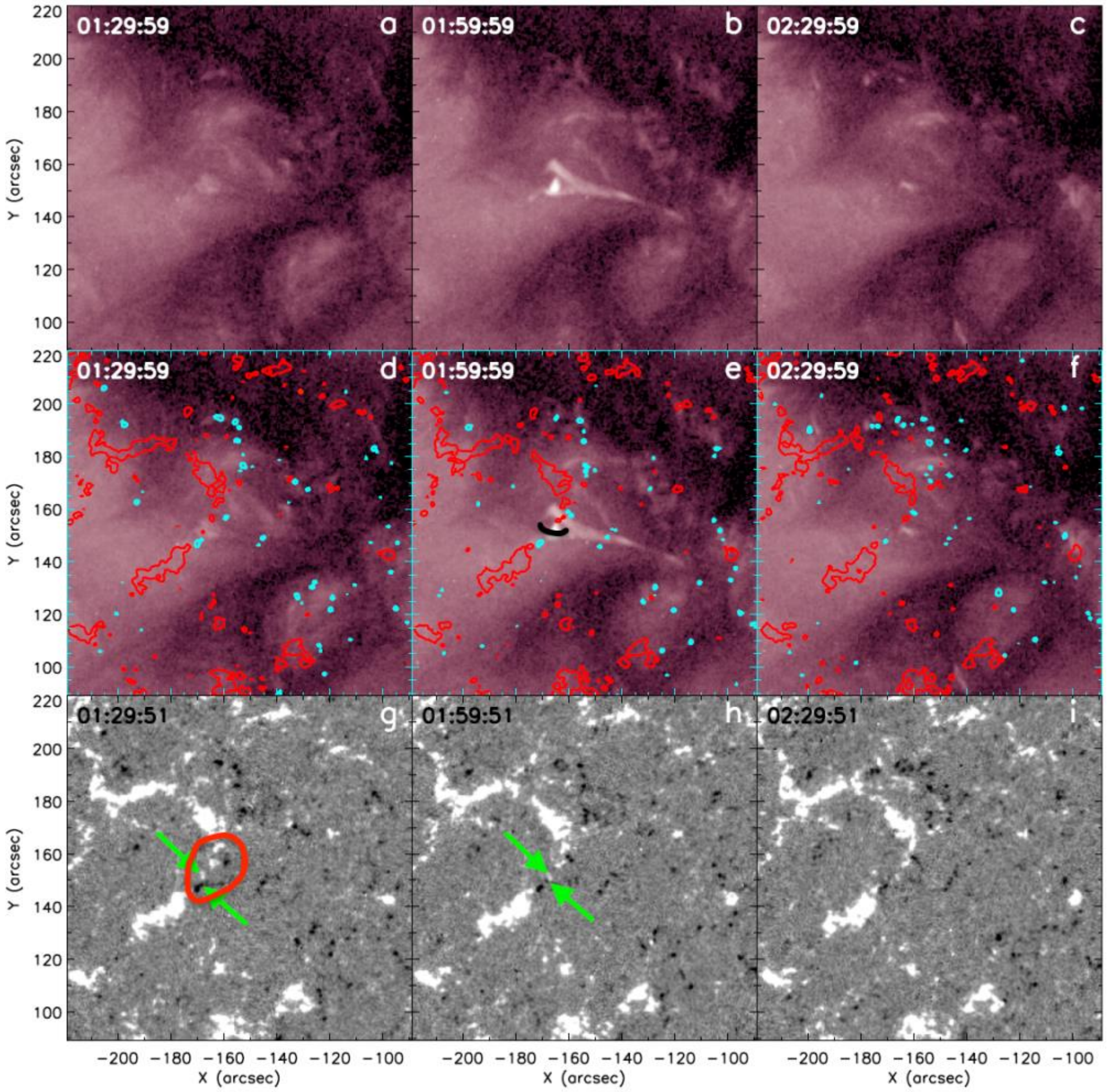

**Figure 16.** Last of BR9's three microflare events. The format is the same as in Figure 4. This event is a mostly-confined magnetic explosion. It explodes from the same inside interval of BR9's PIL from which BR9's next-to-last microflare event exploded (Figure 15), after another hour of flux cancellation at that PIL. The three AIA 211 Å images (a, b, c) and their co-temporal HMI magnetograms (g, h, i) are (a, g) half an hour before, (b, h) at, and (c, i) half an hour after the time of the event's maximum visibility. The closed red curve in panel (g) encloses BR9's positive and negative flux. Panel (b) shows that this event produces a single-strand narrow jet spire. This is the evidence that this event is a mostly-confined magnetic explosion. Panel (e) shows that the JBP microflare arcade sits on the same inside interval of BR9's PIL as does the JBP of BR9's previous microflare event, the blowout-explosion event shown in Figure 15. The black curved line in panel (b) is that interval of inside PIL. In panels (g) and (h), the upward arrow points to a remnant clump of BR9's negative flux and the downward arrow points to a fragment of BR9's positive flux. In panel (i), these two opposite-polarity flux clumps have cancelled and are gone. From comparison of panels (g), (h), and (i) it is seen that before and during this microflare event, the two opposite-polarity flux clumps merge and diminish. In this way, these three magnetograms and the magnetogram movie in BR9MOVIE (BR9.mp4) show that there is flux cancellation at the underlying PIL before and during this mostly-confined microflare-making magnetic explosion.

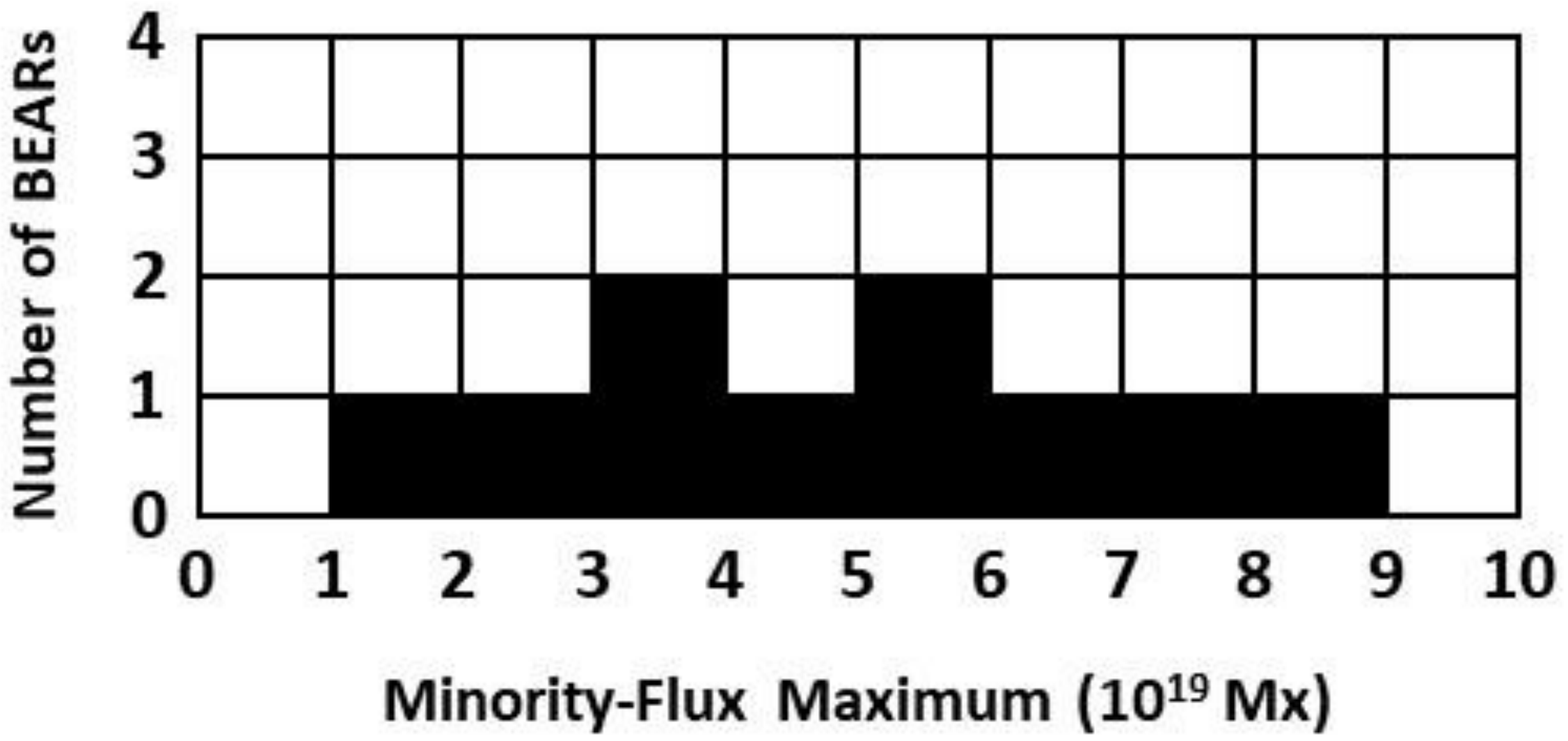

**Figure 17.** Histogram of the maxima of the flux-time plots of minority-polarity flux for the 10 BEARs.

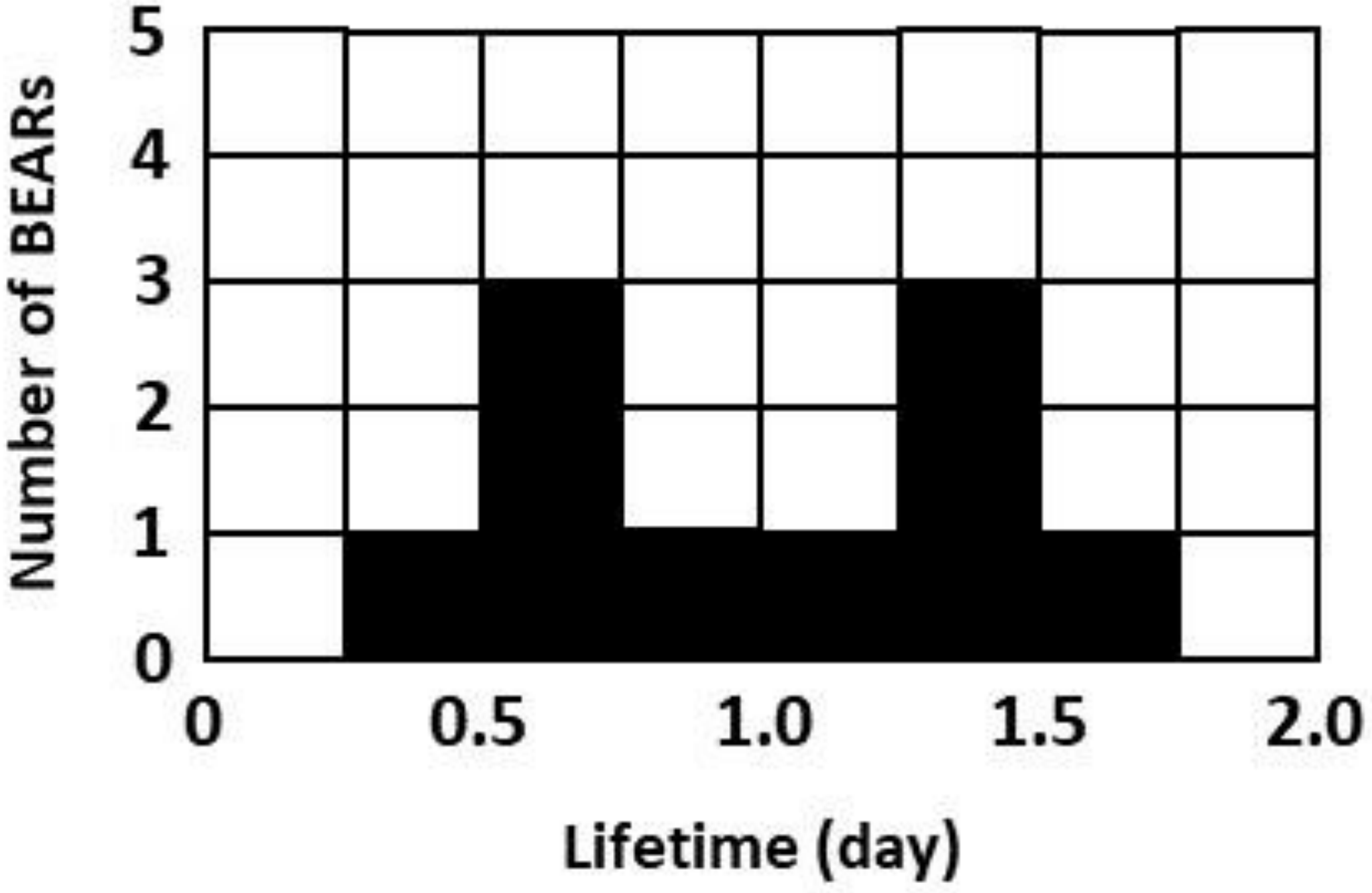

**Figure 18.** Histogram of the lifetimes of the 10 BEARs.

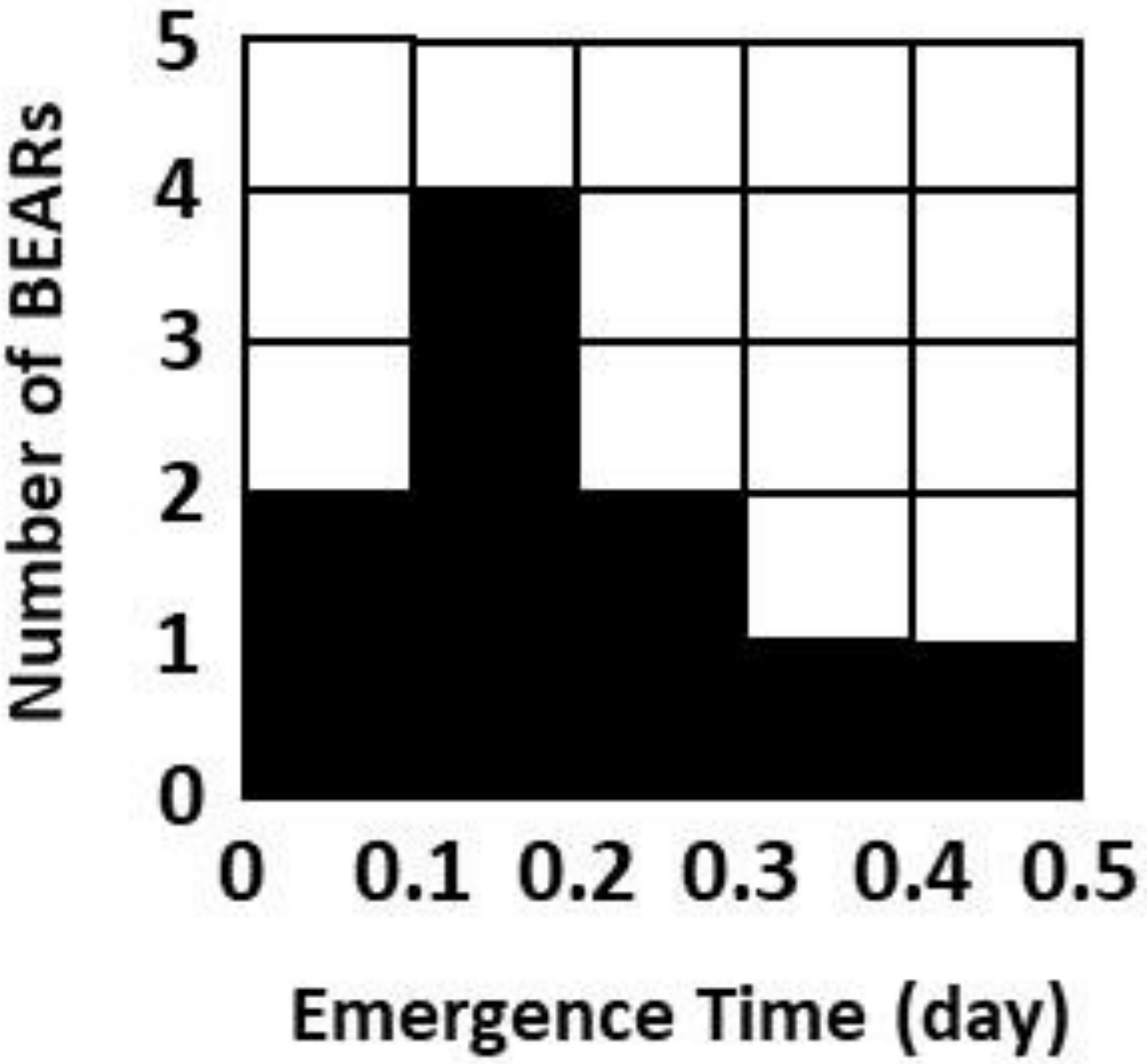

**Figure 19.** Histogram of the emergence times of the 10 BEARs.

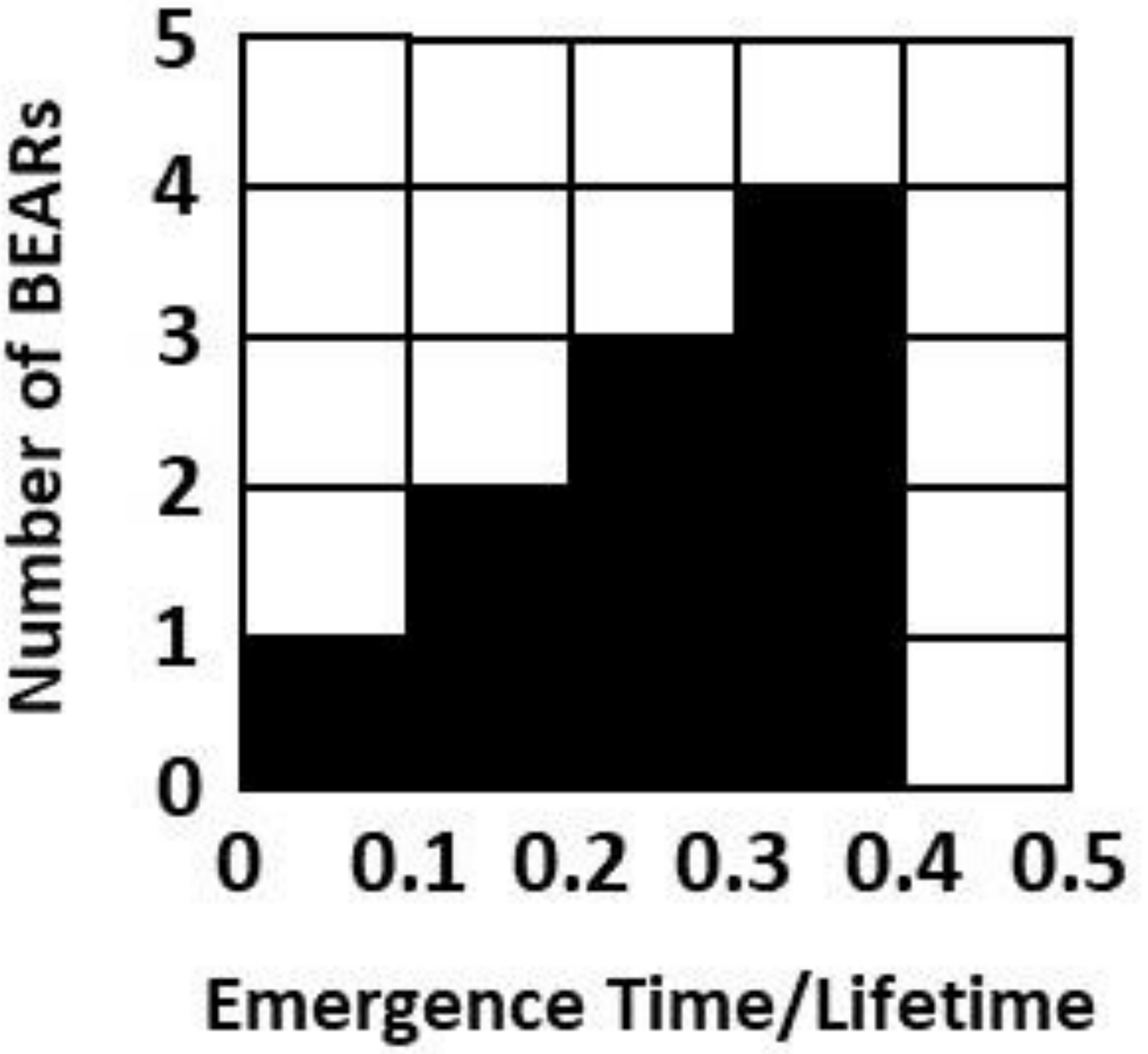

**Figure 20.** Histogram of the ratio of emergence time to lifetime for the 10 BEARs.

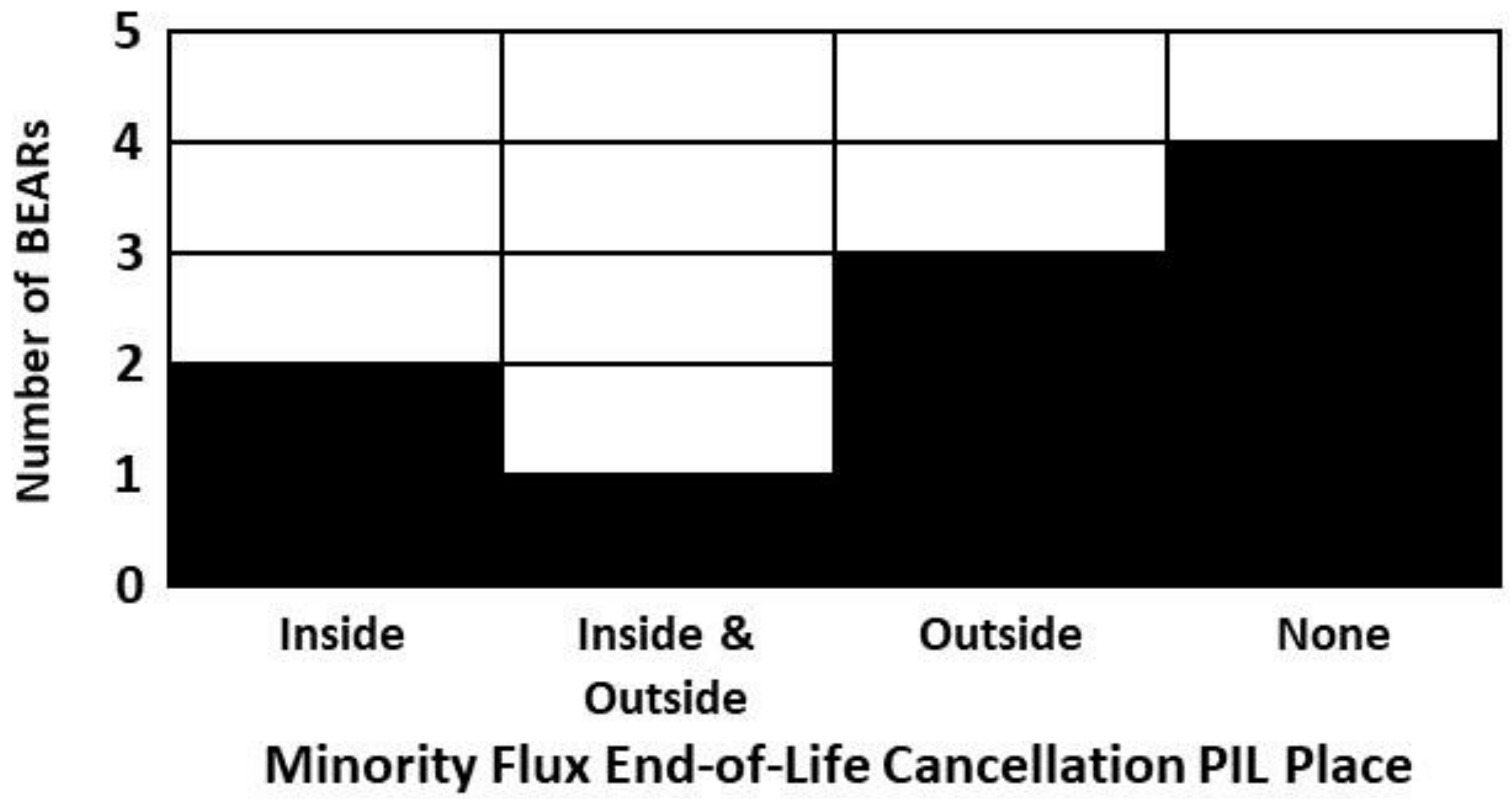

**Figure 21.** Distribution of the 10 BEARs among the four categories of where and how a BEAR's minority-polarity flux ends its trackable life in HMI magnetograms.

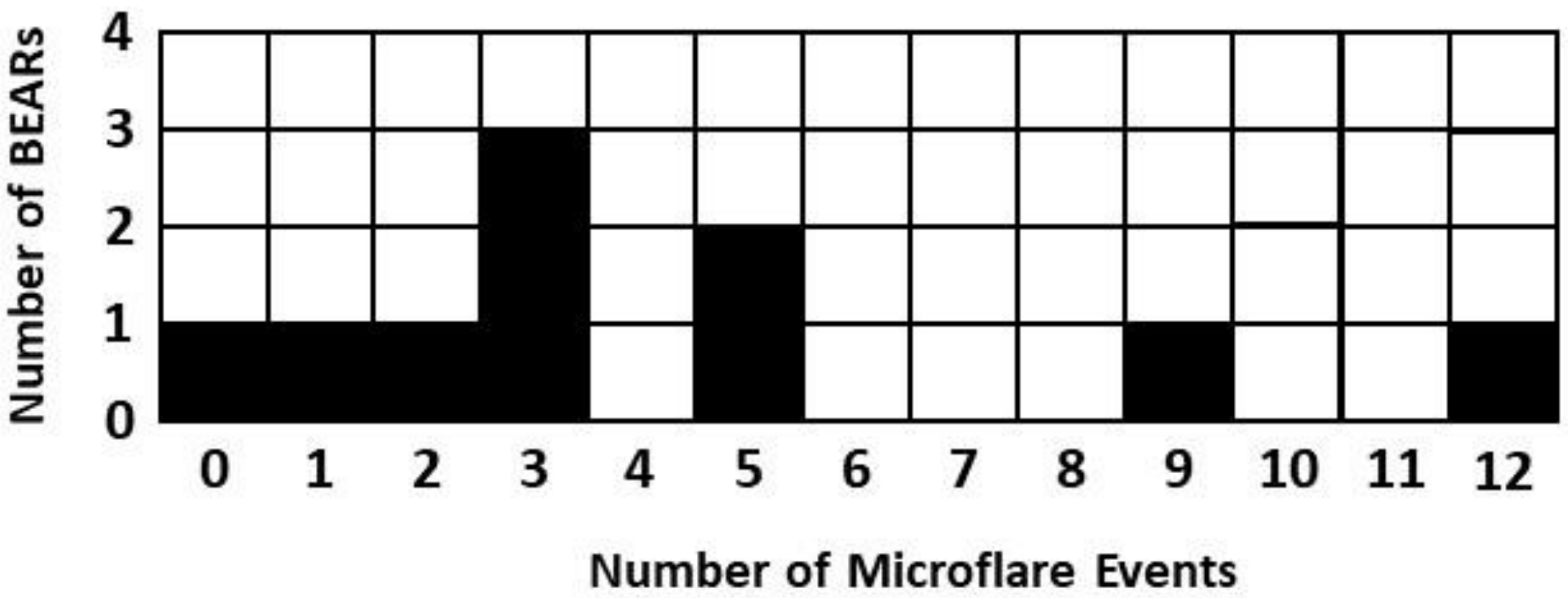

**Figure 22.** Histogram of the number of microflare events produced by a BEAR during the life of the BEAR, for our 10 BEARs.

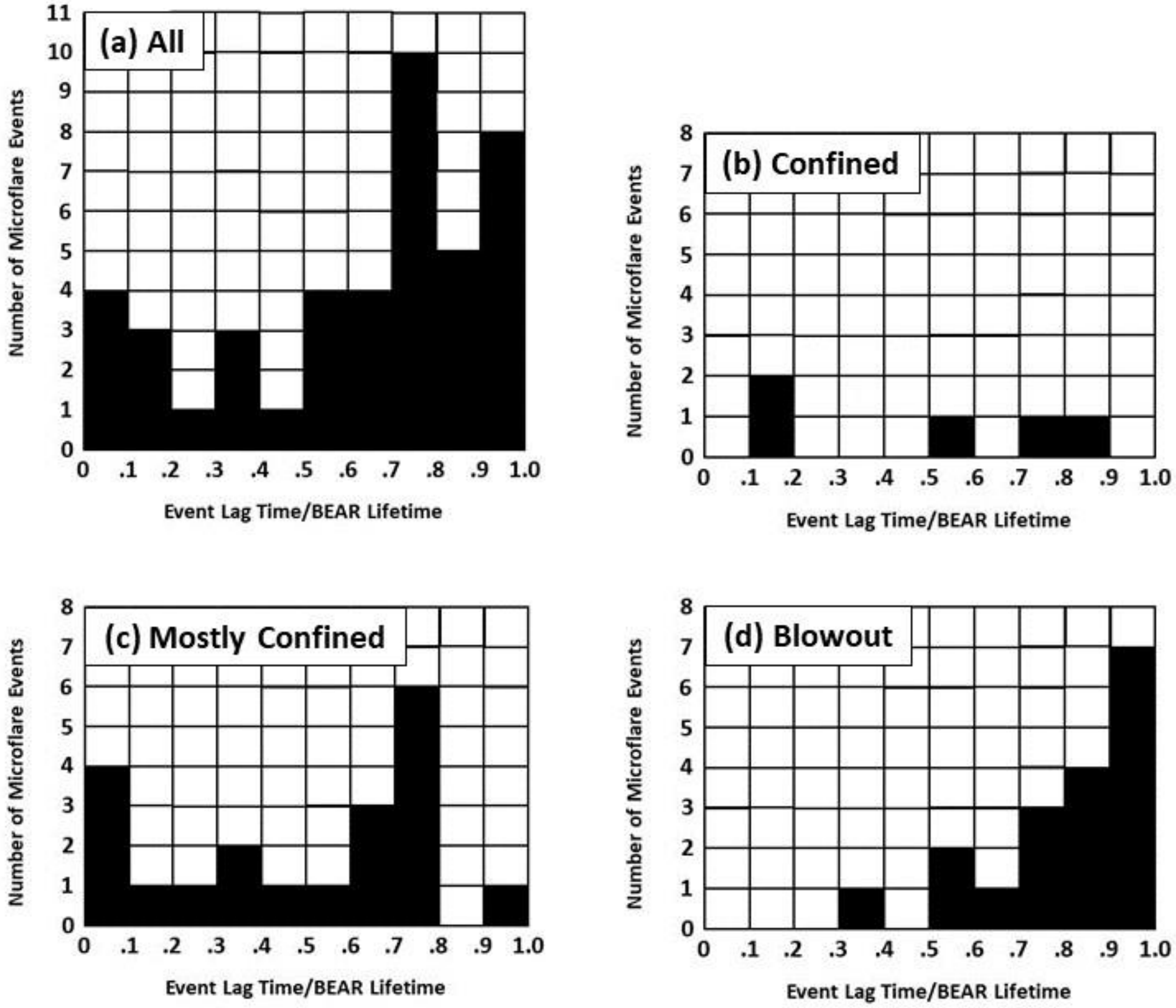

**Figure 23.** Histogram of the ratio of microflare event lag time to BEAR lifetime for (a) all 43 microflare events, (b) the 5 confined-explosion microflare events, (c) the 20 mostly-confined-explosion microflare events, (d) the 18 blowout-explosion microflare events.

Movie Legends

**BR2MOVIE**. Animation for Figure 2, for BEAR BR2. Left panel: AIA 211 Å images of BEAR BR2 and its surroundings. Middle panel: HMI magnetograms of the same field of view as the 211 Å images. Right panel: 40 G field-strength contours of the nearest-in-time magnetogram overlaid on the left panel's 211 Å image; red for positive flux, blue for negative flux. The movie spans the 35 hours spanned by BR2's time-flux plot and time-luminosity plot in Figure 2. The cadence of the 211 Å images is 1 minute. The cadence of the magnetograms is 6 minutes.

**BR7MOVIE.** Animation for Figure 6, for BEAR BR7. The format and cadences are the same as in BR2MOVIE. The movie spans the 16 hours spanned by BR7's time-flux plot and time-luminosity plot in Figure 7.

**BR9MOVIE.** Animation for Figure 13, for BEAR BR9. The format and cadences are the same as in BR2MOVIE. The movie spans the 27 hours spanned by BR9's time-flux plot and time-luminosity in Figure 14.